\documentclass[reprint,superscriptaddress,preprintnumbers,amsmath,amssymb,aps,prd,nofootinbib]{revtex4-2}

\usepackage{placeins} % Float Barrier
\usepackage[normalem]{ulem} % strikeout
\usepackage{graphicx}  % Include figure files
\usepackage{xcolor}
\usepackage{dcolumn}   % Align table columns on decimal point
\usepackage{bm}        % bold math
\usepackage{hyperref}  % add hypertext capabilities
\usepackage{multirow}
\usepackage{siunitx}
\usepackage{ulem}
\usepackage{float}
\usepackage{soul}
\graphicspath{
  {figures/}
}

\soulregister\cite7
\setstcolor{purple}

\def\al{\alpha} 
\def\be{\beta} 
\def\ga{\gamma}
\def\de{\delta}
\def\ep{\epsilon}
\def\ze{\zeta}

\def\th{\theta}

\def\ka{\kappa}
\def\la{\lambda}
\def\rh{\rho}
\def\si{\sigma}
\def\ta{\tau}

\def\ps{\psi}
\def\om{\omega}

\def\La{\Lambda}

\def\pa{\partial}
\def\half{\frac{1}{2}}

\def\bk{{\mathbf{k}}}

\def\bn{{\mathbf{n}}}

\def\br{{\mathbf{r}}}

\def\bv{{\mathbf{v}}}
\def\bx{{\mathbf{x}}}

\def\bX{{\mathbf{X}}}

\def\mcL{{\mathcal L}}

\def\mcP{{\mathcal{P}}}
\def\mcS{{\mathcal S}}

\newcommand{\ben}{\begin{equation}}
\newcommand{\een}{\end{equation}}
\newcommand{\bea}{\begin{eqnarray}}
\newcommand{\eea}{\end{eqnarray}}
\newcommand{\ba}{\begin{array}}
\newcommand{\ea}{\end{array}}
\newcommand{\bit}{\begin{itemize}}
\newcommand{\eit}{\end{itemize}}

\newcommand{\AnaSig}[1]{\check{#1}}

\newcommand{\IniCorLen}{l_{\phi}}
\newcommand{\bew}{\beta_{\rm w}}

\newcommand{\DtauPS}{\Delta\tau_\text{ps}}

\newcommand{\fa}{f_\text{a}} % Axion decay constant
\renewcommand{\Im}[1]{\,\text{Im} #1}

\newcommand{\Jstring}{j}

\newcommand{\Lrest}{\ell_\text{r}}
\newcommand{\Lwind}{\ell_\text{w}}

\newcommand{\ma}{m_\text{ax}} % Axion mass
\newcommand{\ms}{m_\text{s}} % Scalar mass

\newcommand{\mcNax}{\mathcal{N}_\text{ax}}

\newcommand{\nax}{n_\text{ax}}
\newcommand{\nua}{\nu_\text{ax}}

\newcommand{\psiStr}{\psi_\text{str}}
\newcommand{\PSStr}{\mcP_\text{str}}

\renewcommand{\Re}[1]{\text{Re} #1}

\newcommand{\Source}{\mathcal{K}}
\newcommand{\sgn}{\text{sgn}} % Signum

\newcommand{\tStart}{\ensuremath{\tau_{\rm start}}}
\newcommand{\tEnd}{\ensuremath{\tau_{\rm end}}}

\newcommand{\tDiff}{\ensuremath{\tau_{\rm dif}}}
\newcommand{\tcg}{\ensuremath{\tau_{\rm cg}}}
\newcommand{\tRef}{\ensuremath{\tau_{\rm ref}}}
\newcommand{\tFinPS}{\ensuremath{\tau_{\rm fin}}}

\newcommand{\Tc}{T_\text{c}}

\newcommand{\vol}{\mathcal{V}}

\newcommand\vev[1]{\left\langle #1 \right\rangle}

\newcommand{\ws}{w_\text{str}}

\newcommand{\xir}{\xi_{\rm r}}

\newcommand{\xiw}{\xi_{\rm w}}

\newcommand{\zer}{\ze_{\rm r}}
\newcommand{\zew}{\ze_{\rm w}}

\begin{document}

\preprint{}

\title{The spectrum of axions in a scaling string network}

\newcommand{\Oslo}{\affiliation{
Institute of Theoretical Astrophysics,
University of Oslo, P. O. Box 1029 Blindern, N-0315 Oslo,
Norway}}

\newcommand{\Sussex}{\affiliation{
Department of Physics and Astronomy,
University of Sussex, Falmer, Brighton BN1 9QH,
U.K.}}

\newcommand{\HIPetc}{\affiliation{
Department of Physics and Helsinki Institute of Physics,
PL 64, 
FI-00014 University of Helsinki,
Finland
}}

\newcommand{\EHU}{\affiliation{
Department of Physics,
University of the Basque Country UPV/EHU, 
48080 Bilbao,
Spain
}}

\newcommand{\addressEHUQC}{\affiliation{EHU Quantum Center, University of the Basque Country UPV/EHU, Leioa, 48940 Biscay, Spain}}

\newcommand{\Tufts}{\affiliation{
Institute of Cosmology, Department of Physics and Astronomy, 
Tufts University,
Medford, MA 02155,
USA}}

\author{Jos\'e Correia}
\email{joserco@astro.uio.no}
\Oslo
\HIPetc

\author{Mark Hindmarsh}
\email{mark.hindmarsh@helsinki.fi}
\HIPetc
\Sussex

\author{Joanes Lizarraga}
\email{joanes.lizarraga@ehu.eus}
\EHU
\addressEHUQC

\author{Asier~Lopez-Eiguren}
\email{asier.lopez@ehu.eus}
\EHU
\addressEHUQC

\author{Kari Rummukainen}
\email{kari.rummukainen@helsinki.fi}
\HIPetc

\author{Jon Urrestilla}
\email{jon.urrestilla@ehu.eus}
\EHU
\addressEHUQC

\date{\today}

\begin{abstract}
We report on measurements of axion spectra and their unequal time correlators in simulations on $12288^3$ fixed-grids, the largest such measurements to date. We use a novel combination of the fields that allow us to better distinguish the contributions from propagating axions from the contributions from strings, showing that previous measurements of the axion energy spectrum are approximately 30\% derived from the string fields. We conclude that axion emission from a scaling string network is close to scale-invariant ($q \gtrsim 1$ in the standard parametrisation), and that the energy spectrum of sub-horizon modes behaves as $p_\text{ax}\ln( k \tau)$, where $k$ is the comoving wavenumber, $\tau$ the conformal time and $p_\text{ax} \simeq 10$. The number density spectrum evolves towards a single curve for $k\tau \lesssim 10^2$, with higher wavenumber deviations arising from initial conditions and resonant axion production at the string width scale. 
Our results are consistent with scaling framework, in which 
the total number density of axions produced from strings is $\nax=1.66(17) \fa^2 H$, where $\fa$ is the axion decay constant and $H$ the Hubble rate. To interpret these results, we develop a simple analytic model based on an ensemble of string segments. The model reproduces the main features of the measured spectra and unequal-time correlators, and predicts $q \le 1$, leaving $q=1$ as the only value consistent with theory and measurement.  We report on axion production from the final collapse of the network in a  future work.
\\

\end{abstract}

\maketitle

%%%%%%%%%%%%%%%%%%%%%%%%%%%%%%%%%%%%%%%%%%%%%%
%%%%%%%%%%%%%%%%%%%%%%%%%%%%%%%%%%%%%%%%%%%%%%
%%%%%%%%%%%%%%%%%%%%%%%%%%%%%%%%%%%%%%%%%%%%%%
\section{Introduction}
\label{s:Int}
An open problem in cosmology is the fundamental nature of dark matter. A promising candidate which may constitute all or some of the dark matter is the QCD axion \cite{Preskill:1982cy,Abbott:1982af,Dine:1982ah} (and see also \cite{Sikivie:2006ni,Marsh:2015xka,Irastorza:2021tdu,Arza:2026rsl} for reviews). In extensions of the Standard Model which propose to solve the strong CP problem by adding a U(1)-invariant field which undergoes symmetry breaking (the Peccei-Quinn mechanism \cite{Peccei:1977hh,Peccei:1977ur}), the axion arises as angular excitations of this field \cite{Weinberg:1977ma,Wilczek:1977pj}. If the breaking of this symmetry occurs at a post-inflation phase transition, a tangle of one-dimensional objects known as axion cosmic strings \cite{Davis:1986xc,Vilenkin:1982ks,Vilenkin:2000jqa} will form, with at least $\mathcal{O}(1)$ horizon lengths of string formed per horizon volume \cite{Kibble:1976sj,Zurek:1996sj}. 

Throughout the lifetime of this network the strings emit axion radiation, until the temperature of the Universe is comparable to the QCD scale, when the axion becomes massive and domain walls are formed between strings \cite{Dine:1982ah,Sikivie:1982qv}. At this point, the string-wall system collapses into even more axions. The collapse depends on the number of domain walls attached to each string. If this number is $N_\text{dw} = 1$ the strings are drawn together, and the network disappears at the QCD scale. If $N_\text{dw}>1$, explicit breaking of the PQ symmetry must be introduced to prevent the energy density of the universe becoming dominated by the string-wall system \cite{Sikivie:1982qv,Zeldovich:1974uw}. The number of domain walls attached to a string is a model-dependent quantity depending on the PQ charges of the fermions (see e.g. \cite{DiLuzio:2020wdo,DiLuzio:2024xnt} for reviews of models and their associated domain wall number). 

The relic abundance of axions (and therefore the predicted axion mass) is thus 
connected with the long-term evolution of the network and the mechanisms of radiative energy loss from strings.  The study of network evolution is a problem in non-linear classical field theory, necessitating numerical simulations. However, simulations cannot span the entire range of scales between the radius of the string (set by the PQ symmetry breaking scale $\fa$, which is around $10^{11}$ GeV in dark matter axion models) and the size of the horizon at the QCD transition at around $100\,$MeV. 
This means that we must extrapolate the late-time behaviour of relevant quantities extracted from simulations, and have a framework of physical understanding in which to carry out the extrapolation. 
Precise estimates of axion mass are of crucial importance for resonant detectors (examples include \cite{IAXO:2019mpb,ADMX:2024xbv,ADMX:2025vom,DMRadio:2022pkf,HAYSTAC:2023cam,FLASH:2023qed,Melcon:2018dba,ALPHA:2022rxj,McAllister:2017lkb}).

We highlight two quantities which are sources of controversy in literature: the string density parameter $\zeta$ (proportional to the number of horizon lengths of string per horizon volume) and the axion emission spectrum, which is often characterised by the index $q$ of a fit to a power law on scales between the horizon and the string width.
In a previous publication \cite{Correia:2024cpk}, we studied the late-time behaviour of the global string length density parameter $\zeta$ and the root-mean-square (RMS) velocity $v$ in the framework of scaling \cite{VilShe94,Kibble:1976sj} which is well-established for local and semilocal string networks in cosmology \cite{Lopez-Eiguren:2017ucu,Kanda:2025hgi,Correia:2022spe,Lizarraga:2016hpd,Urrestilla:2007sf,Daverio:2015nva,Lizarraga:2016onn,Blanco-Pillado:2011egf,Ringeval:2005kr,Correia:2021tok}. 

There we found more evidence for a convergence of the string density parameter to an $\mathcal{O}(1)$ constant, in agreement with scaling, and with predictions of previous work \cite{Yamaguchi:1998gx,Yamaguchi:2002sh,Hiramatsu:2012gg,Fleury:2015aca,Lopez-Eiguren:2017dmc,Hindmarsh:2019csc}. We measured it to greater accuracy than before, obtaining $\ze_{\text{r},*}=1.491(93)$ for the rest-frame length density parameter. We also investigated a local estimator of the root mean square (RMS) velocity and found it to asymptote to a constant value of $\bar{v}_*=0.5705(93)$.
We further remarked that claims of a long-term evolution in $\zeta$  
\cite{Kawasaki:2018bzv,Gorghetto:2020qws,Buschmann:2021sdq,Kim:2024wku,Saikawa:2024bta,Benabou:2024msj,Kim:2024wku} 
are made in the context of low-density string networks, and are compatible with an approach from below to our estimated string density. 
The string density parameter and RMS velocity of the simulations we present here are consistent with the previous measurement,  but none of the results we present here depend on the results of the previous paper.

In this paper we turn our attention to the spectra of axions emitted by the string network. Historically, there are two proposals for the shape of the axion emission power spectrum $\mcS_\text{ax}$ (defined in Eq.~\eqref{e:SaxDef}) and the associated spectral index $\mcS_\text{ax}(k)\propto k^{1-q}$ (where $k$ is the wavenumber). In one scenario \cite{Harari:1987ht,Chang:1998tb,Hagmann:1990mj,Hagmann:1998me,Hagmann:2000ja}, long strings straighten and loops collapse in one oscillation, which gives rise to a scale-invariant radiative spectrum ($q=1$). In the other \cite{Davis:1986xc,Davis:1988rw,Davis:1989nj,Shellard:1998mi,Battye:1994au}, the typical wavelength of radiated axions is given roughly by the curvature radius of strings, and these radiate axions over multiple oscillations, yielding as a result a power law power spectrum peaked around horizon scales with $q>1$. In the latter, the abundance of low-momenta axions is enhanced with respect to the first scenario, and this leads to a higher number density prediction from a string network \cite{Davis:1988rw,Davis:1989nj,Battye:1993jv,Battye:1998mj}. 

Recent simulations of single loop configurations in flat space 
\cite{Baeza-Ballesteros:2023say,Saurabh:2020pqe} show rapidly evaporating loops and a consistency with $q=1$. On the other hand, simulations of standing waves also in flat space \cite{Battye:1998mj,Drew:2019mzc,Drew:2023ptp}, show strings emitting in harmonics of a fundamental oscillation frequency, which points towards $q>1$. A string network 
is composed of an ensemble of long strings and loops, and thus the axion radiation spectra could be thought of as receiving contributions from both long strings and loops.  In this picture it is natural to expect that there will be a peak at the horizon scale from long strings and a range at higher wavenumber with $q=1$ from loops, with the transition wavenumber depending on the relative importance of the two contributions. 
Evolution of the network towards scaling during which the relative importance changes 
would register as an evolution of the spectral index $q$, thereby contributing a systematic error.

The first network simulations to measure the axion emission spectra from strings \cite{Yamaguchi:1998gx,Yamaguchi:1999yp,Yamaguchi:2002sh,Hiramatsu:2010yu,Hiramatsu:2012gg} found that the power spectra towards the end of the simulation peaked 
at horizon scales, and this was interpreted as consistent with $q \geq 1$.
Subsequently, \cite{Gorghetto:2018myk,Gorghetto:2020qws}
performed fits 
of a power law to the spectra, which showed the spectral index approaching unity from below, 
increasing approximately in proportion to the logarithm of cosmic time $t$.
This time evolution was then extrapolated all the way to QCD phase transition, leading to a prediction of $q > 1$. 

Other groups have since then attempted fits to a power spectrum \cite{Buschmann:2021sdq,Kim:2024wku,Saikawa:2024bta} and subsequent time extrapolations, with varying results ($q=1$ \cite{Buschmann:2021sdq}, $q\geq1$ \cite{Kim:2024wku} or consistent with both \cite{Saikawa:2024bta}), which showed that the time slope of $q$ remained relatively small ($dq/d\ln(t) \sim 10^{-1}$). There are several systematic effects on the estimate of this slope, including lattice discretisation (see \cite{Saikawa:2024bta} for an exploration).
A thorough investigation of a very large simulation with an adaptive mesh code \cite{Benabou:2024msj}, which should avoid systematic effects from the lattice spacing, gives results close to $q=1$. The significance of the slow time evolution of the index is then unclear, both from the point of view of not having a physical model to explain it and in terms of multiple systematic errors obscuring its behaviour. 

In this paper we present the largest measurements to date ($12288^3$ grid points) of axion power spectra in string network simulations. The spectra are obtained from scalar and vector components of a current which can be considered as the derivative of the axion field. The currents are well-defined everywhere and include all the energy-momentum of the axion. The vector current is sourced only by strings, while the scalar currents (timelike and longitudinal) contain contributions from both propagating axions and strings. 
We also investigate for the first time the time correlations of the currents by constructing their unequal time correlators (UETCs), which helps to distinguish between the contributions of the strings and the propagating axion field.

We focus on the scaling properties of the spectra, from which the number density power spectrum can be calculated. 
We test the predictions of the scaling framework, and present an accurate estimate for the number density of axions produced by a scaling string network.

To support our interpretation of the results and our prediction for the late-time scaling behaviour of the spectra, we develop a physical model for the axion field spectra produced by an ensemble of randomly-moving unconnected string segments,  the unconnected segment model (USM) \cite{Vincent:1996qr,Albrecht:1997mz,Pogosian:1999np,Avgoustidis:2012gb,Charnock:2016nzm}.  The model predicts that the vector power spectrum of a scaling string network is itself scale-invariant, that is independent of wavenumber for wavenumbers between the horizon scale and the string width scale. It also predicts that the longitudinal current contains the smallest contribution from the strings. The axion spectrum is predicted to behave as $\mathcal{P}_\text{ax} \propto \ln(k\tau)$, where $k$ is wavenumber and $\tau$ is conformal time, with a coefficient calculable from the string density. 
The spectral index of the axion emission spectrum is bounded by $q \le 1$.
The UETC of the vector current at wavenumber $k$ is predicted to decrease rapidly for time lags greater than $k/v$, where $v$ is the RMS velocity of the strings.

We find that the current power spectra are generally consistent with the expectation of scaling, except in simulations which start with the lowest string density.  That is, when quantities with dimensions of length are appropriately scaled with the horizon distance, power spectra appear to collapse onto a single curve over a wide range of wavenumbers at the horizon scale and beyond, for most initial string densities. The scaling axion energy spectrum is consistent with the predicted logarithmic form, with a coefficient of proportionality roughly $10$, which is roughly half the predicted value. 
The vector UETC fits the USM prediction very well, albeit with a slightly lower RMS velocity than measured from averaged field quantities.
With additional insights from the model we are also able to determine that the universally-used measure of axion number density based on the timelike current is contaminated at a level of approximately 30\% by the non-propagating field carried by the strings (see also Refs.~\cite{Drew:2019mzc,Battye:2026whd}).

The emission spectra are broadly scale-invariant between the horizon scale and the string width scale, but when fitting to a power law we also see a growth towards $q \gtrsim 1$ from below. Combined with the analytical modelling, this indicates that the spectrum is evolving towards $q=1$, and that values above unity are a result of systematic error in a noisy function. The evolution in $q$ is not reflected in an evolution in the overall number density, a much less noisy quantity than the emission spectrum. 
At higher frequencies, the emission spectra of networks with higher initial string densities show a pronounced peak at frequencies corresponding to half the scalar mass, which we ascribe to resonant production from excitations of the string core  \cite{Goodband:1995rt,Blanco-Pillado:2021jad}.

The number density spectra derived from the least contaminated longitudinal current show good scaling up to about $k\ta \lesssim 10^2$. From these we find that the total axion number density is $\nax = 1.66(17) \fa^2 H$, where $\fa$ is the PQ symmetry breaking scale and $H$ the Hubble parameter. This value is estimated from the number density at the end of the simulations, and arises from a set of simulations with initial number densities varying by a factor of two, bracketing the final value. 
We see no evidence for long-term growth of the number density: the measurement is therefore consistent with the scaling framework.

The paper is partitioned as follows: In section \ref{s:ModObs} we summarize the axion model to be simulated and define all of the observables to be studied, in section \ref{s:TheExp} we describe the analytical expectations for the behaviour of the spectrum of radiated axions in a scaling network, we then summarize in our simulation setup and methodology in section \ref{s:SimMet} and in sections \ref{s:Res} and \ref{s:NumDen} we  present and discuss our results. The conclusion (section \ref{s:Con})  summarizes the article and outlines next steps and possible improvements. In a future publication we will report on simulations including the annihilation by domain walls.

Scripts and data needed to generate all figures and tables derived from the simulations can be found at the following location \cite{correia_2026_18351085}.

%%%%%%%%%%%%%%%%%%%%%%%%%%%%%%%%%%%%%%%%%%%%%%
%%%%%%%%%%%%%%%%%%%%%%%%%%%%%%%%%%%%%%%%%%%%%%
%%%%%%%%%%%%%%%%%%%%%%%%%%%%%%%%%%%%%%%%%%%%%%
\section{Axion model and observables}
\label{s:ModObs}

\subsection{Langrangian and field equation}
The axion model we study consists of a complex scalar field $\Phi$ with a global U(1) symmetry, whose dynamics will be given by the following Lagrangian density
\ben
\mcL=  \half \partial_\mu \Phi^* \partial^\mu \Phi - V(\Phi) \,,
\label{eq:lagrangian}
\een
where 
\ben
V(\Phi) = \frac{\la}{4}(|\Phi|^2 - \fa^2)^2.
\label{eq:potential}
\een
With this normalisation of the kinetic term, the vacuum expectation value of the field $\fa$ is the axion decay constant when the field is coupled to fermions, and the fermions are coupled to a non-Abelian gauge field with an anomalous set of charges. The anomaly term in the effective action breaks the U(1) symmetry and lifts the degeneracy of the energy density in the phase of $\Phi$, at a scale of order the confinement scale of the gauge theory. In this paper we are concerned only with the evolution of the string network deep in the radiation era and well above the confinement scale, when the effect of the gauge field can be neglected.

We study the dynamics in a flat Friedmann-Lema\^itre-Robertson-Walker (FLRW) cosmology, with metric 
\ben
g_{\mu\nu} = a^2(\tau) \eta_{\mu\nu},
\label{e:FLRWg}
\een
where $a$ is the scale factor and $\tau$ is the conformal time. In the radiation era $a \propto \tau$. 
The equations of motion of the system are:
\ben
\ddot{\Phi}+2\frac{\dot{a}}{a}{\dot\Phi}-\nabla^2 \Phi + a^2 \lambda (|\Phi|^2-\fa^2)\Phi=0,
\label{eq:eom}
\een
where the dot indicates differentiation with respect to conformal time.

The U(1) symmetry of the Lagrangian (\ref{eq:lagrangian}) is spontaneously broken at a phase transition with critical temperature $\Tc \simeq \fa$, when the field acquires an expectation value.  We study the post-inflationary scenario, in which the phase transition happens after inflation and reheating.

At temperatures well below the critical temperature, the magnitude of the field approaches $\fa$, 
with a massless pseudoscalar fluctuation mode, the axion, and a scalar mode with mass $\ms = \sqrt{2\la}\fa$, the saxion.

During the process of acquiring the vacuum expectation value the direction of the field in field space is chosen at random in uncorrelated regions of the universe.  However, by continuity of the field, there are always lines along which the field remains zero, 
\cite{Kibble:1976sj}, which form the cores of string defects, which have a width of order $\ws \equiv 1/a(\tau)\ms$.
The strings evolve as relativistic line objects, emitting both scalar and pseudoscalar Nambu-Goldstone field modes. 
Axions emitted by decaying strings are a well-motivated dark matter candidate \cite{Davis:1986xc}. 

\subsection{Observables: the saxion and the axion current}

In order to capture the dynamics of the saxion and axion fields separately in a more clear way, we will construct combinations of the fields which represent them directly.  We define a field of unit modulus
\ben
\hat \Phi = \Phi /|\Phi|,
\een
which we use to define  the operator $\hat\Phi^*\pa_\mu\Phi$. We will then project out the different components of that operator.

The saxion field is conventionally defined as the modulus $\psi\equiv|\Phi|$. 
Considering the real part of the operator, 
\bea
\text{Re}\left(\hat\Phi^*\partial_\mu\Phi\right) &=& \pa_\mu\psi.
 \eea
We will also define the canonical momentum of the saxion field $\Pi = \pa_0\psi$, the derivative of the saxion field with respect to conformal time.
 
Likewise, we can consider the imaginary part and define the vector field 
\ben
J_\mu =  \text{Im}\left(\hat\Phi^*\partial_\mu\Phi\right).
\een 
Writing the field as $\Phi = \psi e^{i\alpha}$, we see that $J_\mu = \psi \pa_\mu \alpha$, and in the case of a field with constant magnitude $\fa$ it is the gradient of a canonically normalised scalar $A = \fa \alpha$.  This is the conventional perturbative definition of the axion field.  
Hence $J_0$ is equivalent to the $k=1$ masking procedure of Ref.~\cite{Saikawa:2024bta}, which multiplies $\dot{A}$ with a $(|\Phi|/\fa)^k$.

Using these new fields 
the equations of motion are
\bea
\label{e:ScaEOM}
\half\pa^\mu \pa_\mu \psi &=&   \frac{1}{\psi} J_\mu  J^\mu  - \la (\psi^2 - \fa^2)\psi , \\
\pa^\mu(\psi  J_\mu ) &=&  0 .
\label{e:CurCon}
\eea
The combination $\psi J_\mu$ is a conserved current, with an associated invariant
\ben
N_a = 
\int d^3 x \, \psi J_0 \,.
\een
In the limit that field gradients become much smaller than $\fa$, the equations decouple, and we can write
\bea
\pa^\mu \pa_\mu \psi & \simeq &     - \la (\psi^2 - \fa^2)\psi , \\
\pa^\mu J_\mu & \simeq & 0 .
\eea
In this limit, the quantity $\fa J_\mu$ is a conserved axion number current, with $\fa J_0$ the axion number density.  
The Lagrangian density in terms of the new fields is 
\ben
\mcL = \half J_\mu J^\mu + \half \partial_\mu \psi \partial^\mu \psi - \frac{1}{4} \la (\psi^2 - \fa^2)^2, 
\een
from which it follows that 
the energy-momentum tensor is
\bea
T_{\mu\nu} &=& \pa_\mu \psi \pa_\nu \psi + 
 J_\mu J_\nu - g_{\mu\nu} \mcL.
\eea
Hence by studying $\psi$ and $J_\mu$ we include all contributions to the energy-momentum of the system.

\subsection{Power spectra and correlators}

We will be interested in Fourier transforms of the fields, and we use the conventions that the Fourier transform of a field $f(\bx, t)$ in volume $\vol$ is 
\ben
\tilde{f}(\bk, t) = \int_\vol {\rm d}^3 x\, f(\bx,t)e^{-i \bk\cdot \bx}\, .
\een

We project out the different polarisations of the Fourier transform of the spatial part of the current $\tilde{J}_i(\bk)$ with an orthonormal basis $e_{s}^i$, $e_{A}^i$ ($A=\pm$), where $e_{s}^i = \hat{k}^i$, and the index indicates the eigenvalue under the generator of rotations about $\bk$ or
\ben
i \ep^{ijk} e_{s}^j e_{\pm}^k = \pm e_{\pm}^i .
\label{e:basis}
\een
This way, we obtain one scalar quantity
 \ben
\tilde{J}_\text{s}(\bk,t) = e_{s}^i\tilde{J}_i(\bk,t)\,,
\label{proj1}
\een 
and  vector quantities with two transverse components
\ben
 \tilde{J}_A(\bk,t) = e_{A}^i\tilde{J}_i(\bk,t) \,.
 \label{proj2}†
 \een
The advantage of these projections of $\tilde{J}_i$  is that they more clearly separate the field components which make up the axion $\tilde{J}_{s}(\bk,t)$ and those which make up the string $\tilde{J}_A(\bk,t)$. 
Other groups focus only on $\tilde{J}_0$, which contains contributions from both pseudoscalar modes and strings in unknown proportion.  With our projection technique we are able to measure the relative proportion from the correlation functions of the projected field variables.

Collecting together the quantities of interest, we have two scalar quantities 
$\tilde{J}_0(\bk,t)$ and $\tilde{J}_s(\bk,t)$, 
and two vector quantities
$\tilde{J}_+(\bk,t)$,  $\tilde{J}_-(\bk,t)$. 
They can be combined into scalar and vector unequal time correlation functions (UETCs) given by
\bea
U_{ab}(k, \tau_1, \tau_2) &=& \vol^{-1} \vev{\tilde{J}_{a}^*(\bk,\tau_1) \tilde{J}_{b}(\bk,\tau_2) }_{\hat\bk} , \quad \\
U_{AB}(k, \tau_1, \tau_2) &=& \vol^{-1} \vev{\tilde{J}_{A}^*(\bk,\tau_1) \tilde{J}_{B}(\bk,\tau_2) }_{\hat\bk} . \quad
\label{UETC}
\eea
where $a \in\{0,s\}$ and $A \in \{+,-\}$. Whenever we wish to refer to all components in the polarisation basis we use the indices $\al$, $\be$.
The angle bracket notation in the UETC definition indicates averaging over directions, as we assume that the system follows the isotropy of the FLRW background. Recall that  vector correlators are non-zero for configurations with strings.  

The diagonal entries at equal time define equal time correlators (ETCs) which we denote
\ben
P_\alpha(k, \tau) =  U_{\alpha\alpha}(k, \tau, \tau),
\label{ETC}
\een
where $\alpha$ stands for $a$ (scalar quantities) or $A$ (vector quantities). In other contexts we refer to $P_\al$ as spectral densities. 
From them we define power spectra 
\ben
\mcP_\alpha(k, \tau) = \frac{k^3}{2\pi^2} P_{\alpha}(k, \tau). 
\een
From Parseval's theorem, power spectra have the property
\ben
\int \frac{dk}{k} \mcP_\al(k, \tau) = \langle [J_\al(\bx, \tau)]^2\rangle_\bx
\een
where the angle brackets here denote a volume average. 

The power spectra and UETCs have some symmetry properties which follow from the isotropy and the fact that the fields take values in the real numbers. 
The scalar fields then obey $\tilde{J}^{*}_\mu(\bk, \tau) =  \tilde{J}_{\mu}({-\bk}, \tau)$, so we expect the power spectra and  UETCs to be real after the angle averaging. 

Regarding the vector correlators, note that  under a rotation of $\pi/2$ about $\bk$,  the projectors change  $e_{\pm}^i \to e^{\pm i\pi/2}  e_{\pm}^i$. Hence, statistical isotropy and parity implies that 
\ben
U_{AB}(k, \tau_1,\tau_2) = \delta_{AB} U_\pm (k, \tau_1,\tau_2).
\een
In other words, there is only one independent vector UETC,  $U_\pm = (U_+ + U_-)/2$ which is the average of the two polarisations.

From the UETCs, we define decoherence functions, which measure the time correlations of the currents.  The decoherence function of a field component $\alpha$ is  
\ben
D_\alpha(k, \tau_1, \tau_2) = \frac{ U_{\alpha\alpha}(k, \tau_1, \tau_2) }{\sqrt{P_\alpha(k,\tau_1) P_\alpha(k,\tau_2)} }.
\label{e:DecFunDef}
\een
We will write $D_\pm$ for the decoherence function of $U_\pm$.

\subsection{Observables: length density and velocity}

\begin{figure}[htbp]
    \centering
    \includegraphics[width=0.48\textwidth]{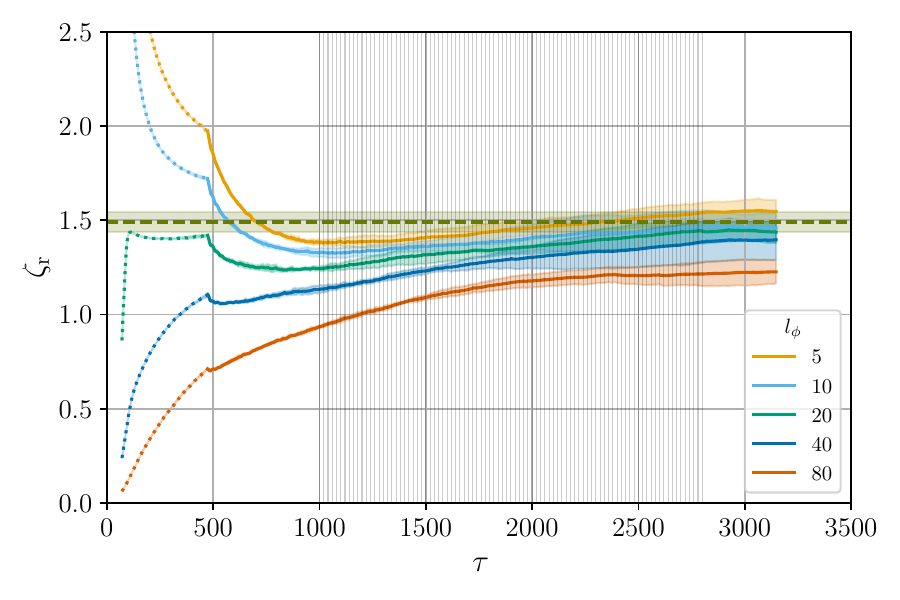}
    \includegraphics[width=0.48\textwidth]{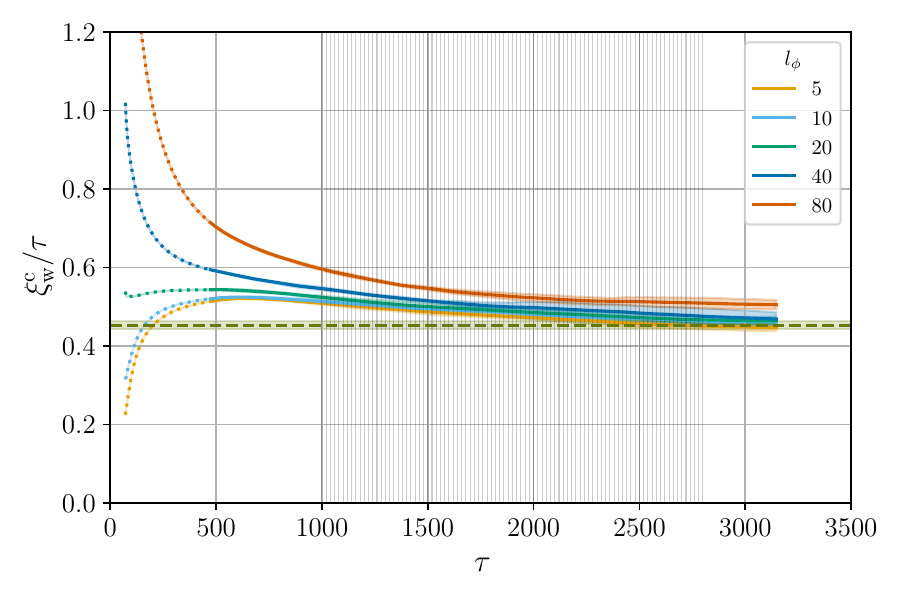}
    \includegraphics[width=0.48\textwidth]{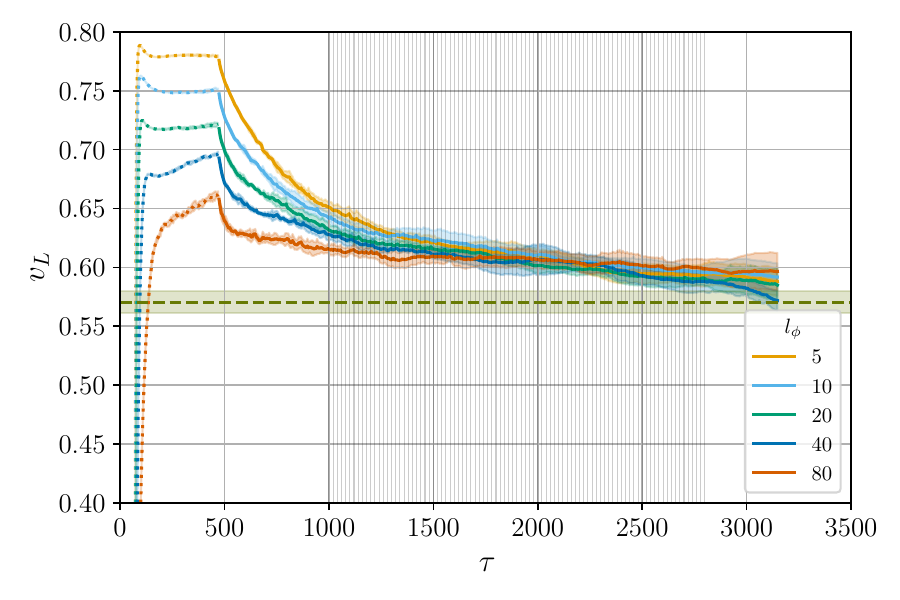}
     \caption{Top: string local rest-frame length density parameter, against conformal time in code units.
     Middle: ratio of the comoving universe-frame mean string separation $\xi^\text{c}_\text{w}$ to comoving horizon length $\tau$.
     Bottom: RMS velocity. 
     Times between the end of diffusive time evolution $\tDiff$ and the start of the physical evolution $\tcg$ are plotted with dotted lines. 
     The fixed point values obtained in Ref.~\cite{Correia:2024cpk}  are shown as the horizontal olive dashed line with 1$\si$ uncertainty denoted by the band. The times in which power spectra and UETCs are calculated are shown as grey vertical lines.  The legend gives the initial field correlation length in code units, in which the string width at the initial time is $\ws = 0.5$.
          \label{f:zetaw}}
 \end{figure}

Finally, we recall the principal average quantities of the string network.  For a complete discussion see Ref.~\cite{Correia:2024cpk}.
The (comoving) length of string can be measured in the local rest frame of the string, denoted $\Lrest$, and in the universe frame, which we denote $\Lwind$. From them we can define dimensionless length density parameters
\ben
\zer \equiv \frac{\Lrest \tau^2}{4 \vol}, \quad
\zew \equiv \frac{\Lwind \tau^2}{4 \vol},
\een
where the factor 4 is conventional. They are related by a Lorentz boost with the local velocity \cite{Correia:2024cpk}. Equivalently, one can define comoving mean string separations 
\ben
\xir^\text{c} \equiv \sqrt{\frac{\vol}{\Lrest}}, \quad
\xiw^\text{c} \equiv \sqrt{\frac{\vol}{\Lwind}}.
\een
The rest-frame length is a measure of the total energy in string, including kinetic, while the universe frame length measures only the apparent length.  Hence $\zer$ keeps track of the invariant length of string in a causal volume, while the universe frame comoving separation $\xiw^\text{c}$ controls the length scale of axion radiation.
The universe frame length is computed from the number of plaquettes pierced by string, and 
the rest frame length from the total energy and Lagrangian of the strings, extracted from averages of the  fields weighted to select string cores.
From the energy and Lagrangian one can also calculate the RMS velocity of the network, which we denote $v_L$.

The evolution of the rest-frame length density parameter, the ratio of the comoving mean string separation to the horizon, and the RMS velocity shown in Ref.~\cite{Correia:2024cpk} from the simulations discussed in this paper are shown in Fig.~\ref{f:zetaw}, with the values during the core growth era with conformal evolution shown in dotted lines. The simulations start from a wide range of initial length densities,  labelled by the initial field correlation length $l_\phi$, given in code units, in which the comoving string width $\ws = 0.5$.
One can see the general tendency of the length density parameter and the RMS velocity 
to approach a fixed point.  The asymptotes are consistent with the fixed point values obtained in $16384^3$ simulations \cite{Correia:2024cpk}, denoted by horizontal dashed lines and uncertainty bands.

We can get approximate figure for numerical constants given above from 
Fig.~\ref{f:zetaw}), which shows that  $\beta_\text{w} \equiv \xiw/\tau$ tends to a constant around $0.45$ 
The final values of $\zew$ are $\mathcal{O}(1)$, and the RMS velocity $\bar{v}$ around $0.6$. Hence we expect to obtain estimates for $r_\pm$ \eqref{e:USMrpm} in the range 10 -- 20.  We give more precise estimates from the simulation data in Table \ref{t:SpeFitPars}. 

Evaluating the parameter from the equations \eqref{e:AxStrTra}, \eqref{e:AxStr0} and \eqref{e:AxStrs}, and 
making the  assumption that the velocity distribution is strongly peaked around its RMS value, so that $\vev{\gamma^2 v^2}_v \simeq 0.5$, $\vev{(\gamma-1)^2}_v  \simeq 0.05$, we find
\ben
\frac{r_{0s}}{r_\pm} \simeq 0.39.
\label{lau}
\een

\section{Theoretical expectations}
\label{s:TheExp}

Our overarching theoretical expectation is that the string configurations in the simulations approach scaling as the network evolves.
When scaling, networks of strings have a self-similar behaviour, which means that all quantities with dimensions of length are proportional to time, without converting mass dimensions into length using Planck's constant.  
For example, a current has dimensions $M/ L$, and so its unequal time correlator has dimensions $M^2 L$, and its power spectrum  has dimensions $M^2/L^2$.  Therefore we expect a power spectrum of a current in a scaling network to behave as
\ben
\mcP_\al(k,\tau) = \frac{\fa^2}{\tau^2} \tilde\mcP_\al(k \tau) .
\label{e:PtilDef}
\een
Hence, when plotting power spectra, we will always plot them multiplied by $\tau^2/\fa^2$. The sign of a scaling network will be that power spectra at different times   collapse onto a single curve, when plotted against $k\tau$.

A second expectation is that the field configurations can be understood as a combination of strings and propagating axions. 
In the next subsections we construct some predictions for the power spectra based on this expectation.

\subsection{Field spectra of axion strings}

A model of a string network which has been successfully used for CMB perturbations from gauge strings is the Unconnected Segment Model (USM)
\cite{Vincent:1996qr,Albrecht:1997mz,Pogosian:1999np,Avgoustidis:2012gb,Charnock:2016nzm} which consists of string segments of comoving length $L$ with random positions and velocities,  comoving density $n_s$ and RMS velocity $\bar{v}$. In the CMB calculations, the segments are sources for metric perturbations.  
In the context of an axion string network, we can treat the segments as sources of an axion field.

The USM models strings much longer than the horizon, while neglecting the distribution of collapsing string loops which carry off much of the long strings' energy. The total length of string is dominated by long strings (see e.g. Fig.~4 of Ref.~\cite{Gorghetto:2018myk} and Fig.~S3 of Ref.~\cite{Buschmann:2021sdq}), as the lifetime of loops is less the one oscillation period \cite{Saurabh:2020pqe,Baeza-Ballesteros:2023say}. Hence neglecting loops in order to estimate field correlation functions is a reasonable first approximation.

In Appendix \ref{s:PowSpeStr} we perform calculations in the USM model of the axion field power spectra, obtaining
\bea
 \mcP_\pm(k, \tau) & \simeq & \half\left(1 + \half \vev{\ga^2v^2}_v\right) \PSStr(k, t), \label{e:AxStrTra}\\
 \mcP_0(k, \tau) & = & \half \vev{\ga^2v^2 }_v  \PSStr(k, t) , \label{e:AxStr0}\\
  \mcP_s(k, \tau)& = & \frac{1}{8} \vev{(\ga - 1)^2}_v \PSStr^s(k, t), \label{e:AxStrs}
 \eea
 where the function $\PSStr$ has asymptotic behavior
\ben
\PSStr(k, \tau) \simeq  2\pi {\fa^2}n_s L,  \quad 1 \ll kL \ll L/\ws, \label{e:AxStrForm}
\een
and $\PSStr^s =  \PSStr\left( 1 + \text{O}(kL)^{-2} \right)$ in this limit. Note that $L$ is the length of the string segments measured in the universe frame, the frame in which Fourier transforms are taken, and thus we take $L = \xiw$, the mean string separation, as an estimate.  In scaling, $\beta_\text{w} \equiv \xiw/\tau$ tends to a constant.
The comoving number density in scaling is proportional to $\tau^{-3}$, and is related to the length density through $n_s L = 4 \zew/\tau^2$. 

These are our expectations for the form of the axion field created by a network of strings.  We see that the dominant contribution of strings is to $\mcP_\pm$, with 
an $\mathcal{O}(v^2)$ contribution to $\mcP_0$ and an $\mathcal{O}(v^4)$ contribution to $\mcP_s$. 

In the wavenumber range $1 \ll k\xiw \ll \xiw/\ws $ we write 
\ben
 \mcP_\pm(k,\tau) =  r_\pm \frac{\fa^2}{\tau^2} ,
 \label{e:StrModPowSpe2}
\een
from which 
we have
\ben
r_\pm \simeq  \left(1 + \half \vev{\ga^2v^2}_v\right) 4\pi \zew.
\label{e:USMrpm}
\een

For future reference we also define constants $r_0$, $r_s$ parametrising the string contribution to $\mcP_0$ and $\mcP_s$, through
\ben
\mcP_a = r_{a}\frac{\fa^2}{\tau^2}.
 \label{e:StrModPowSpe0s}
\een
with $a \in \{0,s\}$ and $1 \ll k\xiw \ll \xiw/\ws $.

One can also derive a prediction for the decoherence functions of the currents \eqref{e:DecFunDef} produced by moving strings.  This lengthy calculation is given in Appendix \ref{s:ModUETC}.  The result is that the transverse decoherence function $D_\pm(k, \tau_1, \tau_2)$ depends on the dimensionless variables $x_1 = k\tau_1$ and $x_2 = k\tau_2$, and tends to zero for $|x_1 - x_2| \gtrsim \pi \bar{v}$. The shape of the function is sensitive to the velocity distribution, for which we use Ref.~\cite{Gorghetto:2020qws}, where it was found that the Lorentz $\ga$ was approximately distributed as a power law. The decoherence function also depends on the parameter $\be_\text{w}$.

\subsection{Freely propagating axions}

The equation for a freely propagating axion can be deduced from the current conservation equation \eqref{e:CurCon} in Fourier space, 
\ben
\dot{\tilde{J}}_0 + 2 \frac{\dot a}{a} \tilde{J}_0 - i k \tilde{J}_s = \tilde{\si}(\bk, \tau),
\label{e:J0dot}
\een
where 
\ben
\si(\bx,\tau) =   \frac{\pa^\mu \psi}{\psi} J_\mu\,.
\een
The equation for $\tilde{J}_\mu$ concerns only the scalar parts $\tilde{J}_0$ and $\tilde{J}_s$. We may write $\tilde{J}_s$ in terms of a scalar field,  $\tilde{J}_s(\bk,\tau) = i k \tilde{A}(\bk,\tau)$.  If the current is small, $A$ is the argument of the complex field (up to a constant of integration). 
Assuming that $\tilde{J}_0$ is generated only by the scalar part of $\tilde{J}_i$, we search for solutions where $\tilde{J}_0 = \dot {\tilde{A}}(\bk,\tau)$, for which 
\ben
\ddot{\tilde{A}}(\bk,\tau) + 2 \frac{\dot a}{a} \dot{\tilde{A}}(\bk,\tau) + k^2 \tilde{A}(\bk,\tau) = \tilde{\si}(\bk, \tau).
\label{e:AxEqn}
\een
In situations where the scalar field is close to its ground state value $\psi = \fa$, we may take $ \tilde{\si}(\bk, \tau) = 0$, and the general solution of the homogenous  equation, in radiation era, is 
\ben
 \tilde{A}(\bk,\tau) = c^{(1)}_\bk h_0^{(1)}(k\tau) +  c^{(2)}_\bk h_0^{(2)}(k\tau),
 \label{e:AxFieSol} 
\een 
where the $h_0^{(1)} $and $ h_0^{(2)}$ are the spherical Hankel functions of order 0. 
The fact that the currents are real means that $\tilde{J}_s^*(\bk,\ta) = - \tilde{J}_s(-\bk,\ta)$, and hence $ \tilde{A}^*(\bk,\tau) =  \tilde{A}(-\bk,\tau)$.
Then we must have 
\ben
 c^{(1)*}_\bk =  c^{(2)}_{-\bk}\,.
\een
In Appendix \ref{appspectra} we show that for a stochastic axion field we expect to see at wavenumbers $k\tau \gg 1$  a spectral density 
\ben
P_s(k,\tau) =  \frac{2}{\tau^2}\vev{|c^{(1)}_\bk|^2}_{\hat{\bk}}\,,
\een
and a decoherence function
\ben
D_s(k,\tau_1,\tau_2) = \cos(x_1 - x_2),
\een
where $x = k\tau$. 
Similar considerations lead to 
\ben
P_0(k,\tau) =  \frac{2}{\tau^2}\vev{|c^{(1)}_\bk|^2}_{\hat{\bk}} \left(1 + \frac{1}{(k\tau)^2} \right)\,,
\een
and
\bea
D_0(k,\tau_1,\tau_2) &=& \frac{ \cos(x_1 - x_2) ( 1 + x_1 x_2) }{\sqrt{( 1 + x_1^2)( 1 + x_2^2) }}  \nonumber\\
&+& \frac{\sin(x_1 - x_2) (x_1 - x_2)}{\sqrt{( 1 + x_1^2)( 1 + x_2^2) }} \,.
\eea
For $k\tau \gg 1$, both decoherence functions tend to $\cos[k(\tau_1 - \tau_2)] $.

Further solutions can be obtained by first noting that
\ben
\pa_\mu(\psi J_\nu) - \pa_\nu(\psi J_\mu)  = 2\Im (\pa_\mu \phi^* \pa_\nu \phi )\,,
\een
which we may write as 
\ben
\pa_\mu J_\nu - \pa_\nu J_\mu  = \Source_{\mu\nu},
\een
where
\bea
\Source_{\mu\nu} &=& \frac{2}{\psi}\Im (\pa_\mu \phi^* \pa_\nu \phi )  
+ J_\mu \frac{\pa_\nu \psi}{\psi} - J_\nu \frac{ \pa_\mu \psi}{\psi}.
\eea
After a Fourier transform and a longitudinal projection,
\ben
\pa_0 \tilde{J}_s - ik \tilde{J}_0 = \tilde\Source_{0s}.
\label{e:Jsdot}
\een
Solutions to this equation can be found with the ansatz 
\ben
J_0 = \pa_k V_k, \quad 
J_i =  \frac{1}{a^2} \pa_0 (a^2 V_i),
\een
where $V_i$ is curl free. The current conservation equation \eqref{e:J0dot} with $\tilde{\sigma} = 0$ is then an identity.

Specialising to the radiation era, the equation for the longitudinal component of the Fourier transform $\tilde{V}_i$ is 
\bea
\ddot{\tilde{V}}_s(\bk,\tau)  + \frac{2 }{\tau}\dot{\tilde{V}}_s(\bk,\tau)   +  \left(k^2 - \frac{ 2}{\tau^2}  \right)\tilde{V}_s(\bk,\tau) = \tilde\Source_{0s}(\bk,\tau). \nonumber\\
\label{e:StrSouEqn}
\eea
The general solution to the homogeneous equation for wavenumber $\bk$ is
\ben
\tilde{V}_s(\bk,\tau) = d^{(1)}_\bk h^{(1)}_1(k\tau) + d^{(2)}_\bk h^{(2)}_1(k\tau),
\label{e:GenSolV}
\een
where $h^{(1)}_1$ and $h^{(2)}_1$ are spherical Hankel functions of order 1.

One can show that a string moving with speed $v$ produces a function $\Source_{0s} = \mathcal{O}(v)$, which acts as a source for the field $V_i$.  We can then understand Eq.~\eqref{e:StrSouEqn} as describing the production of propagating axions by moving strings.  The field $V_i$ can also be written in terms of an antisymmetric tensor field, with $B_{ij} = \ep_{0ijk} V_k$, which can be incorporated into a Kalb-Ramond field $B_{\mu\nu}$ with a gauge symmetry $B_{\mu\nu} \to B_{\mu\nu} + \pa_\mu \La_\nu - \pa_\nu \La_\mu$ \cite{Kalb:1974yc,Davis:1988rw}.

We end this subsection by noting that in a background of an oscillating homogeneous saxion field, we have 
\ben
\Source_{0s} = - J_s \frac{\dot\psi}{\psi}.
\een
Hence the equation for $\tilde{V}_s$ becomes 
\ben
\ddot{\tilde{V}}_s  + \frac{2 }{\tau}\dot{\tilde{V}}_s  +  \left(k^2 - \frac{ 2}{\tau^2}  \right)\tilde{V}_s =  \frac{1}{a^2} \pa_0 (a^2 \tilde{V}_s) \frac{\dot\psi}{\psi}
\label{e:SaxSouEqn}
\een
The term on the right hand side can lead to resonant production of axions with frequency half the oscillation frequency of the saxion, that is, at comoving wavenumber $k = a(\tau)\ms/2 = 1/2\ws$.

\subsection{String-sourced axions}
\label{ssaxions}

In this subsection we give a simple model for the total energy density in axion radiation, which will form the basis of our expectations. We start by defining the axion radiation spectrum as 
\ben
\mcP_\text{ax} = \half \left(\mcP_0 + \mcP_s \right) .
\een
Using the equations \eqref{e:J0dot} and \eqref{e:Jsdot} derived previously we may deduce that 
(see Appendix \ref{s:PaxEqn}) 
\bea
\frac{1}{a^2}\pa_0(a^2 \mcP_\text{ax}) &=& -  \frac{\dot a}{a} (\mcP_0 - \mcP_s) + \frac{k^3}{\pi^2} \Re\tilde{X},
\label{e:PaxEqn}
\eea
where
\bea
\tilde{X} =  \vev{ \tilde{J}^*_0 \tilde\si }_{\hat\bk} + \vev{ \tilde{J}^*_s \tilde\Source_{0s}}_{\hat\bk} .
\eea

 We can recast that expression into a number density power spectra, by dividing by the physical frequency of the mode $k/a$, and convert to physical energy density by dividing by $a^2$,
\ben
\mcNax(k,\tau) =  \frac{1}{2ak}  \left(\mcP_0(k,\tau) +\mcP_s(k,\tau) \right).
\een
which obeys the corresponding equation to \eqref{e:PaxEqn}
\bea
\frac{1}{a^3}\pa_0(a^3 \mcNax) &=& -  \frac{\dot a}{a^2 k} (\mcP_0 - \mcP_s) + \frac{k^3}{\pi^2} \frac{1}{ak} \Re\tilde{X}\,.
\eea
For source-free axions, $\tilde{X}=0$ and $\mcP_0 = \mcP_s$, so we may write
\ben
\mcNax(k,\tau) =  \frac{1}{ak}  \mcP_0(k,\tau), 
\een
which obeys 
\bea
\frac{1}{a^3}\pa_0(a^3 \mcNax) &=&0.
\eea
Hence $\mcNax$ redshifts as a number density power spectrum, 
with total number density
\ben
\nax(\tau) = \int \frac{dk}{k} \mcNax(k,\tau).
\een
Hence $\mcNax(k,\tau)$ behaves as a number density of particles with momentum $k/a$ and energy $k/a$. 

 The definition can be extended to modes which have a temperature-dependent mass $\ma(T)$, 
and adiabatic solutions exists for the mode functions with wavelengths much less than the horizon \cite{Davis:1986xc}, provided the axion field remains perturbative ($A/\fa \ll 1$).
For these modes the number density power spectrum is approximately conserved, in the covariant sense. Of course, the dynamics of domain wall formation and the subsequent collapse of the string network are non-perturbative, but the number density spectrum of a scaling network is still an interesting quantity which sets an order of magnitude against which the importance of the annihilation process can be assessed.

Bearing in mind our discussion following Eqs.~\eqref{e:AxStr0} and \eqref{e:AxStrs} about the relative importance of the string contribution to the power spectra $\mcP_0$ and $\mcP_s$, we  use the latter to
estimate the number density of propagating axion modes with the power spectrum  
\ben
\mcNax^s(k,\tau) = \frac{1}{ak}  \mcP_s(k,\tau).
\label{f:NumDenPSDef}
\een
We can normalise the expression by $\fa^2 H$, where $H$ is the physical Hubble rate, and plot it against $k\tau$. This normalisation is chosen so that the number density spectrum in a scaling string network collapses to a single line.  To see this, we recall that in a scaling system, 
\ben
\mcP_s(k,\tau) = \frac{\fa^2}{\tau^2} \tilde\mcP_s(k\tau), 
\een
and hence in the simulations, carried out with massless axions in the radiation era, 
\ben
\mcNax^s(k,\tau) = {\fa^2} H  \frac{\tilde\mcP_s(k\tau)}{k\tau},
\label{numberdensity}
\een
with physical Hubble rate $H = \dot{a}/a^2$.

Returning now to Eq.~\eqref{e:PaxEqn}, one notices that the first term on the right hand side  contains the difference of two power spectra which in the USM are both flat in the wavenumber range $1 \ll k\xiw \ll \xiw/\ws$, and hence we may define a difference spectrum of the form 
\ben
\mcP_{0s} \equiv \mcP_0 - \mcP_s = r_{0s} \frac{\fa^2}{\ta^2}
\label{e:POsUSM}
\een
where $r_{0s} = r_0 - r_s$, and the constants $r_0$, $r_s$ are defined in Eq.~\eqref{e:StrModPowSpe0s}.

For a stationary string segment in the USM, $\si \equiv  J_\mu\pa^\mu\psi/\psi = 0 $ vanishes. As $\si$ is a Lorentz scalar, it vanishes for moving segments as well.  To understand the vanishing of $\si$ for a stationary segment, we first note that $J_0$ and $\dot\psi$ both vanish, while (after choosing axes appropriately) $J_i \propto \hat{\theta}_i$ and $\pa_i \psi \propto \hat{x}^i$. Hence $J_i \pa_i\psi =0$.

The average $\vev{\tilde{J}^*_s \tilde\Source_{0s}}_{\hat\bk}$  also vanishes in the USM (see Appendix \ref{s:PaxEqn}), and to calculate its true value requires an extension to the model to take into account correlations between neighbouring segments. Even without calculating, however, we can gain insight into its functional form with the following observations.

The left hand side of Eq.~\eqref{e:PaxEqn} is closely related to the emission power spectrum,
\bea
\mcS_\text{ax}(k,\tau) &\equiv& \frac{1}{a^2} \frac{d}{d\ln(k\tau)} \left( a^2 \mcP_\text{ax}(k,\tau) \right), 
\label{e:SaxDef}
\eea
and we obtain the relation, in the radiation era, 
\ben
\mcS_\text{ax}(k,\tau) = - (\mcP_0 - \mcP_s)  + \frac{k^3}{\pi^2} \tau \Re\tilde{X}.
\label{e:SaxEqn}
\een
Assuming a leading power-law behaviour $(k\xiw)^{1 - q}$ for both sides of Eq.~\eqref{e:SaxEqn},  and noting that the string contribution to the spectra gives  $\mcP_0 - \mcP_s \sim (k\xiw)^0$, we must have $q \le 1$, unless there is a special symmetry which exactly cancels the two terms in the right hand side, 
leaving behind a sub-leading term in $\tilde{X}$ with $q > 1$.  We regard this possibility as unlikely. 

Hence we can conclude that the fact that $\mcP_0 - \mcP_s$ is flat in the USM implies that the emission power spectrum has $q \le 1$.  

With the assumption\footnote{This assumption will be supported by our numerical work presented later.} that $q=1$, we can integrate Eq.~\eqref{e:PaxEqn} to obtain
\ben
\mcP_\text{ax}(k,\tau) =  \frac{\fa^2}{\tau^2}\left(p_\text{ax}\ln(k\ta) + c_\text{ax} \right),
\label{e:PaxDef}
\een
where $c$ is an integration constant.  In order to be consistent with Eq.~\eqref{e:POsUSM}, 
the individual components in the axion energy power spectrum must behave as 
\bea
\mcP_0(k,\tau) =  \frac{\fa^2}{\tau^2}\left(p_\text{ax}\ln(k\ta) + c_0 \right), \\
\mcP_s(k,\tau) =  \frac{\fa^2}{\tau^2}\left(p_\text{ax}\ln(k\ta) + c_s \right).
\eea
with
\ben
c_0 - c_s = r_{0s}.
 \label{e:r0s}
\een
This implies that fits to a logarithm plus constant for both spectra will result in coefficients for the logarithm which are the same, and taking the 
difference gives an estimate of $r_{0s}$.

\subsection{Transverse decoherence function}

\begin{figure}[htbp]
\begin{center}
\includegraphics[width=\columnwidth]{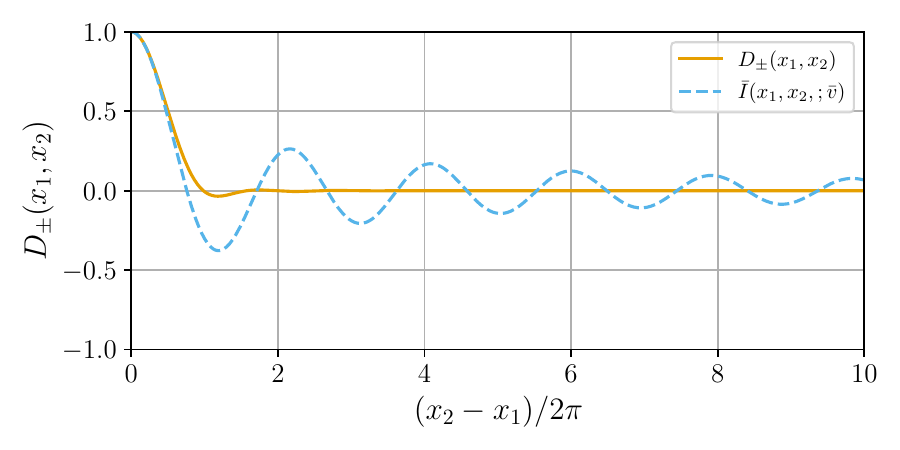}
\caption{Decoherence function for the transverse currents $J_\pm$ in the Unconnected Segment model \eqref{e:DpmUSM}, with the speed distribution given in Eq.~\eqref{e:vDist}, evaluated at $x_1 = 25$, where $x_1 \equiv k\tau_1$. Also shown is the unaveraged expression \eqref{e:DpmUSMnoav}, evaluated at the RMS velocity $\bar{v} = 0.52$.}
\label{f:USMdec}
\end{center}
\end{figure}

Our final expectation is for the time correlations of the transverse vector current $J_\pm$, which we recall is entirely due to the strings. In Appendix \ref{s:ModUETC} we calculate the UETC in the USM, and we plot the corresponding decoherence function in Fig.~\ref{f:USMdec}, for the parameter choices $\be_\text{w} = 0.45$ and $\bar{v} = 0.52$, at $x_1 = 25$.  We also plot the function evaluated at $v = \bar{v}$, to show the effect of the velocity averaging.  We see that the vector decoherence function is predicted to vanishes quickly as a function of the difference in $x = k\tau$, and that the rapid vanishing is a result of the velocity averaging. 
 
\subsection{Summary}
\label{summary}

We summarise our theoretical expectations as follows.

\begin{enumerate}
\item  The dominant contribution of strings is to power spectra is to $\mcP_\pm$, whereas string contribute less  to $ \mcP_0$ $\mathcal{O}(v^2)$ and $\mcP_s$ $\mathcal{O}(v^4)$.

\item The power spectra $\mcP_\pm$ are equal, independent of wavenumber in the range $1 \ll k\xiw \ll \xiw/\ws$ and have an amplitude $r_\pm $.  

\item The difference $\mcP_0 - \mcP_s$  is also independent of wavenumber in the same range.

\item The contribution of axion radiation to the decoherence functions $D_0(x_1, x_2)$ and $D_s(x_1, x_2)$, where $x = k\tau$, are both $D_a(x_1, x_2) = \cos(x_1 - x_2)$ for large $x_1$, $x_2$.

\item  The number density spectrum $\mcNax^s(k,\tau)$ \eqref{numberdensity}  in a scaling string network  should collapse to a single line when multiplied by $k\tau/H$ and plotted against $k\tau$.

\item Barring an unexpected cancellation, the emission power spectrum $\mcS_\text{ax} \propto (k\xiw)^{1-q}$ with $ q \le 1$. 

\item The axion energy power spectrum $\mcP_\text{ax} = (\mcP_0 + \mcP_s)/2$  increases logarithmically with wavenumber $k$ in the range $1 \ll k\xiw \ll \tau/\ws$.

\item Only strings contribute to the transverse decoherence function $D_\pm(x_1,x_2)$, which is concentrated in the region $|x_1 - x_2| \lesssim \pi v$, and tends to zero as $|x_1 - x_2|$ goes to infinity.

\end{enumerate}

When comparing to the simulations, one should keep in mind the crudeness of the model, which has much room for improvement.  For example, we have taken the mean segment spacing to be the same as the segment length.  Furthermore, there are correlations between segments as they are connected, and segments may be curved.   We have also been inconsistent in using the segment velocity distribution \cite{Gorghetto:2020qws} only to derive the decoherence function $D_\pm$ and not to evaluate $\vev{\gamma v^2}$.

\section{Simulation methods}
\label{s:SimMet}

To investigate the dynamics of axion string networks and their scaling properties, we perform large volume 
numerical simulations of a complex scalar field 
$\Phi$ evolving in an expanding  FLRW universe according to Eq.~\ref{eq:eom}. 
The procedure of discretization is described in \cite{Correia:2024cpk}. In that paper we studied the scaling properties of the network of axion strings in the radiation era  using a number of simulations, of various sizes, up to cubic grids with 16384 sites per side.  In this work, we analyze the results of a subset of the simulations, specifically those performed on a cubic lattice with 12288 sites per side (hereafter referred to as  $N=12k$, with $k=1024$). These are the largest Fourier transforms performed on the axion string system to date. These large simulations are possible thanks to HILA \cite{HILA}, a
simulation framework designed for very large-scale simulations on a variety of supercomputing architectures,
including GPGPU machines. 

A novel aspect of this work is that we calculate unequal time correlations (and the corresponding power spectra) of the system, which require a large number of Fourier transforms. This is a computationally very costly procedure, and that is why we only performed it in one of the box sizes analyzed in \cite{Correia:2024cpk}.  With HILA built-in FFT, on a 12$k$ lattice,  one complex-to-complex FFT takes about $2$  seconds using 4096 AMD MI250X GPUs (8192 MPI processes) on LUMI supercomputer \cite{lumi}.  For the analysis presented in this work, about 50\% of the computing time is spent in FFTs.

For reference, we briefly explain the numerical procedure, but direct the interested reader to the details in \cite{Correia:2024cpk}:
We discretise the field equations on a cubic lattice of $N=12k$ points per direction using periodic boundary conditions. The spatial derivatives are computed using a 
 finite-difference scheme with a 7-point stencil for the Laplacian, while time evolution is handled via a leapfrog integration method.  We chose units in which  $\fa=1$. The lattice spacing
and timestep are $\Delta x= 0.5$ and $\Delta \tau = 0.1$ respectively, which satisfy the Courant-Friedrichs-Levy condition for  hyperbolic systems. The initial condition is given by $\Pi=0$, with the field $\Phi$  
 set to be a Gaussian random field of unit variance, with the correlation length in comoving coordinates given by $\IniCorLen$.   We performed five simulations per $\IniCorLen= 5,10,20,40,80$, varying the initial random seed.

We start the simulation at $\tStart$. Since the initial configuration is very energetic, we evolve the system with a non-linear diffusion equation until time $\tDiff$, at which point the field configuration consists of smooth, almost stationary strings, and no radiation.  During diffusion, the effective mass term $\sqrt{2\la}\eta$ is kept constant and $\mathcal{O}(1)$, to speed up the relaxation.

At the end of diffusion, the strings are much thinner than their true comoving value $1/\sqrt{2\la}a(\tDiff)\eta$, and so we adjust the coupling so that the comoving width grows linearly in conformal time, meeting the true value at $\tcg$.  During this period, strings are ``conformal'' \cite{Klaer:2019fxc}: their width grows in proportional to the horizon distance.
At $\tcg$ the physical simulation starts. 
The simulation finishes at $\tEnd$, when the scale factor is normalised so that $a=1$, and the comoving string width returns to its initial value. The initial and final comoving string widths were $\ws(\tDiff) = \ws(\tEnd) = 0.5$ in code units. 
The value of the different simulation time parameters can be found in Table~\ref{simparams}.
 
\begin{table}[h!]
\begin{ruledtabular}

\begin{tabular}{c D{.}{.}{-1}}
Parameter & \multicolumn{1}{c}{Value}  \\
\hline
{$\tStart$} &	50   \\
{\tDiff}    &	70   \\
{\tcg}      &	470  \\
$\tRef$     &	1000 \\
$\tFinPS$   &	2800 \\
${\tEnd}$   &	3150 \\
$\DtauPS$   &	20
\end{tabular}

\end{ruledtabular}
 \caption{\label{simparams}
 Time parameters of the simulation:  $\tStart$ (the begining of the simulation), 
 $\tDiff$ (the end of diffusive evolution and the start of string core growth), 
 $\tcg$ (conformal time at the start of physical evolution),  $\tRef$ (reference time for UETC calculations),  $\tFinPS$  (last time of UETC calculations), $\tEnd$ (conformal time at the end of the evolution) and $\DtauPS$ (lapse between UETC calculations). The value  of conformal $\tcg$  correspond to physical time $t_\text{cg}=70/\ms$, or $\ln(2t_\text{cg}\ms) = 4.9$. The conformal time  $\tEnd$ corresponds to physical time $t_\text{end} = 1575/\ms$, or $\ln(2t_\text{end}\ms) = 8.75$.}
\end{table}

As mentioned above, we want to measure power spectra of the axion field and their time correlations of the axion field in Fourier space, the UETCs.  
We start taking Fourier transforms of $J_\mu(\bx,\tau)$ at a reference time $\tRef$, projecting them to $\tilde{J}_\al(\bk, \tRef)$, and storing them in memory. 
We continue taking Fourier transforms every $\DtauPS$ until $\tFinPS$ (the numerical values can be found in, Table~\ref{simparams}).  At each measure we save spectra $\mcP_\al(k,\tau)$ and UETCs $U_{ab}(k,\tRef,\tau)$ and $U_{AB}(k,\tRef,\tau)$.

The choice of $\tRef$ is important, since we would like the simulation to be close to scaling, but early enough in order to be able to obtain a good dynamical range. 
In Fig.~\ref{f:zetaw} we denote the times at which UETCs are taken by grey lines, which shows that there are departures from scaling which we expect to see reflected in the power spectra and UETCs.  The simulations which are closest to scaling in $\zeta$ during the measurement period have initial field correlation lengths $l_\phi = 5, 10, 20$.  However, the simulations with $l_\phi = 5, 10$ have a high string density during the core growth period, during which axions are also emitted, and we should also expect to see signs of this early radiation in higher wavenumbers.

In order to do the discreet Fourier transform, we first  define the discrete variables $\bx=(x_1,x_2,x_3)$, with $x_i$  running from $-L/2$ to $L/2$   by increments of  $\Delta x$.   Likewise, we define $ \bk=(k_1,k_2,k_3)$ with    $k_i$ running from $-\pi/\Delta x$ to $\pi/\Delta x$ in increments of $\Delta k = 2 \pi / L$. The maximum wavenumber on the lattice is 
$k_{\rm max} =  \sqrt{3} \pi/\Delta x$. We can now define the discreet Fourier transformation as
\ben
\tilde{f}(\bk) = (\Delta x)^3\sum_{\bx} f(\bx) e^{-i x_i k_i} \,.
\een
We bin the momentum space ETCs and UETCs by adding the contribution at momentum $\bk$ into bin number $i$ if $i\Delta k \le |\bk| < (i+1)\Delta k$, where we choose $\Delta k = 2\pi/L$. The value of $k_i$ associated with the bin $i$ is the average of $|\bk|$-values falling into bin $i$. 

For the projections of $\tilde{J}_i(\bk, \tau)$, we construct the basis as follows.   Starting with the Fourier representation of the nearest-neighbour discrete derivative
\ben
K^i  \equiv \frac{2}{\Delta x} \sin \left( \frac{k_i \Delta x}{2}\right)
\een
we construct basis vectors 
\bea
e_s^i &=& \hat{K}^i, \\ 
e_1^i &=&  \frac{\left( 1 - (\hat{K}^1)^2, - \hat{K}^1\hat{K}^2, -\hat{K}^1\hat{K}^3 \right)^i}{\sqrt{1 - (\hat{K}^1)^2}}  \\
e_2^i &=& \ep^{ijk} e_s^j e_1^k.
\eea
In the case that  $\hat{K}^i = (1,0,0)$, we take $e_1^i = (0,1,0)$ and $e_2^i = (0,0,1)$.

  In order to be able to fit and thereby extrapolate the spectra, we need a wide range of wavenumbers which start out at $\tDiff$ much less than the inverse of the comoving string spacing $\xiw^\text{c}$, or $k\xiw^\text{c}(\tDiff) \lesssim 1$, cross $k\xiw^\text{c} \sim 1$ when the network is close to scaling, and have wavenumbers much smaller than the inverse string spacing at the final power spectrum, or  $k\xi^\text{c}(\tFinPS) \gg 1$.  For such a wavenumber, the axion spectrum ought to have its asymptotic behaviour towards the end of the simulation. 

\begin{figure}[htbp]
\begin{center}
\includegraphics[width=\columnwidth]{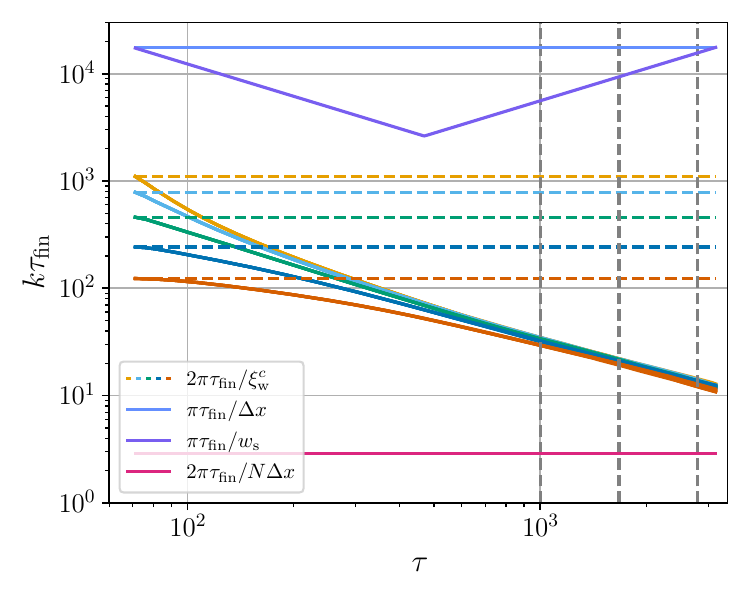}
\caption{Wavenumbers for the analysis of power spectra derived from significant length scales, in units of the final time at which power spectra are recorded $\tFinPS$: lattice spacing $\Delta x$,  string width $\ws$ and the simulation box side length $N\Delta x$. The coloured solid lines are mean universe frame string separation $\xiw^\text{c}$, with the same colour key for initial field correlation lengths as Fig.~\ref{f:zetaw}. The coloured dashed lines are their values at $\tDiff$ (see Table \ref{simparams}). 
}
\label{f:WavNum}
\end{center}
\end{figure}

The difficulty of finding a wide range of such wavenumbers is illustrated in Fig.~\ref{f:WavNum}, where the relevant length scales are converted to wavenumbers, in units of the horizon at the time of the final power spectrum $\tau_\text{fin}$. There we plot the scales $k = 2\pi/\xiw^\text{c}$, which is a more accurate estimate of the wavenumber of the network length scale, along with wavenumbers derived from the lattice spacing $\Delta x$, the string width $\ws$ and the simulation box side length $N\Delta x$. 

In order to satisfy the conditions for all initial string densities, we choose the range $25 < k\tau < 100$ for fitting power spectra.  
One can see that for the lower initial string densities $l_\phi = 40, 80$ the wavenumber crosses the network length scale when the network is at a much lower density than at scaling. 
In low-density networks, we therefore expect to see systematically lower power in wavenumbers at the upper end of the chosen range.

In order to mitigate the difficulty of finding a range of wavenumbers in which power spectra are exhibiting their asymptotic behaviour, one can also examine the growth of the power spectra, which does not have a record of the earlier, non-scaling, phases of the network evolution.  Here, the upper bound on the wavenumber is set by the requirement $k\ws \ll 1$, that is, that the spectrum is not affected by the microphysics of the string width at any time during the simulation.  We also choose the range $25 < k\tau < \tau/(16\ws)$ for fits of the growth of the power spectrum during the evolution of the network.

\section{Results and discussion: power spectra and decoherence functions}
\label{s:Res}

In this section we show the results for power spectra and decoherence functions obtained from the simulations described above, 
and compare them to previous results in the literature. We also discuss the results of fitting to the model described in Section \ref{s:TheExp}.

\subsection{Power spectra}
\label{ss:PowSpe}

\begin{figure*}[htbp]
    \centering
    \includegraphics[width=\columnwidth]{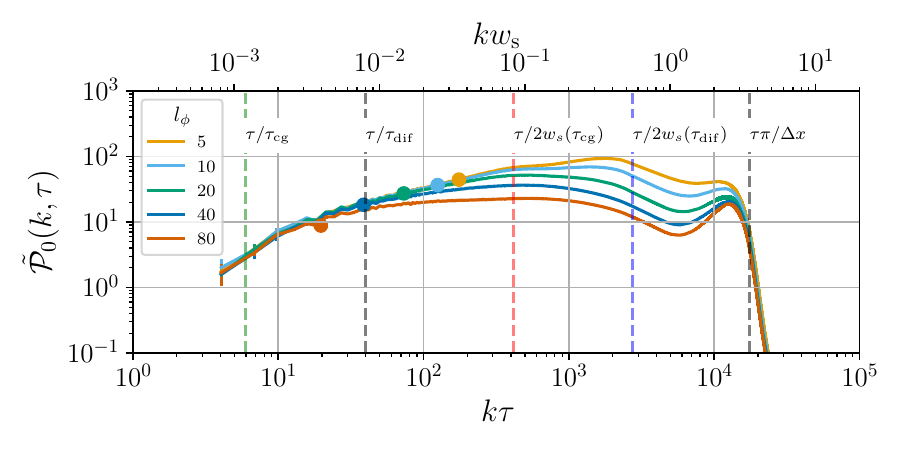}
    \includegraphics[width=\columnwidth]{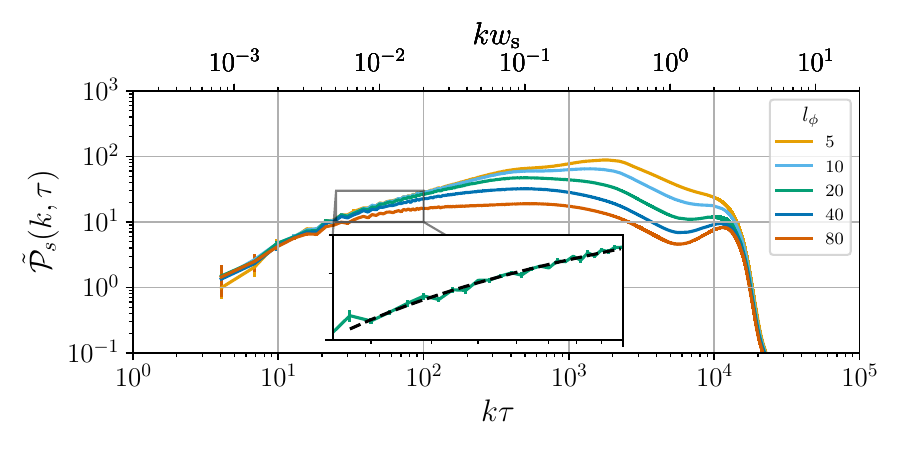}
    \includegraphics[width=\columnwidth]{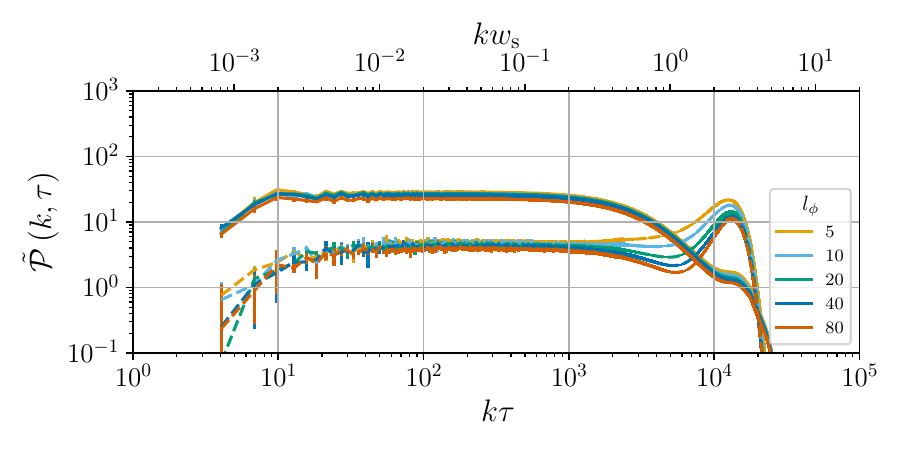}
    \includegraphics[width=\columnwidth]{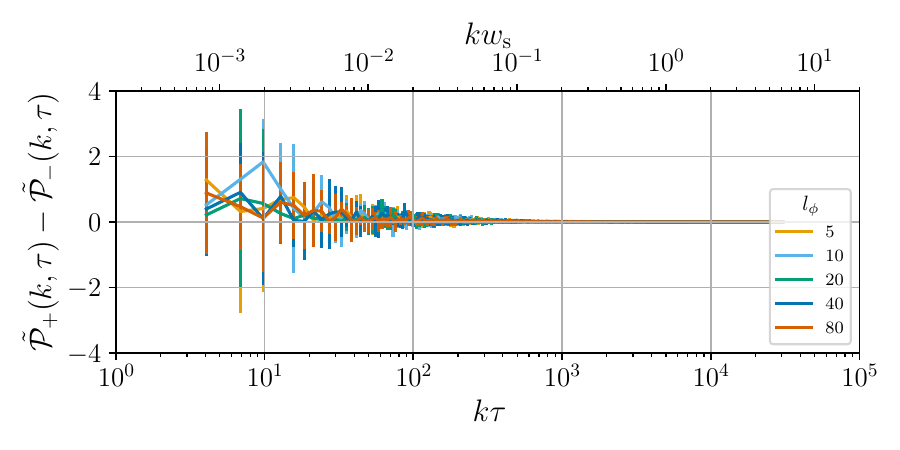}
     \caption{Final $\tau$-scaled power spectra. Top left: $\mcP_0$, the power spectrum of $J_0$. Top right: $\mcP_s$, the power spectrum of $J_s$. Bottom left: the sum $\mcP_+ + \mcP_-$ (solid) and the difference of $\mcP_0 - \mcP_s$ (dashed). Bottom right: the difference $\mcP_+ - \mcP_-$.   The power spectra are plotted against comoving wavenumber $k$ in units of the comoving horizon $\tau$ (bottom axis) and in units of the comoving string width (top axis).
 The top left plot is annotated with wavenumbers corresponding to several relevant length scales. Vertical dashed lines mark the horizon at the start of the second order time evolution $\tDiff$, the horizon at the start of physical evolution $\tcg$, half the comoving string width at $\tcg$, and half the comoving string width at $\tDiff$. Coloured dots mark the wavenumbers corresponding to the universe-frame mean string separation at the start of physical evolution, $\xi(\tcg)$, in each set of simulations. The black dashed line in the inset in the top right figure is the fit to the mean of the $l_\phi = 20$ power spectra, using Eq.~\eqref{e:PowSpeFitFun}, in the range $25 < k\tau < 100$. 
     \label{f:PS_all_tau_final}}
 \end{figure*}

In Fig.~\ref{f:PS_all_tau_final} we show power spectra of various currents and combinations of currents, scaled by the square of the conformal time $\tau$, against $k\tau$, evaluated at the last recorded spectrum $\tFinPS$ (see Table~\ref{simparams}). With these factors, the power spectra of a system in scaling  should  collapse to the same curve, indicating a constant fractional contribution to the energy density in each logarithmic wavenumber interval.

In the top row we show the power spectra of $J_0$ and $J_s$.  The bottom row shows the sum and difference of the power spectra of $J_+$ and $J_-$.   The figure for $J_0$ also shows several relevant length scales: the horizon at the start of the second order time evolution $\tDiff$, the horizon at the start of physical evolution $\tcg$, half the comoving string width at $\tcg$, and half the comoving string width at $\tDiff$. The last two scales are imprinted by emission of axion radiation resonantly produced by interactions with saxion radiation.  

As stated in section~\ref{summary},  only strings contribute to $J_+$ and $J_-$, and parity symmetry requires that the power spectra of $J_+$ and $J_-$ should be statistically the same.  Within the uncertainties, which are estimated from the standard deviation of the power spectra over the 5 runs with the same initial string density, the two chiralities are consistent with each other. There is some tendency to positive mean values at low wavenumbers, although with no statistical significance. 

The sum of the power spectra of $J_+$ and $J_-$ shows a flat ($k^0$) power spectrum in the range $10 \lesssim k\tau \lesssim 10^3$.  We can identify the wavenumber range as between the string separation scale ($\xiw^\text{c} \simeq   0.4\tau $) and the inverse string width $\ws^{-1}$ at the end of the simulation 
This is expected for string-like objects, as summarized in section~\ref{summary}, and explained in more detail  in Appendix \ref{s:PowSpeStr}.  The small differences in the  power spectra between different initial string densities are a result of the small differences in the string densities in the final state.

In the same plot we show the difference between the $J_0$ and $J_s$ power spectra.  Here,  the expectation obtained from the analytical model in section~\ref{s:TheExp} is that the propagating axion contribution to cancel at wavenumbers $k\tau \gg 1$, leaving behind the difference in the contribution of the strings to $J_0$ and $J_s$, which should be $\mathcal{O}(v^2)$ of the sum of $J_+$ and $J_-$ power spectra.  The similarity in shape of the difference of $J_0$ and $J_s$ spectra to the  sum of $J_+$ and $J_-$ confirms this expectation.  

Focusing next on the $J_s$ power spectrum, which we have argued is less contaminated by string contributions, we see that there is a strong impression of collapsing to a single line, which is very clear at low wavenumbers, and extends up to $k\tau \simeq 10^2$ for initial string correlation lengths $l_\phi = 5, 10, 20$. These simulations are closest to the scaling string density identified in \cite{Correia:2024cpk}, so the collapse of the power spectra is consistent with scaling. 

The wavenumber range where the spectra agree is approximately the range of wavenumbers less than the initial string separation, or $k\xi(\tDiff) \lesssim 1$, for which $k\tau$ takes the values $[175, 125, 73, 39, 20]$

The lowest initial string density (at $\l_\phi = 80$) has the lowest power spectrum. This is understandable as a result of there being a lower density of strings throughout the simulation.  The initial string correlation length is also largest, and we expect there to be a particularly large difference in the power spectra for modes with $k\xi(\tDiff) \gtrsim 1$.

The $J_0$ power spectrum has a similar shape to the $J_s$ spectrum, but is noticeably larger where $k\tau \lesssim 100$. 
We attribute the difference to greater contribution of the strings to $J_0$. We see small oscillations in both spectra in the range $10 < k\tau < 100$, which were discussed around Eq.~\eqref{e:PSosc} (see also \cite{Saikawa:2024bta}, section IV).

The figures for the power spectra of $J_s$ also show fits to a function  
\ben
\mcP_{a,\text{fit}}(k\tau) = p_a \ln (k \tau/10) + c_a, 
\label{e:PowSpeFitFun}
\een
in the range $25 \le k\tau \le 100$ 
as discussed in Section \ref{s:TheExp}, with the reference $k\tau = 10$ chosen to be near the start of the flat part of $\mcP_\pm$.  We recall that this function is motivated by the expectation that the emission spectrum is flat, $q=1$.  Then, if the source power spectrum is flat at high $k\tau$, that is $\mcP_\pm = r_\pm\fa^2/\tau^2$, one would expect to see the logarithmic behaviour of Eq.~\ref{e:PowSpeFitFun}.  

In Table \ref{t:SpeFitPars} we show the parameters $r_\pm$, $p_0$, $p_s$ and $r_{0s}$ from fits of the power spectra $\sum_\pm \mcP_\pm$,  $\mcP_0$ and $\mcP_s$ to forms given in Eqs.~\eqref{e:StrModPowSpe2}, \eqref{e:StrModPowSpe0s} and \eqref{e:PowSpeFitFun}. We also show the USM prediction for  $r_\pm$  ($r_\pm^{\rm usm}$) from Eq.~\eqref{e:USMrpm} with the measured values of $v$ and $\xiw$. 

The fit parameters $p_0$ and $p_s$ for the higher initial density runs, which stay closer to scaling in string density through the simulations, have similar values $p_a \simeq 10$.  We also see that $p_0 \simeq p_s$, within errors, as one would expect for a quantity measuring axion radiation density.  We obtain an estimate for the value of the coefficient in the form for the axion radiation spectrum \eqref{e:PaxDef} by averaging over runs with $l_\phi \in [5,10,20]$,
\ben
p_\text{ax} = 12.5(6).
\een

The unconnected segment model predictions for $r_\pm$ are 40 --  50\% higher than the fitted values.  The ratio $r_{0s}/r_\pm$, which depends only on the velocity distribution, is only about 15\% lower than the expected $\sim 0.4$  \eqref{lau}.

\begin{table}[htp]
\begin{ruledtabular}
\begin{center}
     \begin{tabular}{l c c c c c c}
          $l_\phi$ & $ r_\pm$ & $ r_\pm^{\rm usm}$  & $p_0$ & $p_s$ & $r_{0s}$ & $r_{0s}/r_\pm$ \\ 
          \hline
          \hline
          5 &14.22(7)& 20(1)& 13.3(3)& 12.8(4)& 4.43(9)& 0.311(7) \\
          10 &13.51(7)& 19(1)& 12.7(3)& 13.0(4)& 4.46(5)& 0.330(4) \\
          20 &13.42(7)& 18(1)& 11.8(4)& 11.4(3)& 4.26(5)& 0.317(4) \\
          40 &12.96(7)& 18(2)& 9.4(4)& 8.7(2)& 4.23(7)& 0.326(6) \\
          80 &11.23(6)& 16(1)& 4.8(3)& 4.8(2)& 3.75(4)& 0.334(4) \\
     \end{tabular}

\end{center}
\end{ruledtabular}
\caption{Fit parameters for power spectra $\mcP_\pm$, $\mcP_0$, $\mcP_s$ and $\mcP_0 - \mcP_s$ at conformal time $\tau = 2801$ to forms given in Eqs.~\eqref{e:StrModPowSpe2}, \ref{e:StrModPowSpe0s} and \eqref{e:PowSpeFitFun} 
}
\label{t:SpeFitPars}
\end{table}

\subsection{Emission spectra}
\label{emissionspectra}

\begin{figure*}[hbtp]
    \centering
    \includegraphics[width=\columnwidth]{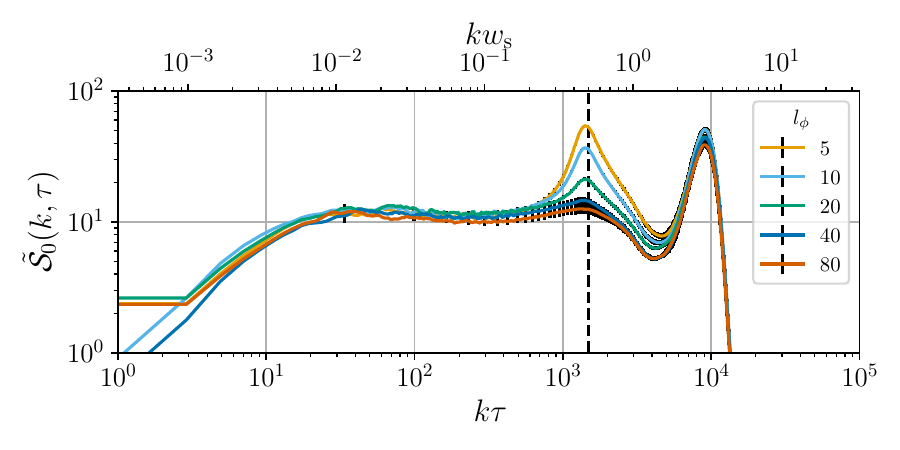}
    \includegraphics[width=\columnwidth]{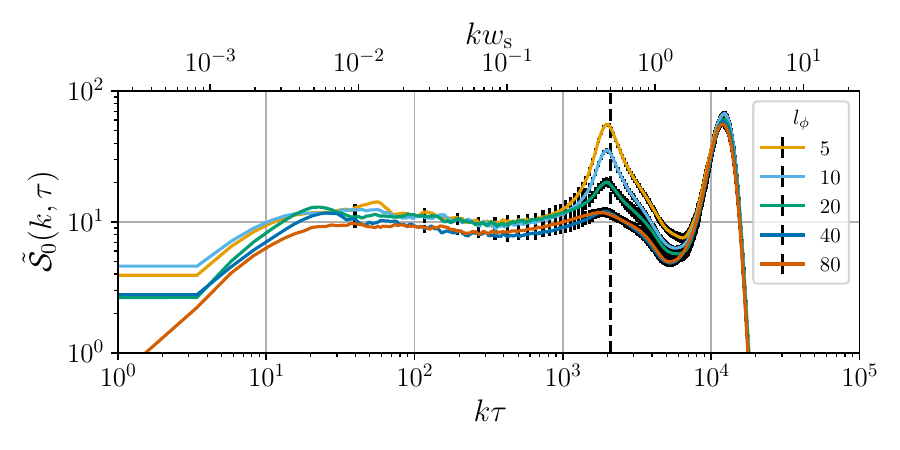}
    \includegraphics[width=\columnwidth]{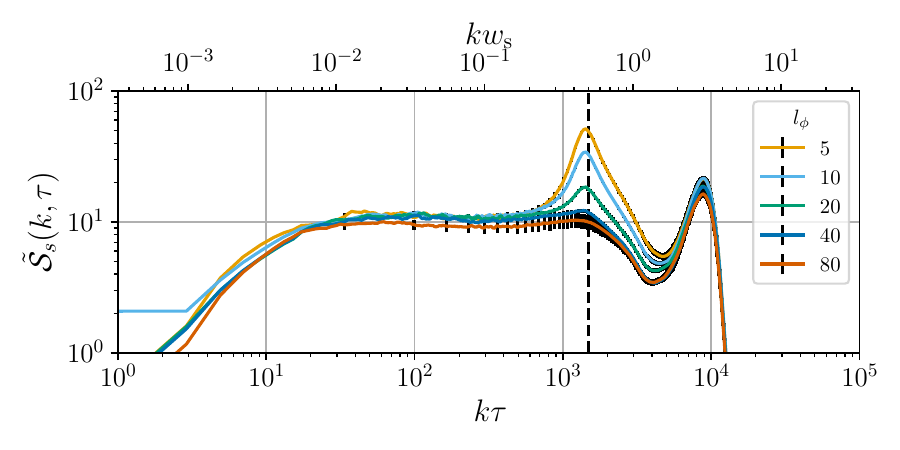}
    \includegraphics[width=\columnwidth]{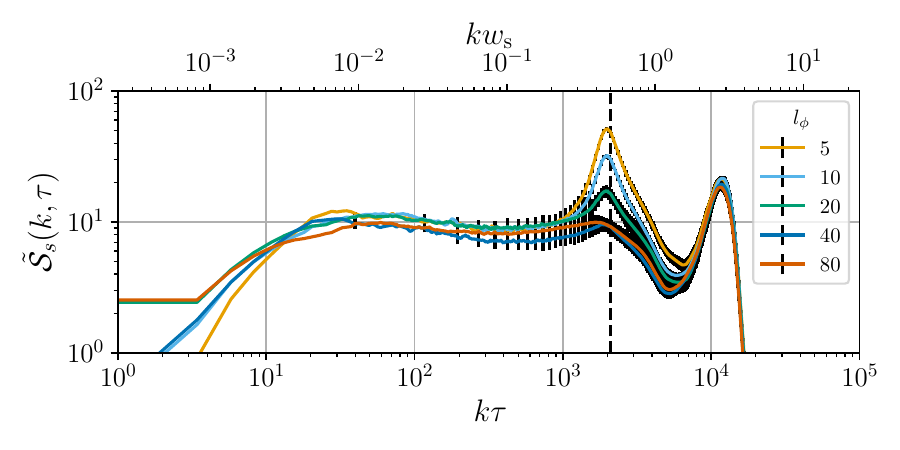}
     \caption{Emission spectra for  (top to bottom)  $J_0$ and $J_s$, between times $\tau = [2000, 2340]$ (left) and $\tau = [2360, 2801]$ (right). The spectra have been smoothed with a Savitsky-Golay filter \cite{savitzky1964smoothing}, with window length 16 and polynomial order 3.  The vertical dashed lines show the comoving  wavenumber corresponding to half the physical saxion mass at a conformal time half way between the end-points.  
           \label{f:PS_diff_all_tau_final}}
 \end{figure*}

In Fig.~\ref{f:PS_diff_all_tau_final} we show a numerical approximation to the logarithmic  rate of change of the scaling power spectra of $J_0$ and $J_s$, or 
\ben
\tilde{\mcS}_a(k,\tau) \equiv   \tau\frac{d}{d\tau}\tilde\mcP_a(k,\tau) ,
\een
where we recall the definition of the scaling power spectrum $\tilde\mcP$ in Eq.~\ref{e:PtilDef}.  
The quantities $\tilde{\mcS}_a$, termed emission spectra,  contain information about the rate of change of the spectrum of the propagating axions, as a system of freely propagating axions without sources has  $J_0$ and $J_s$ behaving as $1/\tau^2$, and therefore $\tilde\mcS_a = 0$. 

The numerical differences are taken between conformal times 2200 and 2340 (left column) and  2360 and 2801 (right column).   We plot against $k\tau$ and show $k\ws$, the wavenumber in string width units, on the top axis. Curves are smoothed with a Savitsky-Golay filter \cite{savitzky1964smoothing} with parameters given in the figure caption. 
The peaks at $k\ws \simeq 3$ correspond to the Nyqvist spatial frequency $\pi/\Delta$ and are therefore a lattice effect. 
In Appendix \ref{s:EmiSpeEvo} Fig.~\ref{f:PSdiffJ0J0JsJs} we show the $\tilde\mcS_a$ spectra for each initial string density over the whole range of times for which the spectra are recorded.
 
Two main features can be underlined: on the one hand, the peak at at $k\ws = 0.5$, and on the other,  in range $k\tau \lesssim 100$, the emission spectra $\tilde\mcS_a(k,\tau)$ are close to a single curve which rises from low values of $k\tau$ to 
an approximately flat line.

The peak at $k\ws = 0.5$ is perhaps the most dramatic feature of $\tilde\mcS_a$. They are largest for the highest initial string density ($l_\phi = 5$). These peaks are a signal of resonant axion production in a background of saxion modes.  Straight global strings are known to possess a countably infinite set of bound saxion modes with angular frequencies $\omega < \ws^{-1}$ \cite{Blanco-Pillado:2021jad}, with the first two at  $\omega_1 = 0.89\ws^{-1}$ \cite{Goodband:1995rt} and $\omega_2 = 0.99\ws^{-1}$  \cite{Blanco-Pillado:2021jad}. The density of states as $\omega \to \ws^{-1}$ from below therefore diverges. On the other hand, the density of freely propagating modes with $\omega \to \ws^{-1}$ from above remains finite.  Therefore it seems very likely that the resonant production involves the bound states. One expects the bound states to be excited by the initial curvature of the string network, by string intersections, and by regions of high curvature which appear during the evolution. Higher string densities imply both a higher curvature and a larger volume in which the bound modes are oscillating. 

The saxion field around an excited string will be on average smaller than the ground state value: positive excursions are penalised more strongly by the quartic potential. Hence the width, as measured by a field isosurface, will be greater. In Appendix~\ref{oscillation} we show that the effective volume occupied by the most excited strings ($\l_\phi = 5$) is approximately twice that of a string in its ground state.

The same peak in the axion spectrum has also been seen in simulations of collapsing axion string loops \cite{Saurabh:2020pqe}. It was pointed out in Ref.~\cite{Baeza-Ballesteros:2023say} that the peak can be greatly reduced in these simulations by carefully preparing the initial conditions so that the radial string field profile is locally very close to that of an infinite string, and stays close throughout the collapse. This supports the idea that the peak is associated with oscillations in the radial profile of the field, due to excitations of the bound states.

Features in the axion spectrum at $k\ws = 0.5$ have been observed previously in network simulations \cite{Gorghetto:2020qws,Saikawa:2024bta}, although they are much less prominent as both papers focused on low-density string networks, similar to our simulations with $l_\phi = 80$. 
In those papers, because of the larger oscillations in the average saxion field, oscillations are also visible in the spectrum of $J_0$.  They extend up to wavenumber $1/2\ws$ (see e.g. Fig.~15 of Ref.~\cite{Saikawa:2024bta}) where the relative amplitude is around 5\%.  Our spectra are smooth at this level for $100 \lesssim k\tau \lesssim \tau/2\ws$.  

Turning now to the wavenumber range $k\tau \lesssim 100$, the emission spectra $\tilde\mcS_a(k,\tau)$ are close to a single curve which rises from low values of $k\tau$ to 
an approximately flat line, with $\tilde\mcS_a \simeq 10$ for all initial string densities. In Fig.~\ref{f:PSdiffJ0J0JsJs} one can see that this collapse onto a single curve is already present in the first emission spectra.  This is a clear sign of scaling behaviour in the emission spectrum, and the flatness is consistent with the logarithmic behaviour of the scaled spectra $\tilde\mcP_a$.

 \begin{figure*}[htbp]
    \centering
    \includegraphics[width=0.48\textwidth]{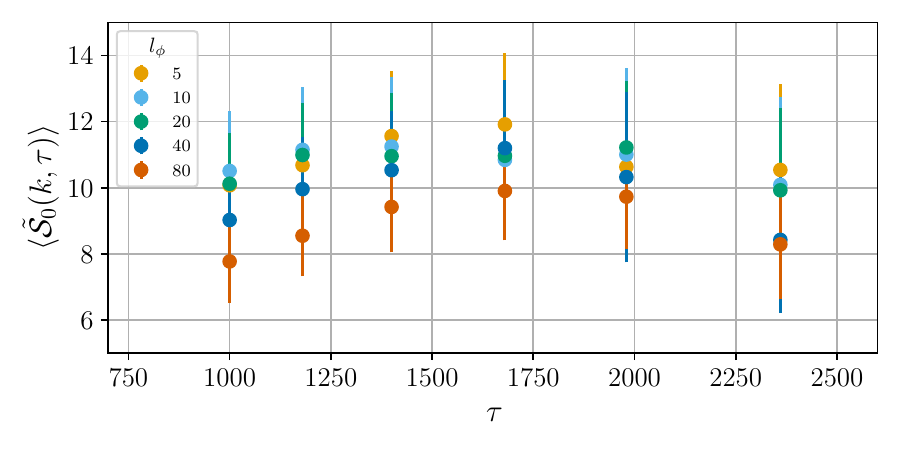}
    \includegraphics[width=0.48\textwidth]{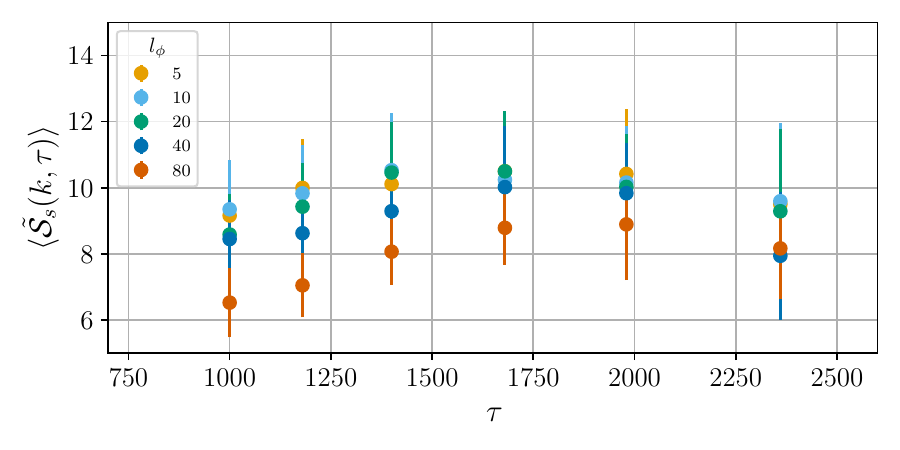}
    \includegraphics[width=0.48\textwidth]{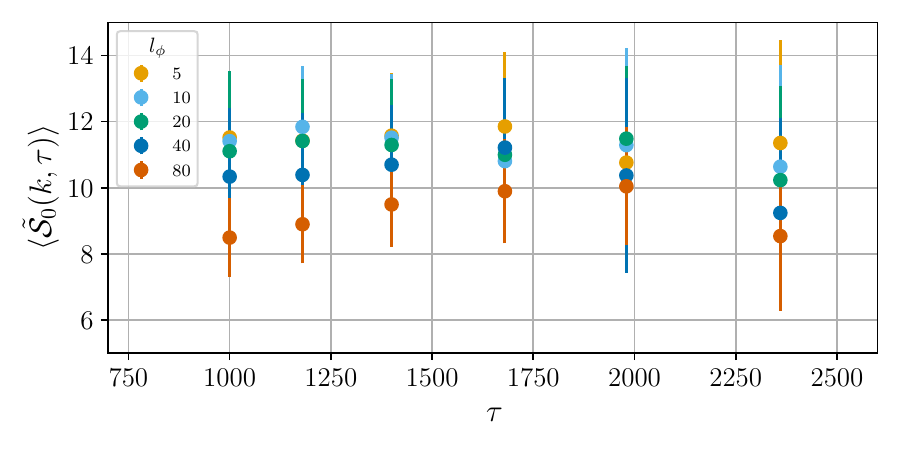}
    \includegraphics[width=0.48\textwidth]{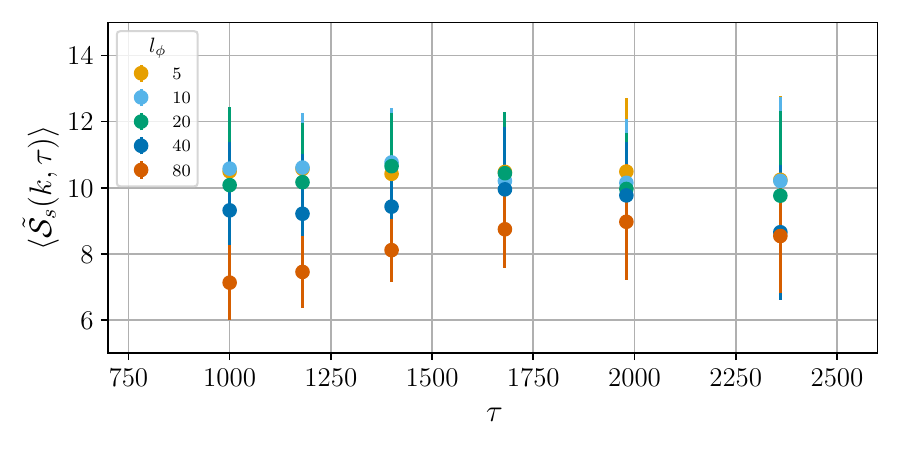}
     \caption{Mean and standard deviation of axion emission spectra (Fig.~\ref{f:PSdiffJ0J0JsJs}). 
     in range $25  < k\tau < \tau/(16 \ws)$ (top) and  $25 < k\tau < 100$ (bottom). Left:  $J_0$; right: $J_s$.  
     \label{f:emitfits_mean}}
 \end{figure*}

We plot the mean and standard deviation of the scaled emission spectra $\tilde{\mcS}_a$ in two different wavenumber ranges:  $25 < k\tau < \tau/(16 \ws)$ and   $25 < k\tau < 100$ in Fig.~\ref{f:emitfits_mean}.  We see that in these ranges the emission spectra stay within about 20\% of $\tilde\mcS_a \simeq 10$, for all initial string densities. Our estimate for the asymptotic value of the plateau in the emission spectrum in the range $25 < k\tau < 100$ is taken from the last value of $\tilde\mcS_s$, averaged over all initial string densities.
\bea
\vev{\tilde{\mcS}_0}_{25\le k\tau \le 100}(\tau = 2360) &=&  10.0(2.8), \\
\vev{\tilde{\mcS}_s}_{25\le k\tau \le 100}(\tau = 2360) &=&  9.5(2.3).
\eea

\begin{figure*}[htbp]
    \centering
    \includegraphics[width=0.48\textwidth]{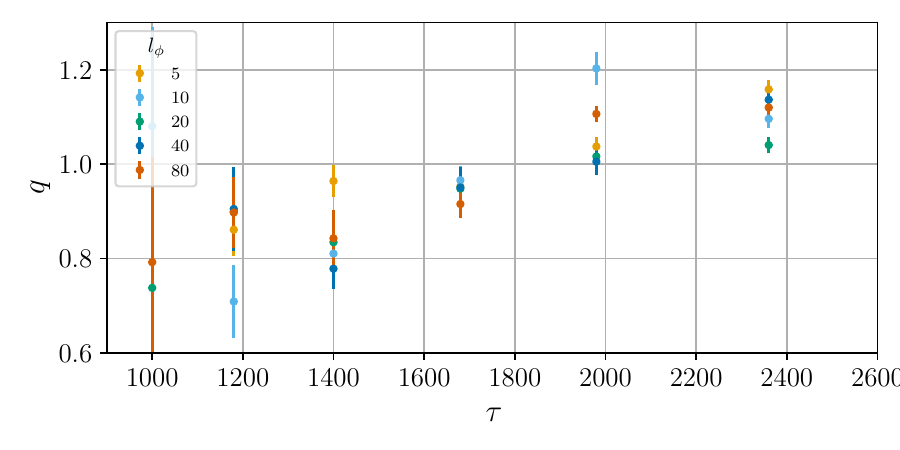}
    \includegraphics[width=0.48\textwidth]{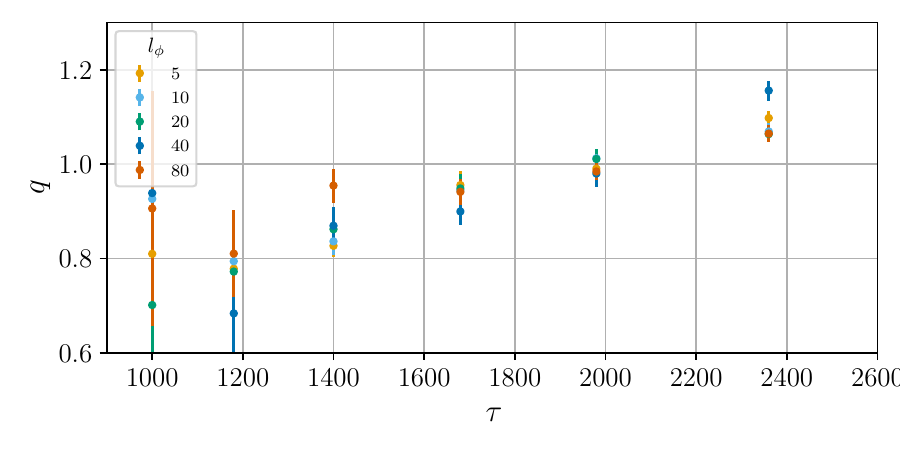}
    \includegraphics[width=0.48\textwidth]{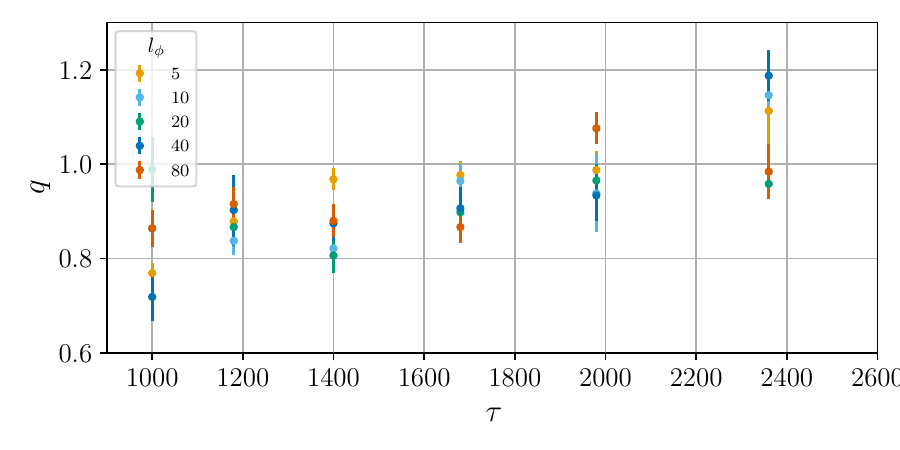}
    \includegraphics[width=0.48\textwidth]{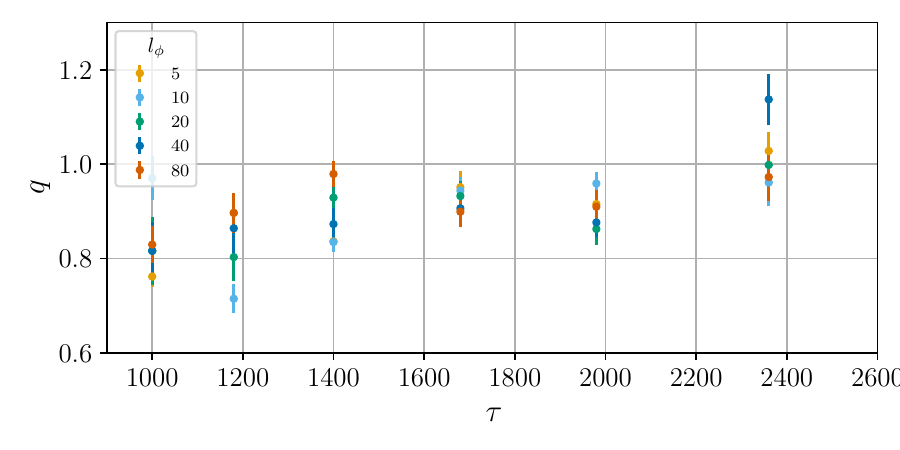}
     \caption{Slopes of power law fits to emission spectra in Fig.~\ref{f:PSdiffJ0J0JsJs}.  Left:  $\tilde{\mcS}_0$; right: $\tilde{\mcS}_s$.  
     Top: fit range $25  < k\tau < \tau/(16 \ws)$; 
     Bottom: fit range $25  < k\tau < 100 $.
     \label{f:emitfits}}
 \end{figure*}

It is important to know whether the plateau extends to higher wavenumbers. 
One can parametrise the shape by fitting estimates of the the $J_0$ emission spectrum to a power law
\ben
\tilde{\mcS}(k,\tau) \propto (k\tau)^{1 - q} .
\label{e:SFitFun}
\een
Other groups find results clustering around $q \simeq 1$, and a tendency for $q$ to increase throughout the simulation \cite{Gorghetto:2020qws,Saikawa:2024bta,Kim:2024wku}.
The fitted value of $q$ depends sensitively on the fitting range \cite{Saikawa:2024bta,Kim:2024wku,Benabou:2024msj}. 
In Fig.~\ref{f:emitfits} we show the results of fits to the form \eqref{e:SFitFun} to our emission spectrum data for both $J_0$ and $J_s$, over various wavenumber ranges, as a function of time. We confirm the general tendency of $q$ to increase throughout the simulation, and the sensitivity to the fitting range.

A feature of all our simulations is that universe-frame length density $\zew$ approaches its scaling value from below. Hence, a scaled power spectrum of the form $\tilde\mcP_\text{ax} \propto \zew(\tau) \ln(k\ta)$ will have an extra contribution to the emission spectrum of the form $\tilde\mcS_\text{ax} \propto \tau \dot \zew \ln(k\ta)$, which acts to increase the slope, and hence decrease $q$, in a scale-dependent way.  As the scaling fixed point is approached, and $\dot\zew \to 0$, the slope of the emission spectrum $1- q$ tends to unity. 

The fitted value of $q$ also depends on the lattice resolution: a coarser grid tends to increase $q$ by depressing the spectrum at higher wavenumbers, although using a higher-order Laplacian operator improves the stability \cite{Saikawa:2024bta}. This may be playing a role in our simulations, which uses the simplest Laplacian, accurate to $\mathcal{O}(\Delta x^2)$. The string width is progressively less well resolved as the simulation proceeds, manifesting as a growth of the emission spectrum at the Nyqvist frequency. The worsening resolution could be an explanation for the dip in the final emission spectrum.  The careful investigation in Ref.~\cite{Saikawa:2024bta} indicates that the $\mathcal{O}(\Delta x^2)$ Laplacian depresses the spectrum by up to 20\% at $k \simeq 1/2\ws$.

In simulations with adaptive mesh refinement \cite{Benabou:2024msj}, 
when the fitting range is chosen to be well away from the inverse string separation and inverse string width, the value of $q$ is consistent with 1.

In our simulations, the final value of $q$ for the $J_s$ emission spectrum fitted in the range $25 < k\tau < 100$ is within one standard deviation of 1 for all initial string densities,  except for $l_\phi = 40$.

\begin{figure*}[htbp]
    \centering
    \includegraphics[width=\columnwidth]{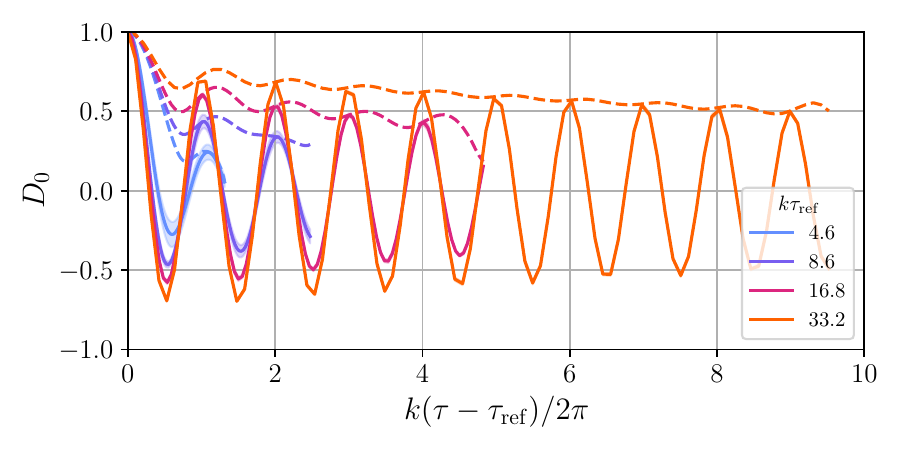}
    \includegraphics[width=\columnwidth]{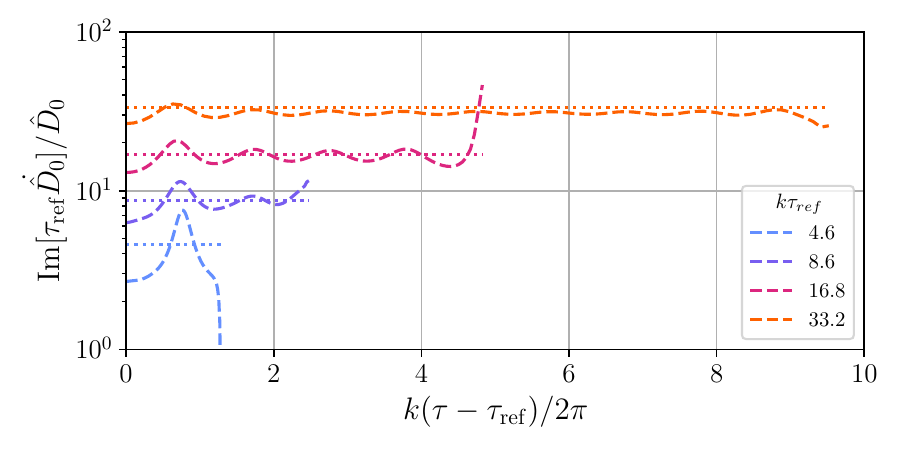}
    \includegraphics[width=\columnwidth]{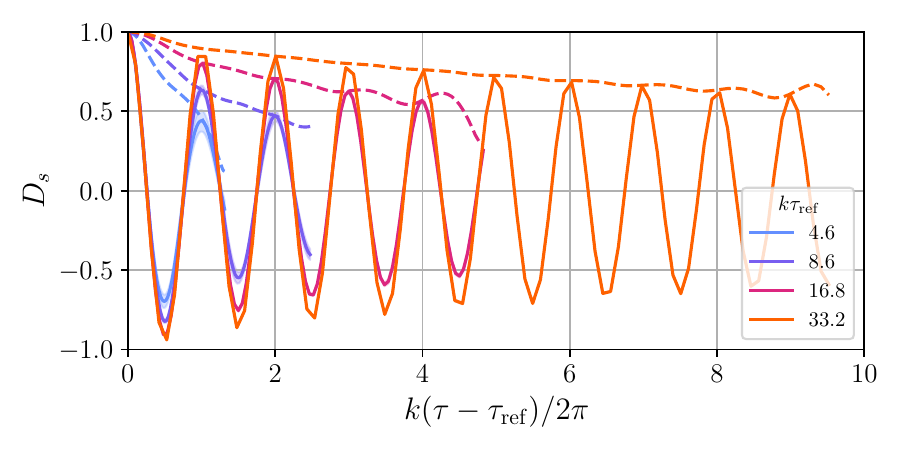}
    \includegraphics[width=\columnwidth]{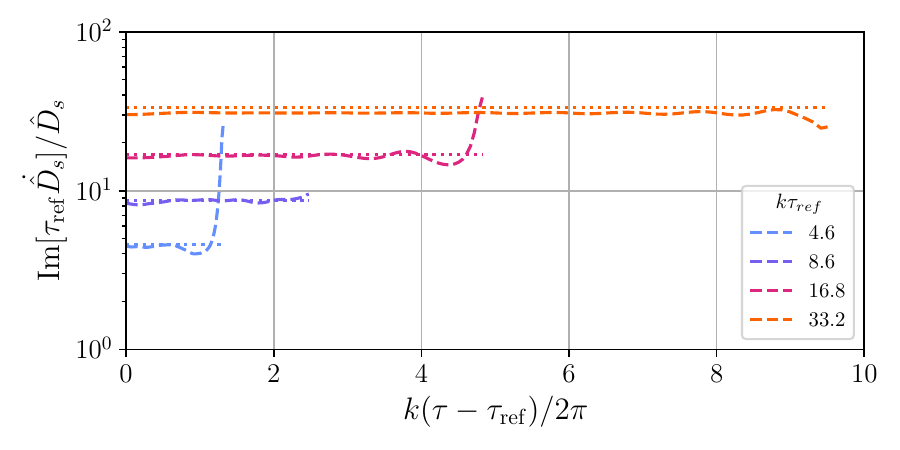}
     \caption{Left: decoherence functions $D_0$,  $D_s$,  for UETCs of currents $J_0$, $J_s$ at $l_\phi = 20$ (solid), along with the amplitude of the analytic signal (dashed), for selected $k\tRef$.  
     Right: the instantaneous frequency of the analytic signal (dashed), with $k\tRef$ plotted as a horizontal dotted line. Error bands showing 1-$\si$ errors on the mean are plotted for the decoherence functions. The analytic signal is computed for the mean decoherence function. See Fig.~\ref{f:uetcs_axion} of Appendix \ref{s:DecFun1} for the decoherence functions $D_0$ and $D_{s}$ for all initial field correlation lengths.
     \label{f:uetc_all_lp20}}
 \end{figure*}

\subsection{Unequal time correlators}
\label{ss:UETCs}

We study the unequal time correlations in the form of the decoherence functions \eqref{e:DecFunDef}, which are normalised to unity at zero lag ($\tau - \tRef=0$) by the square root of the equal time correlators at the two times.
We plot the decoherence functions $D_0$ and $D_s$ for runs with initial field correlation length $l_\phi = 20$ as solid lines in Fig.~\ref{f:uetc_all_lp20}.  We plot against $k(\tau - \tRef)/2\pi$, which shows that $D_0$, $D_s$ oscillate with period $2\pi/k$, as one would expect for time correlations of the massless axion field.   

\begin{figure}[htbp]
    \centering
    \includegraphics[width=\columnwidth]{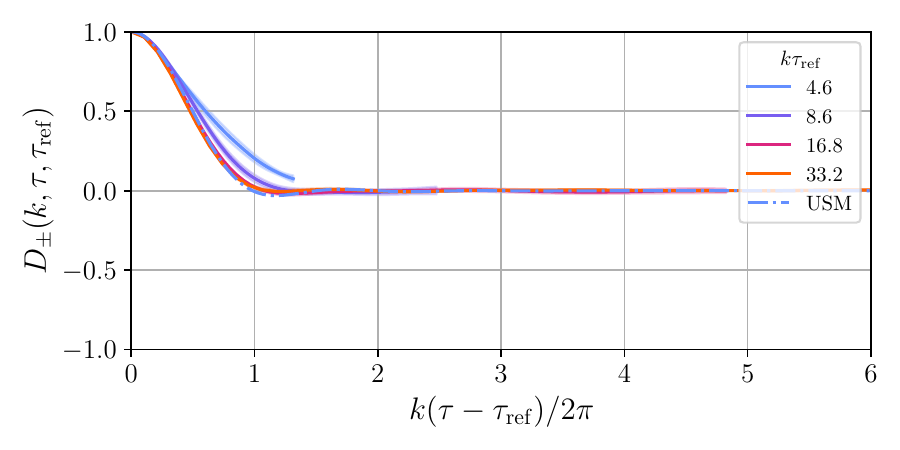}
    \includegraphics[width=\columnwidth]{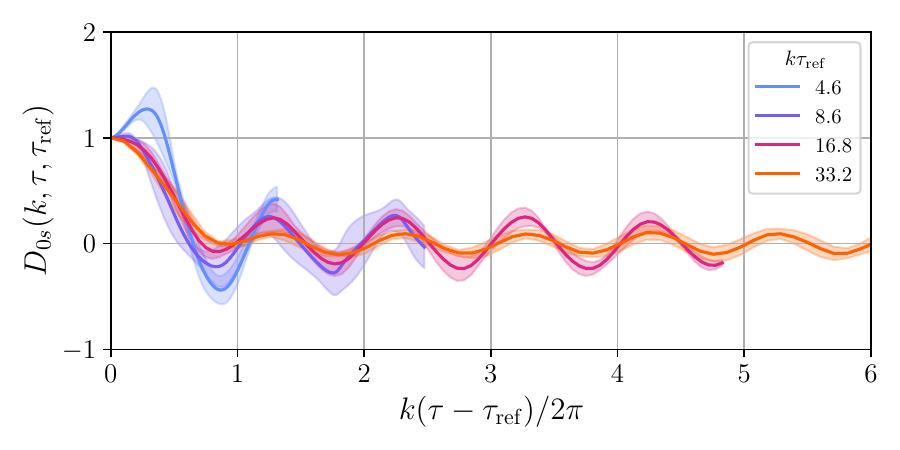}
     \caption{Decoherence functions $D_\pm$ and $D_{0s}$ for initial field correlation length $l_\phi=20$, along with the unconnected segment model (USM) prediction with $\bew = 0.45$ and RMS velocity $\bar{v} = 0.52$ (see Appendix \ref{s:ModUETC}). See Fig.~\ref{f:uetcs_string} of Appendix \ref{s:DecFun2} for the decoherence functions $D_\pm$ and $D_{0s}$ for all initial field correlation lengths.
     \label{f:Dpmlp20usm}}
 \end{figure}

In Fig.~\ref{f:Dpmlp20usm} (top) we plot the decoherence function of the average transverse UETC, $D_\pm$ for runs with initial field correlation length $l_\phi = 20$. 
One sees that $D_\pm$ decorrelates rapidly, and is very close to zero (with small oscillations) for $k|\tau - \tRef| > 2\pi$. 
In Appendix \ref{s:ModUETC} we compute the prediction for this decoherence function in the USM, and plot it for the segment length parameter $\bew = 0.45$ and RMS velocity $\bar{v} = 0.52$, parameters chosen to achieve a good visual fit.  

The result is very close to the data from the simulation in the range $k|\tau - \tRef|< 2\pi$, which lends high confidence to the USM as capturing the essential physics for high wavenumbers. The RMS velocity is somewhat lower than the value obtained from weighted field operators, which ranges between $0.63$ and $0.59$ in the range over which UETCs are measured (see Fig.~\ref{f:zetaw}).

In Fig.~\ref{f:Dpmlp20usm} (bottom) we plot the difference spectrum, $D_{0s}$, against $k(\tau - \tRef)/2\pi$, 
where we define 
\ben
D_{0s} = \frac{ U_{00}(k, \tau_1, \tau_2) - U_{ss}(k, \tau_1, \tau_2) }{\sqrt{P_{0s}(k,\tau_1) P_{0s}(k,\tau_2)} },
\een
where 
\ben
P_{0s}(k,\tau) = P_{0}(k,\tau) - P_{s}(k,\tau) = U_{00}(k, \tau, \tau) - U_{ss}(k, \tau, \tau). 
\een
The decoherence function $D_{0s}$ is similar to $D_\pm$, with an oscillatory component which becomes smaller at higher wavenumbers.  This adds evidence that the difference spectrum $\mcP_0 - \mcP_s$ is dominated by strings at high wavenumbers.   At wavenumbers $k\tau \sim 1$ freely propagating axion radiation does not obey 
$\mcP_0 - \mcP_s = 0$, and one expects to see some oscillatory behaviour.  One can also expect to see a more complicated decoherence than either $D_0$ or $D_s$ due to the combined behaviour of the $D_0$, $D_s$ and the relative amplitude of $P_0$ and $P_s$. 

We show decoherence functions for all initial correlation lengths in
 Fig.~\ref{f:uetcs_axion} of Appendix \ref{s:DecFun1} ($D_0$ and $D_s$) 
and  Fig.~\ref{f:uetcs_string} of Appendix \ref{s:DecFun2} ($D_\pm$ and $D_{0s}$).

Returning to $D_0$, $D_s$, we can investigate the functions using the analytic signal (see e.g. \cite{hlawatsch2013time}). 
The analytic signal of a real-valued function $D(t)$ with Fourier transform $\tilde{D}(\omega)$ can be defined as the inverse Fourier transform of 
twice the positive frequency part, 
or 
\ben
\AnaSig{D}(t) = \int \frac{d\omega}{2\pi} \left[ \tilde{D}(\omega) + \sgn(\omega)\tilde{D}(\omega) \right] e^{-i\omega t}.
\een
The analytic signal is a complex function with only a positive frequency component.  Its imaginary part is the Hilbert transform of the original function. 
The instantaneous amplitude is defined as 
\bea
A_D(t) &=& |\AnaSig{D}(t)|,
\eea
while the instantaneous angular frequency is 
\bea
\om_D(t) &=& \Im \left(  \frac{1}{\AnaSig{D}(t)} \frac{d \AnaSig{D}(t) }{dt} \right).
\eea
The analytic signal is useful for investigating the envelope of a modulated periodic signal.

We compute the analytic signal of the decoherence function at a particular wavenumber by symmetrising it around $\tau - \tRef = 0$, in order to improve the behaviour near the origin.  
The results for selected values of $k\tRef$ are plotted as dashed lines in Fig.~\ref{f:uetc_all_lp20}, with the instantaneous amplitude in the left column, and the instantaneous frequency on the right. Dotted lines in the right column show the frequency $k$, which is the value for a propagating axion with wavenumber $k\tau \gg 1$. 

The plots confirm that $D_0$ and $D_s$ have a strong signal with angular frequency $k$ for $k\tRef \gg 1$,  consistent with the dominant source of the correlations being the propagating axion field.  The envelopes show a bump at the origin of width $k\tRef \simeq 1$, particularly prominent for $\AnaSig{D}_0$, which supports the proposal that there is a contribution from the moving strings.  

\begin{figure}[htbp]
    \centering
    \includegraphics[width=\columnwidth]{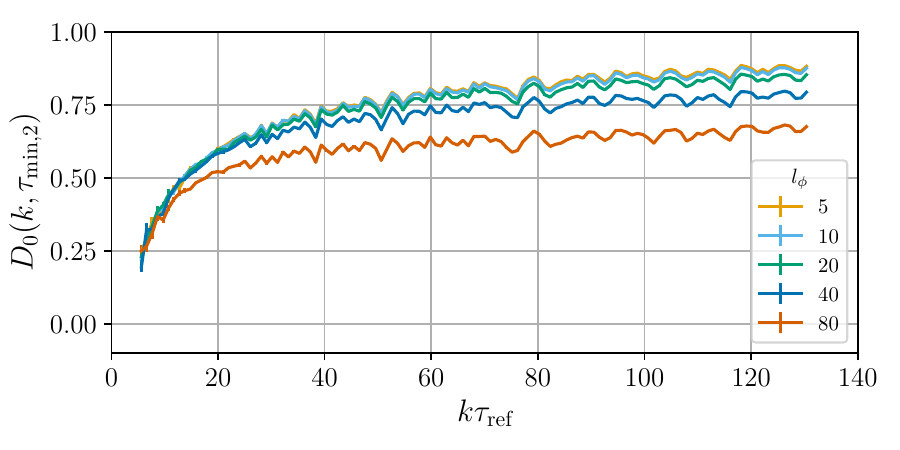}
    \includegraphics[width=\columnwidth]{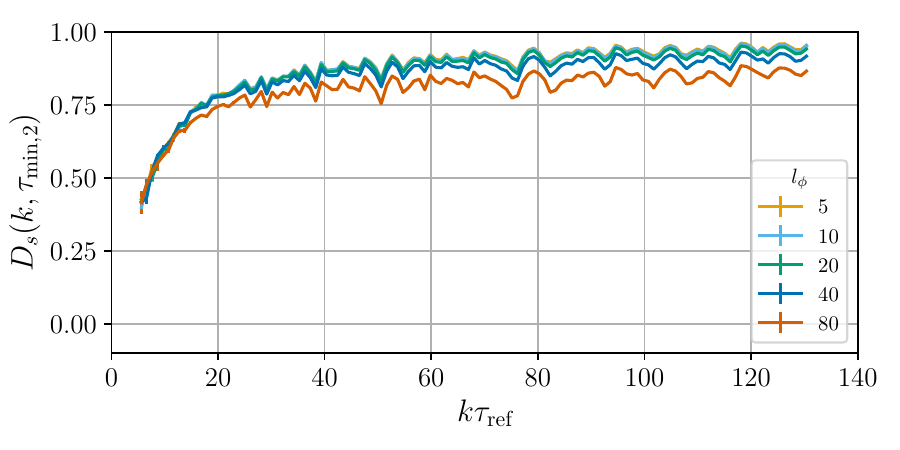}
     \caption{The value of decoherence functions $D_0$ and $D_s$ at the second minimum as a function of $k\tRef$. Error bars show 1-$\si$ errors on the mean are plotted, but are smaller than the linewidth. 
     The oscillations are due to the finite sampling rate of the UETC, which is itself oscillating. The sampling rate and the decoherence oscillation frequency are the same for all initial string densities.
     \label{f:D_min2}}
 \end{figure}

The envelope decays more slowly at larger $k\tRef$. We do not have a good model for this decay, and so we cannot be very quantitive about the relative contribution from the strings. However, as an approximate lower bound on the contribution from propagating axions, we can take the second local maximum of the function at approximately $k(\tau - \tRef)/2\pi = 3/2$, where the envelope of $D_+$ is very small.
We plot the results in Fig.~\ref{f:D_min2}. The maximum value of $D_+$ at the time of the second minima of $D_s$ is approximately $0.02$ for $k\tRef > 20$.

The results show that the  amplitude of the second local maximum is higher in $D_s$ than in $D_0$, which is consistent with the proposal that the propagating and oscillating axion field  contributes a higher proportion.  In both graphs, the  amplitudes are higher for higher initial  string  densities. This can be explained as a result of an initial period of extra radiation as the higher string densities evolve to be closer to scaling.   
Recalling that the string contribution is of order $v^2 \simeq 0.33$, we note that, at high $k\tRef$, the amplitude for the second local maximum of $D_0$ at  $l_\phi = 80$ is between 0.6 and 0.7, which is consistent with the expected value in the model of $1 - v^2 \simeq 0.67$.

\section{Results and discussion: number density power spectra}
\label{s:NumDen}

\begin{figure}[htbp]
    \centering

\includegraphics[width=0.47\textwidth]{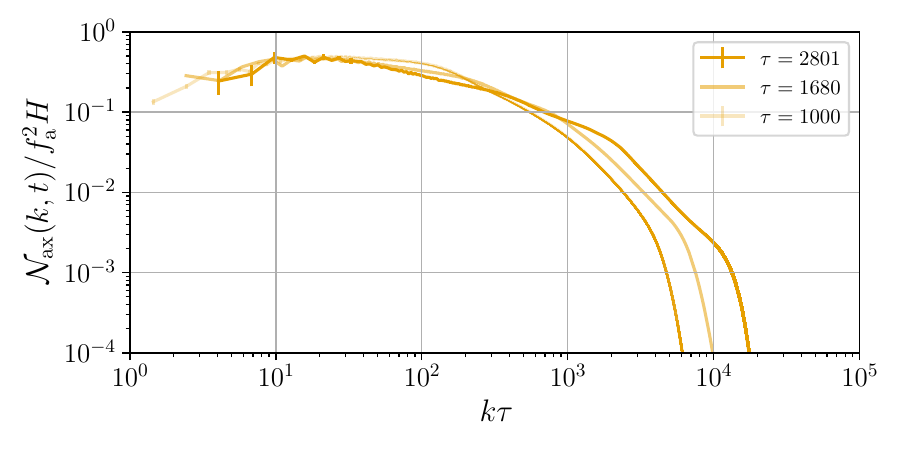}
\includegraphics[width=0.47\textwidth]{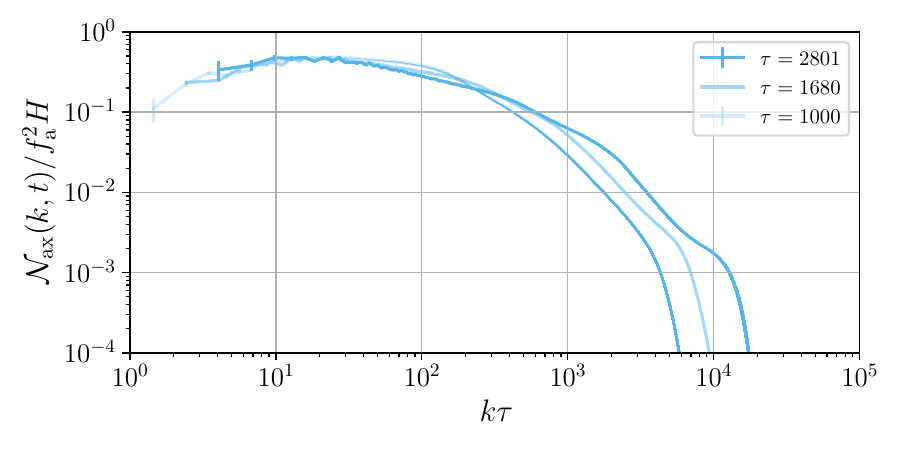}
\includegraphics[width=0.47\textwidth]{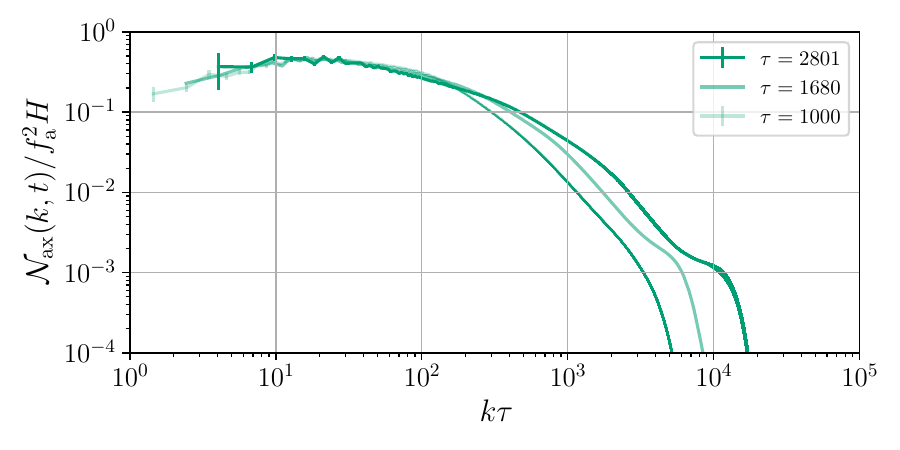}
\includegraphics[width=0.47\textwidth]{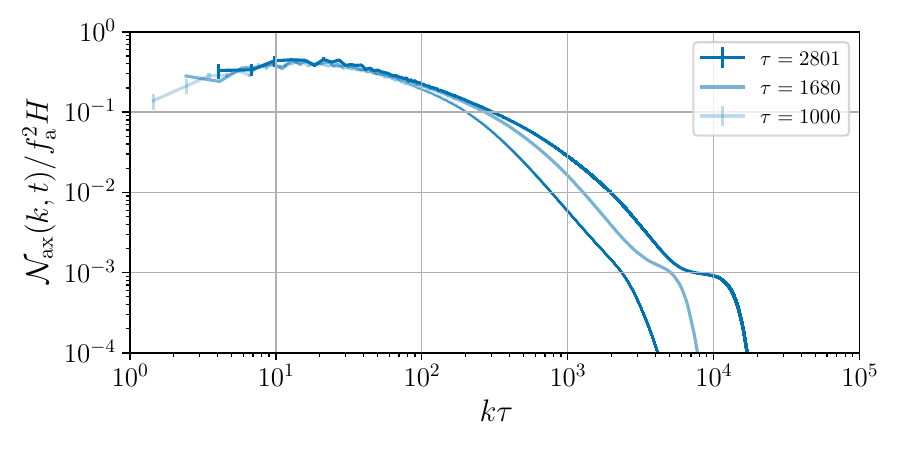}
\includegraphics[width=0.47\textwidth]{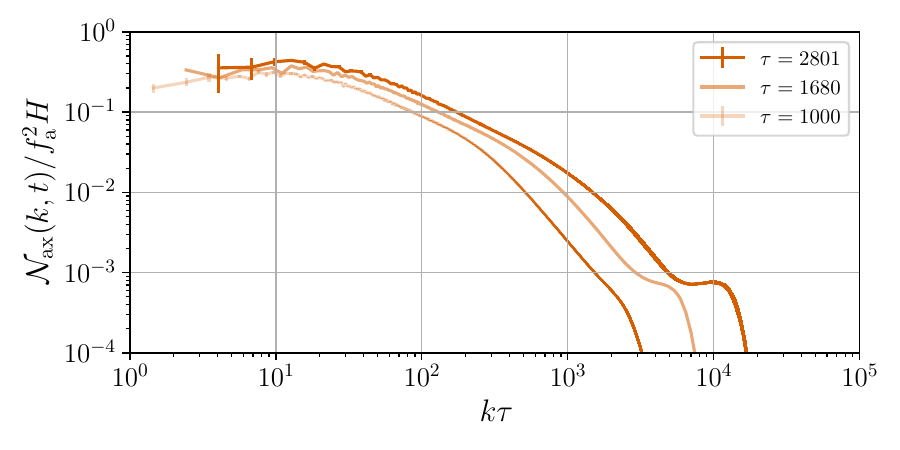}

\caption{ \label{f:nax_by_lp}
Number density power spectra, defined in Eq.~\eqref{f:NumDenPSDef}, at selected conformal times given in the legends. Top to bottom: initial field correlation lengths $\l_\phi = 5, 10, 20, 40, 80$. 
}
 \end{figure}

In this section we show our results for the number density power spectra $\mcNax^s(k,\tau)$  \eqref{numberdensity}. 
These spectra are shown in Fig.~\ref{f:nax_by_lp}, normalised by $\fa^2 H$, where $H$ is the physical Hubble rate, and plotted against $k\tau$. With this normalisation the number density spectrum in a scaling string network collapses to a single line. 

In fact, we see that simulations with higher initial string densities (initial field correlation length $l_\phi = 5,10,20$) collapse onto a single line in the wavenumber range $k\tau \lesssim 100$, while the lowest initial density increases throughout the simulation.  At higher wavenumbers the number density spectra differ, but the differences are in the wavenumber range where axion radiation emitted during the initial acceleration phase can be important. The evolution of the spectra towards the line of collapse is noticeable in the range $25 < k\tau < 100$, where fits to the emission spectrum are carried out, and the evolution of the index $q$ can be understood as a signal of this evolution.

\begin{figure}[htbp]
\begin{center}
\includegraphics[width=\columnwidth]{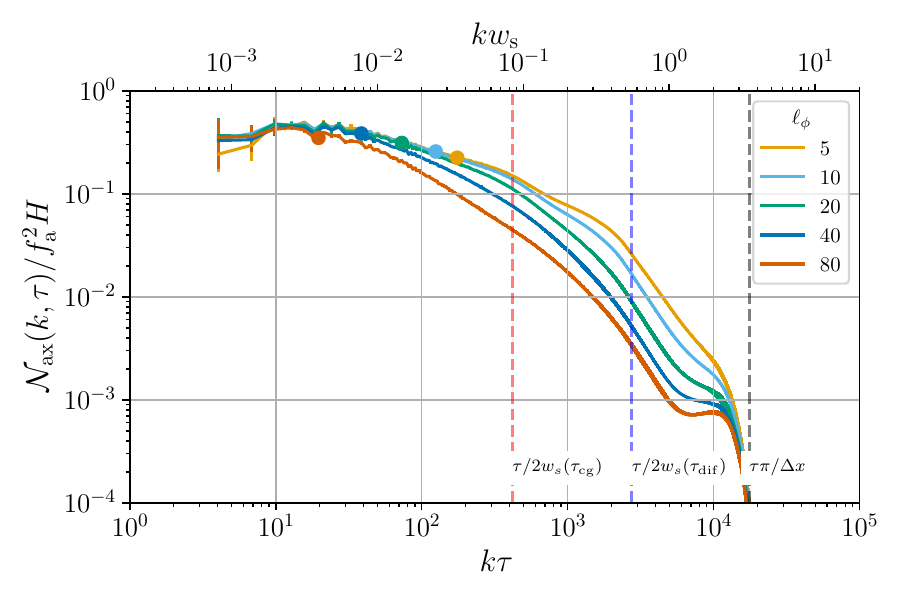}
\caption{Final number density power spectra, with scale annotations. Coloured dots mark the wavenumbers corresponding to the universe-frame mean string separation at the start of physical evolution, $\xi(\tcg)$, in each set of simulations.
}
\label{f:nax_all_final}
\end{center}
\end{figure}

The spectra at the end of the spectrum-measuring interval is shown in Fig.~\ref{f:nax_all_final}, where we see agreement amongst higher initial string densities in the range $k\tau \lesssim 100$, while the spectrum of lowest initial string density is substantially below the others.  This is strong evidence that the low initial density simulations have not converged. Other groups base their number density results on an extrapolation of the amplitude evolution in low-density string networks, which our results indicate will eventually cease. 

\begin{figure}[htbp]
    \centering

\includegraphics[width=\columnwidth]{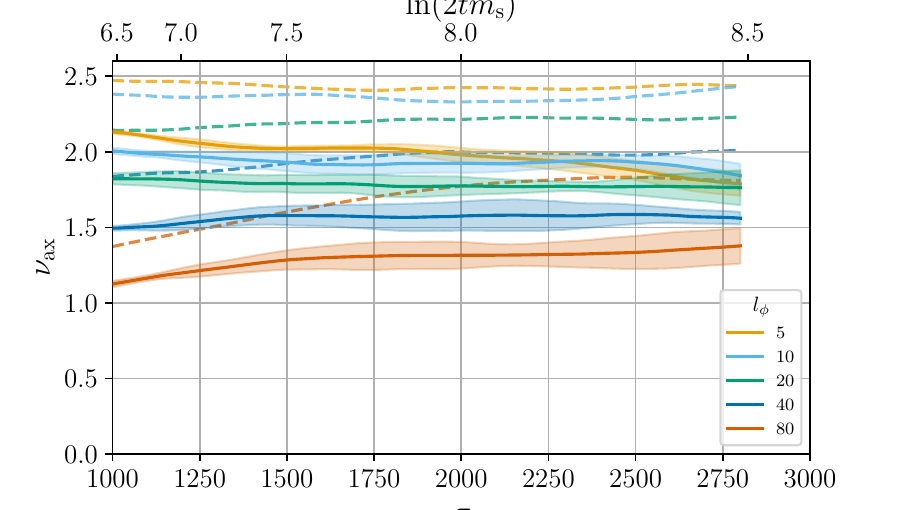}

\caption{ \label{f:nax_tot}
Dimensionless number density parameter Eq.~\eqref{e:NuaDef} against conformal time $\tau$, with the natural logarithm of the Hubble length in inverse scalar mass units as a secondary axis,  for all initial correlation lengths $\l_\phi$. 
Solid lines with 1-$\si$ error bands show the number density from the spectrum evaluated with $\mcP_s$ \eqref{f:NumDenPSDef}, the power spectrum of the longitudinal component of the current. The dashed lines show the mean value of the number density spectrum from the spectrum evaluated with $\mcP_0$, the timelike component of the current. The latter has a string contribution of $\mathcal{O}(v^2)$, while the string contribution is only $\mathcal{O}(v^4)$ in the former.
}
 \end{figure}

Let us 
define a dimensionless number density parameter 
\ben
\nu_\text{ax} = \frac{\nax}{\fa^2 H}.
\label{e:NuaDef}
\een
We plot this quantity against conformal time in Fig.~\ref{f:nax_tot}. It is computed by summing the spectral density $P_s(k,\tau)/k$ over all Fourier modes, with $k$ the modulus of the wavevector. For convenience of comparison with other groups (see e.g.~Fig.~12 of Ref.~\cite{Gorghetto:2018myk}), we give $\ln(2 t \ms)$ as a secondary $x$ axis.

We also show, as dashed lines, what the mean number density would be if computed with $\mcP_0$ instead of $\mcP_s$, which includes an $\mathcal{O}(v^2)$ contribution from the axion field of the string.  This estimate, used by all other groups,  is significantly higher.  The simulations with $l_\phi = 80$ have similar string densities to the the initial string densities favoured by other groups, and the number density indicated with the dashed line is broadly compatible with an extrapolation of Fig.~12 of Ref.~\cite{Gorghetto:2018myk}, which finishes at  $\ln(2 t \ms) = 6$ with $\nu_a \simeq 1.6$.

There is a clear convergence in $\nua$, which starts out in the range $1.12(2) < \nua(\tRef) < 2.13(1)$  and finishes in the range $1.38(12) < \nua(\tFinPS) < 1.79(8)$. 
Taking the mean and standard deviation of all final central values, even those which have not yet reached scaling,  leads to our estimate of the scaling axion number density, 
\ben
\nu_{\text{ax},*} = 1.66(17) .
\label{e:NuVal}
\een
The simulations with number density power spectra closest to scaling are $l_\phi = 20$ and $l_\phi=40$, and their final values $1.56(4)$, $1.76(12)$ lie within 1-$\si$ of the inferred scaling number density. The number density is a more stable quantity than the high-wavenumber slope, as it is dominated by a lower-wavenumber part of the spectrum which is much closer to its asymptotic form. 

The number density can be converted to an equivalent axion mass $m_{\text{ax},*}$ using standard methods \cite{Wantz:2009it,Marsh:2015xka}, and conservation of axion number.  This axion mass equivalent is defined by the assumption of number conservation, and is not a prediction of the axion mass. 

It is most convenient to use the relationship between $\nu_{\text{ax},*}$ and axion mass displayed in Fig.~29 of Ref.~\cite{Saikawa:2024bta}.  The $x$ axis of that figure shows $K = \nu_{\text{ax},*}/c_\text{mis}$, where  $c_\text{mis} = 2.31$, and the $y$ axis the equivalent axion mass, for various temperature dependences of the QCD topological susceptibility $\chi(T)$.  The susceptibility gives the axion its mass through $\ma^2(T) = \chi(T)/\fa^2$. The constant $c_\text{mis}$ is an estimate of the axion number density resulting from the misalignment mechanism in the pre-inflationary scenario. The graph takes a reference temperature $T_*$ defined by $\ma(T_*) = H(T_*)$, and the high-temperature susceptibility measurement of Ref.~\cite{Borsanyi:2016ksw}.

We have $K = 0.71(7)$ from the scaling axion string network, for which the axion mass equivalent with a temperature dependence $\chi \propto T^{-8}$ is $m_\text{ax,*} \simeq 4\,\mu$eV.  We emphasise that this is not a prediction of the axion mass, as the number density does not include the contribution from the network annihilation. Rather, it is a scale-setting number, which quantifies the importance of axions produced by a scaling network down to the temperature $T_*$. We will give our axion mass prediction elsewhere, based on simulations with a time-dependent mass and network annihilation by domain walls \cite{Correia:2026wnl}.

\section{Conclusions}
\label{s:Con}

In this paper we presented an analysis of field power spectra, unequal time correlators (UETCs) and axion number density in cosmological simulations of axion string networks in the radiation era. We leave for the future matter era simulations, appropriate for scenarios with early matter domination by heavy quarks  \cite{Cheek:2023fht,DiLuzio:2024xnt}.

The simulations were performed on grids of side $12288$, with initial conditions spanning a wide range of network length densities, consisting of widely separated smooth strings with otherwise negligible field excitations. 
%Our initial conditions reduce the spurious oscillations in the power spectrum due to unphysically large oscillations in the scalar field, observed in other simulations. 

The analysis tests the standard scaling framework beyond the simple measures of string density and RMS velocity, finding strong evidence in favour of scaling in the power spectra.
With scaling, we are able to extrapolate our simulations from soon after the formation of the network at the Peccei-Quinn phase transition up to the beginning of its annihilation at the QCD transition, and accurately estimate the contribution of scaling axion string network to the axion number density. 

In order to better separate the contribution to the axion field from propagating axions from that of the string network, we introduced a current $J_\mu$, equal to the PQ symmetry current divided by the field modulus, and studied projections of spatial parts in Fourier space.  
%With this definition of the current, the energy density in the axion field is $[(J_0)^2 + \bJ^2]/2$.

Other groups have studied observables close to or equivalent to $J_0$, which is proportional to the time derivative of the axion field in string-free field configurations.  
%We show that the power spectrum of this observable $\mcP_0$ is ``contaminated'' by an $\mathcal{O}(v^2)$ contribution from the string network, where $v \simeq 0.57$ is the network RMS velocity. 
We introduced new observables $J_s$ and $J_\pm$, the longitudinal and transverse components of the spatial part of the current.  
%whose power spectrum $\mcP_s$ has only an $\mathcal{O}(v^4)$ contribution from strings. 
The transverse components of the current $J_\pm$ receive contributions only from strings, and so the corresponding power spectra $\mcP_\pm$ enable the monitoring of the scaling of the network at all wavenumbers. 

We took 91 axion spectra in a conformal time range 
%$1000 \le \tau \le 2800$, a range 
chosen to allow the network time to establish scaling, terminating before the half light-crossing time of the simulation box, $\tau_{1/2} = 3072$, and maintaining the lattice spacing above half the inverse scalar mass. 
In practice, the networks with the two lowest initial densities had not yet reached scaling, as indicated by both the mean string density and the power spectra. High initial densities generated a significant non-scaling component of axion radiation at high wavenumber, but up to wavenumber of order $10^2$ inverse horizon lengths the power spectra $\mcP_0$ and $\mcP_s$ collapsed onto similar curves, a signature of scaling. The transverse power spectra $\mcP_\pm$  exhibited collapse onto the same shape curve over the entire wavenumber range, with amplitude differences ascribable to the differences in string density.

In order to have a quantitative understanding of the observations, we introduced a model of the axion field generated by the string network based on the unconnected segment model (USM) \cite{Vincent:1996qr,Albrecht:1997mz,Pogosian:1999np,Avgoustidis:2012gb,Charnock:2016nzm}, in which the string-sourced axion field is generated by a statistical ensemble of string segments with length, number density, and RMS velocity chosen to reproduce the measurements of length density and RMS velocity of the string network.

The transverse power spectra agreed well with the USM, which predicts them to be flat in the range $k\xiw^\text{c} \ll k\tau \ll k\ws$, with an amplitude proportional to the length density $\xiw^{-2}$. We found that the measured value was approximately 40 -- 50\% below the value predicted by the model on the basis of the string density alone, which we count as a success for such a simple model. The disagreement can be accounted for by introducing a parameter which models the segment length separately from the number density. We leave this modelling improvement for a future work. Extensions of the USM incorporating dynamics of loops and kinks can already be found in the literature \cite{Rybak:2021scp,Silva:2023diq}.

We showed that in the USM strings contribute at $\mathcal{O}(v^2)$ to $\mcP_0$ and  $\mathcal{O}(v^4)$ to $\mcP_s$, while axion radiation contributes equally to both for $1 \ll k\tau$. The difference in the spectra is therefore a tracer of the string component, and we demonstrated from the data that the difference spectra had a similar shape to the transverse $\mcP_\pm$, which is unambiguously due only to strings. The combination of model and data therefore enables an estimate of the string contribution to the  $\mcP_0$  and  $\mcP_s$ power spectra.  

Given the flat axion field power spectrum for the string-sourced field, we argue for a flat spectrum for the emission of propagating modes of the axion field, or $q=1$ in the common parametrisation. The argument is based on the equation of motion for the axion energy power spectrum, which shows that the emission spectrum has a manifestly negative flat contribution proportional to $\mcP_0 - \mcP_s$, plus an extra term which must at least cancel it. Direct measurement of the emission spectrum shows that $q \gtrsim 1$, which leaves $q=1$ as the most likely value. Support for this value also comes from the fact that collapsing loops also have a $q=1$ emission spectrum.

We show that $q=1$ implies that the axion radiation power spectrum, the dominant part of $\mcP_0$  and  $\mcP_s$,  is proportional to $\ln(k\tau)$ in the range $\tau/\xiw^\text{c} \ll k\tau \ll \tau/\ws$, which is consistent with considerations of the covariant energy conservation of the axion radiation density. 

The logarithmic behaviour emerges most clearly for wavelengths which start out much greater than the mean string separation and end up much less.  To achieve these conditions we focused on wavenumbers in the range $25 < k\tau < 100$. Here we found  the logarithmic coefficients $p_0$ and $p_s$, for power spectra $\mcP_0$ and $\mcP_s$, both converging to a value of around 10, bracketed by simulations with initial string density parameters $l_\phi = 20$ and $l_\phi = 40$. The consistency of the coefficients is expected for axion radiation, which contributes equally to both spectra.

At high wavenumbers in the emission spectrum we observed a strong feature peaking at half the inverse mass of the scalar mode.  The higher the initial string density, the stronger the feature was.  
We interpret this feature as resonant axion production from oscillating states of the massive mode bound to the string \cite{Goodband:1995rt,Saurabh:2020pqe,Blanco-Pillado:2021jad,Baeza-Ballesteros:2023say}.

We have measured for the first time the unequal time correlators (UETCs) of the axion field, defined in Eq.~\ref{UETC}.  We exhibited the decoherence functions, which are obtained by normalising with the square roots of the power spectra (see Eq.~\ref{e:DecFunDef}), in Figs.~\ref{f:uetc_all_lp20}, \ref{f:uetcs_axion} and \ref{f:uetcs_string}. The decoherence function of the transverse components of the axion field $D_\pm(x_1, x_2)$ is well described by the USM with a velocity distribution which is a power-law in the Lorentz $\ga$-factor \cite{Gorghetto:2020qws}. It predicts that the decoherence function becomes very small for time differences $\Delta\tau \gtrsim 1/(2\pi k \bar{v})$, where $\bar{v}$ is the RMS velocity.  Good fits are obtained with an RMS velocity $\bar{v} = 0.52$ at conformal time $\tRef = 1000$, for all initial string densities. 
This value is lower than the values obtained from averaged weighted field operators \cite{Correia:2024cpk} at that time, which are in the range $0.62 < \bar{v} < 0.65$.
This suggests that the excitations of the scalar field around the string, in evidence from the emission spectrum, are also contributing to the operator-based estimates of the velocity displayed in Fig.~\ref{f:zetaw}, especially at high initial string densities. All local field-based estimates of the velocity are likely to be affected in this way. Further work is needed to understand the excitations and their impact on the string velocity.

The decoherence functions $D_0$ and $D_s$ confirm that the axion field has contributions from both propagating modes and the string, with a smaller contribution to $D_s$.  
%The string components die out for time differences $\Delta \tau > 1/(2\pi k\bar{v})$, leaving behind an oscillating function of $k\Delta\tau$.  
Analysis of the analytic signal shows that they both have an oscillating component of frequency $k$, as expected for propagating axions, and that its  amplitude of slowly decays.  The decay is perhaps the result of the radiation emitted from the string being out of phase with the background axion field. We note that the oscillatory behaviour of $D_0$ and $D_s$  and the rapid decay of $D_{\pm}$ are qualitatively similar to decorrelation of compressional \cite{Correia:2025qif} and vortical modes \cite{kraichnan:1964,Auclair:2022jod} in a fluid.

When considering the number density power spectrum, the observation that the power spectrum $\mcP_0$ has an $\mathcal{O}(v^2)$ contribution from strings motivates the use of $\mcP_s$ to extract the number density of axions emitted by the string network, according to Eq.~\eqref{f:NumDenPSDef}. In a scaling string network, the spectra evolve towards a common function of $k\tau$, multiplied by $\fa^2 H$, where $H$ is the physical Hubble rate.  The measured spectra, displayed in Figs.~\ref{f:nax_by_lp}, show a clear evolution towards a common function, with all but the lowest initial string density collapsing onto the same curve in the range $k\tau \lesssim 100$ by the end of the simulation (see Fig.~\ref{f:nax_all_final}). In the range $10 \lesssim k\tau \lesssim 100$ the curve $\ln(k\tau)/k\tau$ follows from the fitted form of $\mcP_s$ and the definition of the number density spectrum.

By summing all Fourier modes in the number density spectrum we obtain a measurement of the axion number density, which in a scaling network is proportional to  $\fa^2 H$.  The constant of proportionality $\nua$, plotted in Fig.~\ref{f:nax_tot}, is remarkably constant throughout the period of measurement and across all initial string densities, and converges towards an $\mathcal{O}(1)$ value \eqref{e:NuVal}.  This is our headline result for the axion number density in a scaling string network. The stability of this number suggests that attention may move away from fits of the emission spectrum to a more precise characterisation of the number density spectrum.

We note that computing the number density with the power spectrum $\mcP_0$, as done by all other groups up to now,  gives values approximately 30\% higher, and with a less clear convergence, because of the contamination by the ``Coulomb" field of the strings. We note also that other groups have in recent years focused almost exclusively on low-density initial conditions, whose number density parameter $\nua$ increases throughout their simulations.  This has been taken as a sign that the number density evolution should be extrapolated over many e-foldings of cosmic expansion. In our simulations, which have a wide range of initial string densities, and use a less contaminated measure of number density, scaling in the number density is observed.  This supports our earlier conclusion that the apparent scaling violation is in fact a transient \cite{Correia:2024cpk}, and the extrapolation is therefore unwarranted.

The axion number density measurement can be translated into an equivalent axion mass, by assuming number density conservation and equating to the dark matter density today.  While not a prediction of the axion mass itself, it is nonetheless a useful scale-setting estimate. Thus defined, and using the high-temperature axion mass computed in Ref.~\cite{Borsanyi:2016ksw}, 
the axion mass equivalent for a scaling string network is approximate $4\,\mu$eV.

In a forthcoming publication we will report on simulations which include the dynamics of axion production during the annihilation phase of the string network, and thus a full prediction of the dark matter axion mass in this model, which is the simplest realising the Peccei-Quinn solution of the strong CP problem.\\

\section*{Acknowledgments}
JC (ORCID ID 0000-0002-3375-0997) acknowledges support from Research Council Finland grant 354572, ERC grant CoCoS 101142449 and from the European Union’s Horizon Europe research and innovation programme under the Marie Skłodowska-Curie grant agreement No. 101126636; MH (ORCID ID 0000-0002-9307-437X) from Academy of Finland grant 333609, Research Council of Finland grant 363676 and STFC grant ST/X000796/1; KR (ORCID ID 0000-0003-2266-4716) from the ERC grant CoCoS 101142449 and the Research Council of Finland grant 354572.
JL (ORCID ID 0000-0002-1198-3191), ALE (ORCID ID 0000-0002-1696-3579) and JU (ORCID ID 0000-0002-4221-2859) acknowledge support from Eusko Jaurlaritza IT1628-22 and by the PID2024-156016NB-I00 grant funded by MCIN/AEI/10.13039/501100011033/ and by ERDF; “A way of making Europe”. JC and MH acknowledge support from COST (European Cooperation in Science and Technology) Action COSMIC WISPers CA21106. Our simulations made use of Lumi at CSC Finland under pilot access project AxCESS with 1.1 MGPU-hours.

\appendix

\section{Axion current from strings in the USM}
\label{s:StrFieUSM}

In this appendix we use the unconnected segment model (USM) to produce expression for the axion current $J_\mu$ produced by segments of string, which will used in subsequent appendices to compute power spectra and decoherence functions.

We start with the observation that the field equation \eqref{eq:eom} possesses vortex or string solutions \cite{Davis:1986xc}.  
Consider a field with the form,  
\ben
\Phi(\bx) = \ps(\rho) e^{i \varphi},
\een
where $\rho$, $\varphi$ are cylindrical coordinates and $\psi$ is real.  This is a solution to the field equations if $\psi$ satisfies
\bea
\ddot \psi + \frac{\dot a}{a} \dot \psi - \frac{1}{\rho} \frac{d}{d\rho} \left( \rho \frac{d\ps}{d\rho}\right) 
 + \frac{ \ps}{\rh^2}  +  a^2\la(\ps^2 -\fa^2)\ps = 0, \nonumber \\
\eea
with boundary conditions $\psi = 0$ at $\rho = 0$, and $\psi \to \fa$ as $\rho \to \infty$.
The time dependence of the equations may be neglected in the limit $\ws/\tau \to 0$, where 
\ben
\ws =  \frac{1}{a(\tau) \ms}
\label{e:StrWidDef}
\een
is the comoving width of the string, and $\ms = \sqrt{2\la}\fa$. 
There is then an approximate solution $\psiStr(\rho/\ws)$, with 
\ben
\psiStr(s) \sim 
\left\{
\ba{ll}
\fa s, &  s \to 0, \\
\fa \left( 1 - {s^{-2}} \right), & s \to \infty,
\ea\right.
\een
and $s = \rho/\ws$. 
The current of this configuration is 
\ben
\Jstring_i(\bx) = - \ep_{ij3} \frac{x^j}{\rho^2} \psiStr(\rho/\ws). 
\label{e:JstrForm}
\een
Hence, at large distances from the string, the energy density is dominated by the current, and the fully covariant energy density becomes 
\ben
T_{00} \to \half \frac{\fa^2}{\rho^2}. 
\label{e:EneDenStr}
\een
The fully covariant energy per unit length of a single static string $\mu \to \pi \fa^2 \ln(R/\ws)$  is therefore logarithmically divergent as the upper limit of the integral $R$ tends to infinity.  In the early universe, strings are formed at phase transitions, separated by finite distances \cite{Kibble:1976sj,Zurek:1996sj}, acting as an upper cut-off to the integral. Hence the energy per unit length of the field associated with string remains finite. 

It will be convenient to use a spherical basis $\mathbf{e}_s, \mathbf{e}_\pm, $ for quantities in Fourier space. 
Labelling a Cartesian basis with $\mathbf{e}_1$,  $\mathbf{e}_2$,  $\mathbf{e}_3$, we can choose
\ben
\mathbf{e}_{s} = \mathbf{e}_3, \quad
\mathbf{e}_{\pm} = \pm \frac{1}{\sqrt{2}} \left( \mathbf{e}_1 \mp  i \mathbf{e}_2\right)
\een
Then for each wavevector $\bk$ we can align the basis as
\ben
e_{s}^i \equiv \hat{k}^i, \quad
i \ep_{ijk}e_{s}^je_{\pm}^k = \pm e_{\pm}^i.
\een
We consider comoving coordinates and wavevectors, and we do not distinguish between raised and lowered spatial indices.

The Fourier transform of the current \eqref{e:JstrForm} is 
\bea
\tilde{\Jstring}_i(\bk) &=& - \ep_{ij3} \fa \int d^3x  \frac{{x}^j}{\rho^2} F(\rho/\ws)  e^{-i \bk\cdot\bx} ,
\eea
where $F(s) = \psiStr(s)/\fa$. 
Evaluating the integrals in cylindrical coordinates, 
\bea
 \tilde{\Jstring}_i(\bk) &=& 2\pi i \ep_{ij3} \hat{k}_\perp^j \tilde{G}(k_\perp\ws) \int dz e^{-ik_z z}
 \label{e:JStrFT}
\eea
where $k^j_\perp = (k^1, k^2, 0)$ $k_\perp = \sqrt{k_1^2 + k_2^2}$, and $\hat{k}_\perp^j = k^j_\perp/k_\perp$.
After using a Bessel function identity we get 
\bea
\tilde{G}(k_\perp\ws) &=&  \fa  \int_0^\infty d\rho  F(\rho/\ws) J_1(k_\perp\rho),
\eea
and $J_1$ is the Bessel function of order 1. 
Then 
\bea
\tilde{G}(k_\perp\ws) &=&  \frac{\fa}{k_\perp} \tilde{F}(k_\perp \ws),
\eea
with $\tilde{F}(\ka) = \int_0^\infty ds  F(s/\ka) J_1(s)$, $s = k_\perp \rho$ and $\ka = k_\perp\ws$. 
For $\kappa \ll 1$ we may crudely approximate the function as $F(s) = \theta(s - 1)$, and we have the exact integral
\ben
\tilde{F}(\kappa) = J_0(\kappa) \to 1, 
\een
where $J_0$ is the Bessel function of order 0.  
When $\ka \gtrsim 1$, the integral is sensitive to the structure of the string core. 

Inspecting the form of the Fourier transform, we can already see that the scalar and vectors components are 
\ben
\tilde{\Jstring}_s(\bk) = 0, \quad \tilde{\Jstring}_\pm(\bk) = 2\pi ({\bf e}_\pm \cdot \hat{\bf z}) \frac{\fa}{k_\perp} \tilde{F}(k_\perp\ws) \int dz e^{-ik_z z}.
\een
where $ \hat{\bf z}$ is a unit vector in the $z$ direction. 
Hence a static straight string contributes only to the vector components of the current, $ \tilde{J}_\pm(\bk) $, and not to the scalar component $ \tilde{J}_\text{s}(\bk) $.

A moving string does contribute to the scalar component.  Consider the current sourced by a string segment moving with velocity $v_i$ in a direction orthogonal to the tangent vector.  
We suppose that the segments have length $L$ in the frame of the FLRW universe, the frame in which Fourier transforms are taken. 
We write the number of segments per unit volume as $N_s/\vol$, 
\bea
\tilde{J}_0 &=&  \gamma v \left( \hat{v}_i  \tilde{\Jstring}_i \right), \\
\tilde{J}_i &=& \gamma \hat{v}_i \left( \hat{v}_j  \tilde{\Jstring}_j \right)  + \left (\de_{ij} - \hat{v}_i \hat{v}_j \right) \tilde{\Jstring}_j , 
\eea
Then the scalar component of the boosted current is 
\ben
\tilde{J}_\text{s} = \left( \gamma - 1 \right) \left( \hat{k}_i \hat{v}_i\right) \left( \hat{v}_j  \tilde{\Jstring}_j \right) .
\een
which is $\mathcal{O}(v^2)$.  The timelike component of the current $J_0$ is, as expected, of order $v$.

\section{Power spectra in the USM}
\label{s:PowSpeStr}

Now we compute the power spectrum of a set of randomly placed  string segments of length $L$ in a volume $\vol$, such that their number density is $L^{-3}$.
We write their coordinates 
\ben
\bX_n(\si_n, t) = \bx_n + \bX_n' \si_n + \dot{\bX}_n t 
\een
where the tangent vectors $\bX_n'$ and velocities $\dot\bX_n$ are Gaussian random variables with the constraints $(\bX')^2 = 1 - \dot\bX^2$ and $\dot\bX \cdot \bX' = 0$.
For each segment we write $\bX'_n = \hat\bn |\bX'_n|$, where $\hat\bn$ is a unit vector, and we integrate radially only up to the average distance between segments, $L$.
Then, for a single static segment, 
\bea
|\tilde{\mathbf{\Jstring}}(\bk, t)|^2 &=& 4 \pi^2\frac{\fa^2}{k_\perp^2} [\tilde{F}(k_\perp\ws, L/\ws) ]^2    L^2 \text{sinc}^2(k_\parallel L/2)  \nonumber\\
\eea
where $k_\parallel = \bk\cdot \hat \bn$ and $\bk_\perp = \bk - (\bk\cdot \hat \bn) \hat \bn $, and 
\ben
\tilde{F}(\ka, Y)  = \int_0^Y ds  F(s/\ka) J_1(s), 
\een
where 
$Y = L/\ws$. 
In the limit 
$\ka \ll 1 $ we have $\tilde{F}(\ka, Y) \to 1 - J_0(Y) $. 
Therefore, combined with $Y \gg 1$ we have $\tilde{F}(\ka, Y) \to 1$.
For $1 \ll \ka \ll Y$, the Fourier transform explores the core of the string and  $\tilde{F}(\ka, Y) \to 0 $. 

 Averaging over the string ensemble in a volume $\vol$, and dropping the explicit notation for the time dependence, 
 \bea
\vev{|\tilde{\mathbf{J}}(\bk)|^2} &=& 
\vev{\ga^2 v^2 |  \hat{\bv} \cdot  \tilde{\mathbf{\Jstring}}  |^2 } + \vev{ |  \tilde{\mathbf{\Jstring}}  |^2}.
 \nonumber\\
 \eea
 The second term involves only an average over directions of the tangent vector $\hat\bn$. 
 Recalling that $\tilde{\Jstring}_i$ is proportional to $\ep_{ijk} \hat{k}^j \hat{n}^k$ and 
 using the vector identity
 \ben
 \ep_{ijk}\ep_{ilm} \hat{k}^j\hat{n}^k \hat{k}^l\hat{n}^m = 1 - \hat{k}_\parallel^2 = \hat{k}_\perp^2
 \een
 we find 
 \bea
 \vev{|\tilde{\textbf{\Jstring}}(\bk)|^2}_{\hat\bn} &\simeq& 4 \pi^2\fa^2 N_s L^2 \frac{1}{k^2}\nonumber\\
 &\times&  
 \vev{ [\tilde{F}(k_\perp\ws, L/\ws) ]^2 \text{sinc}^2(k_\parallel L/2)  }_{\hat\bn} \nonumber\\
\eea
where $N_s$ is the number of segments in the volume $\vol$ and the angle brackets indicate an average over tangent vector orientations. 
To evaluate the first term we note that averaging over velocity directions for fixed $\hat{\bn}$, we have
\ben
\vev{\hat{v}_i\hat{v}_n} = \half \left( \de_{in} - \hat{n}_i\hat{n}_n \right).
\een
Then
 \ben
 \ep_{ijk}\ep_{nlm} \vev{\hat{v}_i\hat{v}_n} \hat{k}^j\hat{n}^k \hat{k}^l\hat{n}^m = \half \left(1 - \hat{k}_\parallel^2 \right), 
 \een
and we have that
\ben
\vev{\ga^2 v^2 |  \hat{\bv} \cdot  \tilde{\mathbf{\Jstring}}  |^2 } = \half \vev{\ga^2 v^2}_v \vev{ | \tilde{\mathbf{\Jstring}}  |^2}_{\hat\bn} .
\een
where the first angle brackets denotes an average over the distribution of speeds $v$.
Thus 
\bea
\vev{|\tilde{\mathbf{J}}(\bk)|^2} = \left( \half \vev{\ga^2 v^2}_v + 1 \right)\vev{ | \tilde{\mathbf{\Jstring}}(\bk)  |^2}_{\hat\bn} , \label{e:JiStrPowSpe}
\\
\vev{|\tilde{J}_0(\bk)|^2} = \half \vev{\ga^2 v^2}_v \vev{ | \tilde{\mathbf{\Jstring}}(\bk)  |^2}_{\hat\bn} ,. 
\eea
With similar considerations, we find that the scalar component spectral density is 
\ben
\vev{|\tilde{J}_\text{s}(\bk)|^2} =  \vev{(\ga - 1)^2}_v \frac{1}{8}\vev{ | \hat{k}_\perp^2 \tilde{\mathbf{\Jstring}}(\bk)  |^2}_{\hat\bn} ,
\label{e:JsStrPowSpe}
\een
which is $\mathcal{O}(v^4)$. One can obtain an expression for the vector ($\pm$) power spectra by subtracting \eqref{e:JsStrPowSpe} from \eqref{e:JiStrPowSpe} and dividing by two, but already one can see from the smallness of the coefficient and the fact that $\hat{k}_\perp^2 \le 1$, that
\bea
\vev{|\tilde{J}_{\pm}(\bk)|^2}  &\simeq& \half \left( 1 + \half \vev{\ga^2 v^2}_v \right)  \vev{ | \tilde{\mathbf{\Jstring}}(\bk)  |^2}_{\hat\bn}\label{e:JpmStrPowSpe}
\eea
Therefore the sum of the vector components dominate the current $\tilde{\mathbf{J}}$ unless the string is moving relativistically ($v \simeq 1$).

To gain insight into the form of the power spectrum, we note that the polar angle average 
\ben
\frac{1}{2}\int_{-1}^{1} d\mu = \frac{1}{k L} \int_{-kL/2}^{kL/2} d(k_\parallel L/2)
\een
where $k_\parallel = k\cos\th$ and $k_\perp = k\sin\th$. 
Hence 
\bea
 &&\vev{|\tilde{\mathbf{\Jstring}}(\bk)|^2}_{\hat\bn} \simeq 
 {\fa^2}  4\pi^3 \frac{N_s L}{k^3}, \qquad   1 \ll kL \ll L/\ws \nonumber\\
 \eea
We define power spectra,
\bea
\PSStr(k, t) &\equiv& \frac{k^3}{2\pi^2 \vol} \vev{ | \tilde{\mathbf{\Jstring}}(\bk)  |^2}_{\hat\bn}, \\
\PSStr^s(k, t) &\equiv& \frac{k^3}{2\pi^2 \vol} \vev{ | \hat{k}_\perp^2\tilde{\mathbf{\Jstring}}(\bk)  |^2}_{\hat\bn}, 
\eea
where $\PSStr^s(k, t) = \PSStr(k, t)[1 + \text{O}((kL)^{-2})]$ for $kL \gg 1$.
Then
\bea
 \sum _\pm \mcP_\pm(k,\tau)  &\simeq& \left( 1 + \half \vev{\ga^2 v^2}_v \right)  \PSStr(k, \tau) .
 \eea
The power spectra are proportional to the string length density $N_s L /\vol$. We will make the approximation that the (universe frame, comoving) segment length and density is such that 
\ben
\frac{N_s L}{\vol} = \frac{1}{L^2},
\een
Restoring numerical factors, and noting that the average over the direction cosine $\hat{k}_\parallel$ of $\mathrm{sinc}^2(k_\parallel L/2)$  tends to $\pi$ as $kL$ tends to infinity, we have 
\ben
\PSStr(k, \tau) \simeq  
 \displaystyle  
 2\pi \frac{\fa^2}{L^2},  \qquad 1 \ll kL \ll L/\ws
 \een
 In the limit $kL \to \infty$, $\vev{ | \hat{k}_\perp^2 \tilde{\mathbf{\Jstring}}(\bk)  |^2}_{\hat\bn} \to \vev{ | \tilde{\mathbf{\Jstring}}(\bk)  |^2}_{\hat\bn}$, as the dominant contribution comes from $k_\parallel \simeq 0$.
 The quantity $1/L^2$ is equal to $4 \zew /\tau^2$,  where $\zew$ is the length density parameter in the FLRW frame. 
 Hence we expect the sum of $\pm$ power spectra,  in the wavenumber range between the inverse segment spacing and the inverse string width, to be 
 \bea
 \sum _\pm \mcP_\pm(k,\tau)  &\simeq& 8\pi \zew  \left( 1 + \half \vev{\ga^2 v^2}_v \right)  \frac{\fa^2}{\tau^2} .
 \eea

\section{Spectral density and decoherence function for a stochastic axion field}
\label{appspectra}

As we saw in section \ref{s:TheExp},  the equation of freely propagating axion can be solved by writing  $\tilde{J}_s$ in terms of a scalar field,  $\tilde{J}_s(\bk,\tau) = i k \tilde{A}(\bk,\tau)$, with $\tilde{J}_0 = \dot {\tilde{A}}(\bk,\tau)$,
with
\ben
 \tilde{A}(\bk,\tau) = c^{(1)}_\bk h_0^{(1)}(k\tau) +  c^{(2)}_\bk h_0^{(2)}(k\tau),
 \label{e:AxFieSolApp}
\een 
where  
\ben
h_0^{(1)}(x) = -i\frac{e^{ix}}{x}, \quad  h_0^{(2)}(x) = i \frac{e^{-ix}}{x}.
\een
The currents for these plane wave solutions are therefore 
\bea
\tilde{J}_s(\bk,\tau)  &=&  c^{(1)}_\bk \frac{e^{ik\tau}}{\tau}  -  c^{(2)}_\bk \frac{e^{-ik\tau}}{\tau}, \label{e:JsEqOne}\\
\tilde{J}_0(\bk,\tau)  &=&  c^{(1)}_\bk \frac{e^{ik\tau}}{\tau} \left( 1 + \frac{i}{k\tau} \right) \nonumber\\
&+&  c^{(2)}_\bk  \frac{e^{-ik\tau}}{\tau} \left( 1 - \frac{i}{k\tau} \right). \label{e:J0EqOne}
\eea
As the currents are real, we must have 
\ben
 c^{(1)*}_\bk =  c^{(2)}_{-\bk}\,.
\een
The resulting unequal time correlator for the longitudinal current is
\bea
U_{ss}(k, \tau_1, \tau_2) &=& \frac{2}{\tau_1\tau_2}\vev{|c^{(1)}_\bk|^2}_{\hat{\bk}} \cos[k(\tau_1 - \tau_2)] \nonumber\\
&+ &  W (k,\tau_1,\tau_2)\,,
\eea
where
\bea
W (k, \tau_1, \tau_2) &=& -  \frac{2}{\tau_1\tau_2} \Re\left[\vev{c^{(1)}_\bk c^{(2)*}_{\bk}}_{\hat{\bk}} e^{ik(\tau_1 + \tau_2)}\right] \,.\nonumber\\
\eea
For axion radiation from a stochastic source, we expect the phases of $c^{(1)}_\bk$ and  $c^{(2)*}_{\bk}$ to be uncorrelated, both with each other, and between modes with different $\hat{\mathbf{k}}$.  In the stochastic case, therefore, $W (k,\tau_1,\tau_2) = 0$.
However, a field $A$ initialised at $\tau_\text{i}$ with zero time derivative will have, for $k\tau \ll 1$, $c^{(1)}_\bk e^{ik\tau_\text{i}} = c^{(2)}_{\bk}e^{-ik\tau_\text{i}}$. Hence
\bea
W (k, \tau_1, \tau_2) &=& -  \frac{2}{\tau_1\tau_2}\vev{|c^{(1)}_\bk|^2}_{\hat{\bk}} \cos[k(\tau_1 + \tau_2 - 2\tau_\text{i})]. \nonumber\\
\label{e:PSosc}
\eea
For wavenumbers $k(\tau_1+ \tau_2) \sim 1$ we expect to see oscillations in the spectrum as a function both of time and $k$. 
In a numerical setting, one always averages over a certain interval $\Delta k$, and the oscillations will average to zero for $\Delta k (\tau_1 + \tau_2) \gg 1$. In our case, we take 
$\Delta k = 2\pi / L$, where $L = N\Delta x$ is the side length of the simulation box, and oscillations appear only in the first few modes.
A detailed discussion of this effect can be found in Ref.~\cite{Saikawa:2024bta}.

Thus, for a stochastic axion field we expect to see at wavenumbers $k\tau \gg 1$  a spectral density 
\ben
P_s(k,\tau) =  \frac{2}{\tau^2}\vev{|c^{(1)}_\bk|^2}_{\hat{\bk}}\,,
\een
and a decoherence function
\ben
D_s(k,\tau_1,\tau_2) = \cos(x_1 - x_2),
\een
where $x = k\tau$. 

As shown in section \ref{s:TheExp}, another set of solutions can be found with the ansatz $\tilde{J}_0 = k_k V_k$,  $\tilde{J}_i =  a^{-2} \pa_0 (a^2 V_i)$, resulting in 
Eq.~\ref{e:StrSouEqn}, from whose general solution \eqref{e:GenSolV} it follows that 
\bea
\tilde{J}_s &=&  k \left[d^{(1)}_\bk  h^{(1)}_0(k\tau)  + d^{(2)}_\bk h^{(2)}_0(k\tau)\right], \\
\tilde{J}_0 &=&  i k\left[d^{(1)}_\bk  h^{(1)}_1(k\tau)  + d^{(2)}_\bk h^{(2)}_1(k\tau)\right],
\eea
These equations are consistent with current conservation and Eqs.~(\ref{e:JsEqOne}, \ref{e:J0EqOne}), when $c^{(1)}_\bk  = -i k d^{(1)}_\bk$, $c^{(2)}_\bk  = -i k d^{(2)}_\bk$.

\section{Model for unequal time correlators}
\label{s:ModUETC}

In this appendix we construct a theoretical prediction for the functional form of the unequal time correlators of the transverse components of the current $\tilde{J}_\pm$, in the framework of the unconnected segment model.

We start with the Fourier transforms of the currents sourced by a moving string segment, moving away the coordinate origin at $\tau=0$ with 3-velocity $\bv$, 
\bea
\tilde{J}_0(\bk,\ta) &=& {\Lambda_0}^i(\bv) \tilde{\Jstring}_i  e^{i \bk \cdot \bv \ta}, \\
\tilde{J}_i(\bk,\ta) &=&  {\Lambda_i}^j(\bv) \tilde{\Jstring}_j  e^{i \bk \cdot \bv \ta},
\eea
where
\bea
{\Lambda_0}^i(\bv) &=&  \gamma v \hat{v}^i , \\
 {\Lambda_i}^j(\bv) &=&  \gamma \hat{v}_i\hat{v}^j  + (\de_{ij} - \hat{v}_i\hat{v}^j ).
\eea
and
\ben
 \tilde{\Jstring}_i = \frac{2\pi i\fa}{k_\perp} L\, \text{sinc}(k_\parallel L/2) \ep_{ikl} \hat{k}^k \hat{n}^l.
\een
with $k_\parallel = \hat\bn\cdot\hat\bk$, $k_\perp^2 = k^2 - k_\parallel^2$, and we assume $ k\ws \ll 1$.

The unequal time correlator of the spatial components of two currents is 
\ben
U_{ij}(k, \ta_1, \ta_2) = \vev{\tilde{J}_i^*(\bk,\ta_1)\tilde{J}_j(\bk,\ta_2)}_{\hat{\bn},\bv} ,
\een
where we average over velocities $\bv$ and orientations $\hat{\bn}$ of segments. 
Hence
\bea
U_{ij}(k, \ta_1, \ta_2) \simeq \vev{ {\Lambda_i}^k(\bv) {\Lambda_j}^l(\bv) \tilde{\Jstring}_k \tilde{\Jstring}_l    e^{i \bk \cdot \bv (\ta_2 - \ta_1)} }_{\hat{\bn},\bv}  ,.\nonumber\\
\label{e:JiUETC}
\eea
Consider first the velocity average 
\bea
{V_{ij}}^{kl} \equiv \vev{ {\Lambda_i}^k(\bv) {\Lambda_j}^l(\bv) e^{i \bk \cdot \bv (\ta_2 - \ta_1)} }_{\bv}  , 
\label{e:JiUETC0}
\eea
which contains terms of three kinds
\bea
{V^{(1)}_{ij}}^{kl} \equiv \vev{ {\de_i}^k {\de_j}^l e^{i \bk \cdot \bv (\ta_2 - \ta_1)} }_{\bv}   ,  
\label{e:JiUETC1} \\
{V^{(2)}_{ij}}^{kl} \equiv \vev{ {\de_i}^k {\hat{v}_j}\hat{v}^l e^{i \bk \cdot \bv (\ta_2 - \ta_1)} (\ga - 1)}_{\bv}  ,  
\label{e:JiUETC2} \\
{V^{(3)}_{ij}}^{kl} \equiv \vev{ {\hat{v}_i}\hat{v}^k {\hat{v}_j}\hat{v}^l e^{i \bk \cdot \bv (\ta_2 - \ta_1)} (\ga - 1)^2}_{\bv} ,  
\label{e:JiUETC3} 
\eea
Averaging separately over directions $\hat{\bv}$ and magnitudes $v$, and noting that $\hat\bv\cdot\hat{\bn} = 0$, we have 
\ben
{V^{(1)}_{ij}}^{kl}  = {\de_i}^k {\de_j}^l  \vev{W^{(1)}(v k_\perp \Delta \ta)}_v, 
\een
where $\Delta \ta = \ta_2 -\ta_1$, 
\bea
W^{(1)}(z) &=& \frac{1}{2\pi} \int_0^{2\pi} e^{i z\cos\varphi} d\varphi,  \nonumber \\
&=& J_0(z) ,
\eea
and $J_0(z)$ is the Bessel function of order 0. 
The next term in the average over velocity is 
\ben
{V^{(2)}_{ij}}^{kl}  = {\de_i}^k \hat{k}_{\perp,l} \hat{k}_\perp^l  \vev{W^{(2)}(v k_\perp \Delta \ta) (\gamma -1)}_v ,
\een
where
\bea
W^{(2)}(z) 
&=& - \frac{d^2}{dz^2} J_0(z ) .
\eea
The third term is 
\ben
{V^{(3)}_{ij}}^{kl}  = \hat{k}_{\perp,i} \hat{k}_\perp^k \hat{k}_{\perp,j} \hat{k}_\perp^l  \vev{W^{(3)}(v k_\perp \Delta \ta) (\gamma -1)^2}_v
\een
where
\bea
W^{(3)}(z) 
&=&  \frac{d^4}{dz^4} J_0(z ) .
\eea
The expansion of the zeroth order Bessel function at small arguments is 
\ben
J_0(z) = 1 - \frac{1}{4} z^2 + \frac{1}{64} z^4.
\een
We see that the first term $W^{(1)}$ is the most important, as subsequent terms are smaller by powers of $v^2$ and coefficients of the expansion of the Bessel function.

Then
\ben
U_{ij}(k, \ta_1, \ta_2) \simeq U^{(1)}_{ij},
\een
where
\bea
U^{(1)}_{ij} &=& {4\pi^2\fa^2} \ep_{ikl}\ep_{jmn} \hat{k}^k \hat{k}^m \vev{N_{(1)}^{ln}(\bk)}_v
\eea
with
\begin{widetext}
\bea
N_{(1)}^{ln}(\bk) & =  & \vev{\frac{4 \sin(k_\parallel L_1/2)\sin(k_\parallel L_2/2)}{k_\perp^2k_\parallel^2} \hat{n}^l \hat{n}^n J_0(k_\perp v \Delta \ta) }_{\hat\bn} % \nonumber \\
\eea
It is apparent that only the transverse component contributes to the correlator $U^{(1)}_{ij}$ , whereupon
\bea
N_{(1)}^{\perp}(k) & =  & \vev{\frac{4 \sin(k_\parallel L_1/2)\sin(k_\parallel L_2/2)}{k^2 k_\parallel^2} J_0(k_\perp v \Delta \ta) }_{\hat\bn} \nonumber \\
&=&   \frac{1}{2k^2} \int_{-1}^{1} d\mu \frac{ \sin(\mu kL_1/2)\sin(\mu k L_2/2)}{(\mu k/2)^2} J_0((1 - \mu^2)^{1/2} k  v \Delta \ta) % \nonumber\\
\eea
In the USM, the segment length grows in proportion to time, $L(\tau) = \be_\text{w} \tau$.  
Writing $x_1 = k\tau_1$ and $ x_2 = k\tau_2$, we have
\bea
N_{(1)}^{\perp}(k) & =  &  \frac{\sqrt{L_1L_2}}{k^3} I(x_1, x_2; \bew, v)
\eea
where
\ben
I(x_1, x_2;  v) = \frac{\bew \sqrt{x_1x_2}}{2} \int_{-1}^{1} d\mu \frac{ \sin(\mu \bew x_1/2)\sin(\mu \bew x_2/2)}{(\mu \bew x_1/2)(\mu \bew x_2/2)} J_0((1 - \mu^2)^{1/2} v \Delta x) 
\label{e:IfunDef}
\een
and $\Delta x = x_1 - x_2$. We have suppressed the dependence of the function $I$ on $\bew$ for conciseness, and we fix the parameter to $\beta_\text{w} = 0.45$ to approximately match the measured universe-frame mean string separations \cite{Correia:2024cpk}.
\end{widetext}

Then the transverse decoherence function 
\ben
D_\pm(x_1, x_2) = \frac{ \vev{I (x_1, x_2; v)}_v}{\sqrt{\vev{I (x_1; v)}_v\vev{I (x_2; v)}_v} },
\label{e:DpmUSM}
\een
where $I(x_1;v) \equiv  I(x_1,x_1;v)$.

The speed distribution of string has been measured in numerical simulations in \cite{Gorghetto:2020qws}, who found that the Lorentz $\ga$ was approximately distributed as a power law. One can translate this into a distribution for speed $v$, with RMS velocity $\bar{v}$,
\ben
p(v) = 2(\bar{v}^{-2} - 1) v \ga^{4 - 2\bar{v}^{-2}}.
\label{e:vDist}
\een
Averaging the function $I$ \eqref{e:IfunDef} over the distribution \eqref{e:vDist} gives the decoherence function $D_\pm$ plotted in Fig.~\ref{f:USMdec}.  There we also plot the function obtained with a velocity distribution $p(v) = \delta(v - \bar{v})$, 
\ben
\bar{I}(x_1, x_2) = \frac{ I (x_1, x_2; \bar{v})} {\sqrt{I (x_1; \bar{v}) I (x_2; \bar{v}) }},
\label{e:DpmUSMnoav}
\een
to show that the effect of the velocity averaging is to smooth out the oscillations.

\section{Dynamical equations for power spectra}
\label{s:PaxEqn}

We start with equations \eqref{e:J0dot}, \eqref{e:Jsdot}, given again here for convenience
\begin{eqnarray*}
\pa_0{\tilde{J}}_0 &=& - 2 \frac{\dot a}{a} \tilde{J}_0 + i k \tilde{J}_s + \tilde{\si}(\bk, \tau), \\
\pa_0 \tilde{J}_s &=& ik \tilde{J}_0 + \tilde\Source_{0s}.
\end{eqnarray*}
Then we consider spectral densities
\bea
P_0(k,\tau) &=& \vev{ \tilde{J}^*_0(\bk,\tau) \tilde{J}_0(\bk,\tau)}_{\hat\bk}, \\
P_s(k,\tau) &=& \vev{ \tilde{J}^*_s(\bk,\tau) \tilde{J}_s(\bk,\tau)}_{\hat\bk}, 
\eea
Using the equations we can derive  
\bea
\frac{1}{a^2}\pa_0(a^2 P_0) &=& - 2 \frac{\dot a}{a} P_0 + ik\left( \vev{ \tilde{J}^*_0 \tilde{J}_s}_{\hat\bk} - \vev{ \tilde{J}^*_s \tilde{J}_0}_{\hat\bk}  \right) \nonumber\\
&& + \left( \vev{ \tilde{J}^*_0 \tilde\si}_{\hat\bk} + \vev{\tilde{\si}^*\tilde{J}_0}_{\hat\bk}  \right)
\eea
and
\bea
\frac{1}{a^2}\pa_0 (a^2P_s) &=&  2 \frac{\dot a}{a} P_s +  ik\left( \vev{ \tilde{J}^*_s \tilde{J}_0}_{\hat\bk} - \vev{ \tilde{J}^*_0 \tilde{J}_s}_{\hat\bk}  \right) \nonumber\\
&& + \left( \vev{ \tilde{J}^*_s \tilde\Source_{0s}}_{\hat\bk} + \vev{  \tilde{\Source}^*_{0s}  \tilde{J}_s}_{\hat\bk}  \right)
\eea
We add the equations and multiply by $k^3/2\pi^2$ to obtain an equation for $\mcP_\text{ax} = (\mcP_0 + \mcP_s)/2$,
\bea
\frac{1}{a^2}\pa_0(a^2 \mcP_\text{ax}) &=& -  \frac{\dot a}{a} (\mcP_0 - \mcP_s) + \frac{k^3}{\pi^2} \Re\tilde{X},
\label{e:PaxEqnApp}
\eea
where
\bea
\tilde{X} =  \vev{ \tilde{J}^*_0  \tilde\si}_{\hat\bk}  + \vev{ \tilde{J}^*_s \tilde\Source_{0s}}_{\hat\bk}.
\eea
In the USM, $\si \equiv  J_\mu\pa^\mu\psi/\psi = 0 $ vanishes for a stationary string segment, where $J_0$ and $\dot\psi$ both vanish, while $J_i \propto \hat{\theta}_i$ and $\pa_i \psi \propto \hat{x}^i$. As $\si$ is a Lorentz scalar, it vanishes for moving segments as well.
 
The operator $\tilde\Source_{0s}$ does not itself vanish for a moving segment in the USM, being the Lorentz transform 
\ben
\tilde\Source_{0s} = \hat{k}^j{\La_0}^k(\bv) {\La_j}^l(\bv) \tilde\Source_{k l}
\een
where 
\ben
\tilde\Source_{k l} = 2\pi \fa L \text{sinc}(k_\parallel L/2) \left( \hat{k}_l \ep_{kmn} - \hat{k}_k \ep_{lmn}  \right) \hat{k}_m \hat{n}_n ,
\een
in the wavenumber range $1 \ll kL \ll k\ws$. 
However, $\vev{\tilde{J}^*_s \tilde\Source_{0s}}_{\hat{\bk}}$ does vanish for uncorrelated segments. 
To see this, we recall that 
\ben
\tilde{J}_s = \hat{k}^l {\La_l}^i(\bv) (2\pi \fa)  L\, \text{sinc}(k_\parallel L/2) i \ep_{ijk} \hat{k}_j\hat{n}_k.	
\een
Therefore $\tilde{J}_s$ and $\Source_{k l}$ have opposite parity under the operation $\bk \to -\bk$, and their product must vanish when averaged over directions $\hat\bk$.

This conclusion applies only to the contribution to $\tilde{X}$ from autocorrelations of string segments.  
When one takes into account correlations between different segments located at $\bx_1$ and $\bx_2$
\ben
\tilde{X} =  \vev{ \tilde{J}^*_s \tilde\Source_{0s} e ^{i \bk \cdot \br_{12} }} _{\hat\bk} 
\een
one will get a non-zero result.  This would require extending the Unconnected Segment model, which we do not attempt here.

\FloatBarrier

\begin{widetext}
\newpage
\section{Power spectra evolution}
\label{s:PowSpeEvo}
\begin{figure*}[h]
    \centering
    \includegraphics[width=0.4\textwidth]{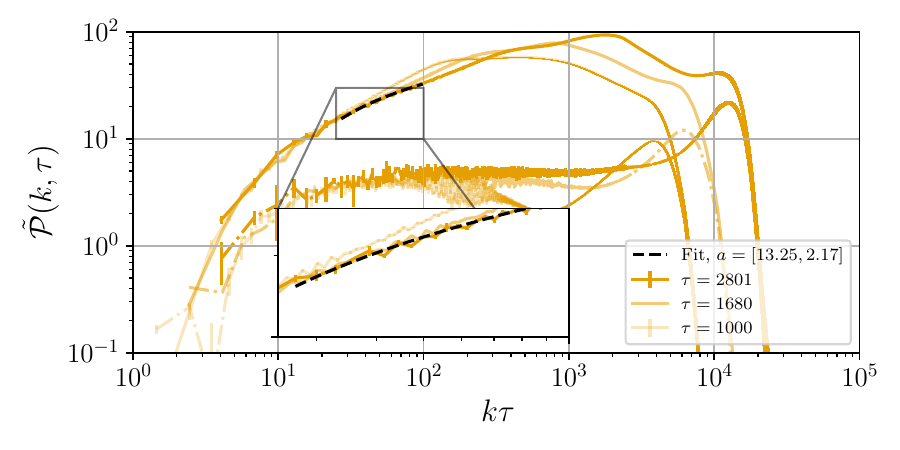}
    \includegraphics[width=0.4\textwidth]{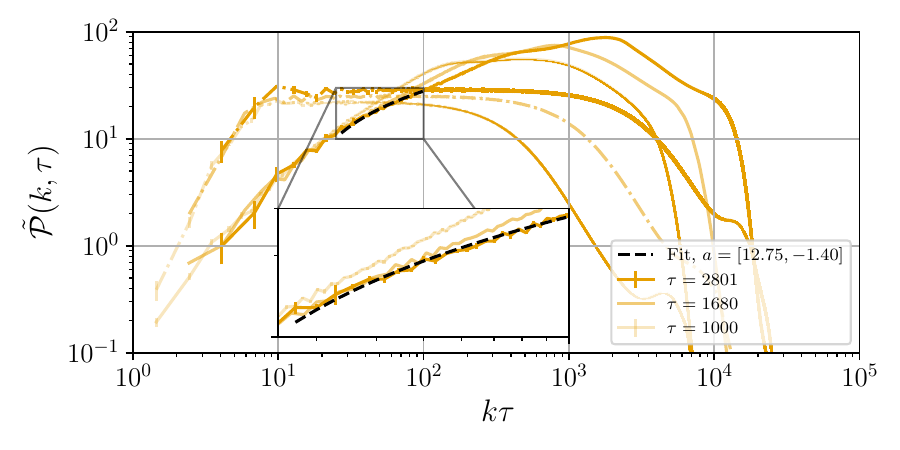}\\
    \includegraphics[width=0.4\textwidth]{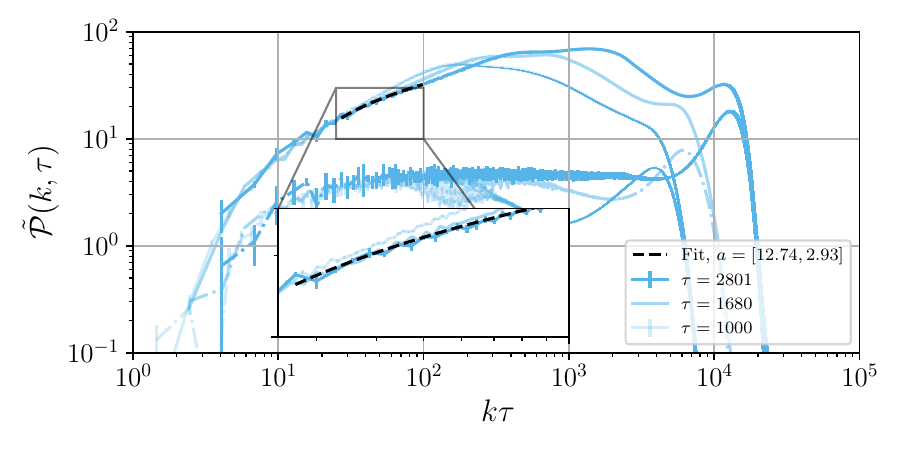}
    \includegraphics[width=0.4\textwidth]{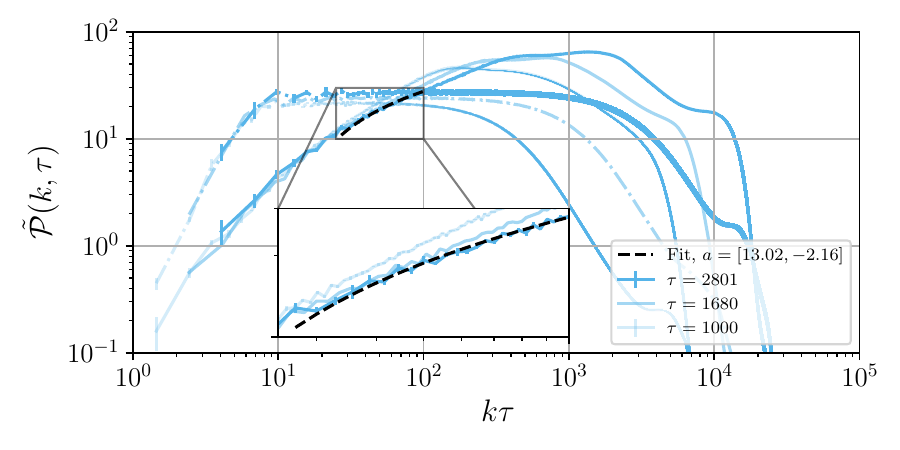}\\
    \includegraphics[width=0.4\textwidth]{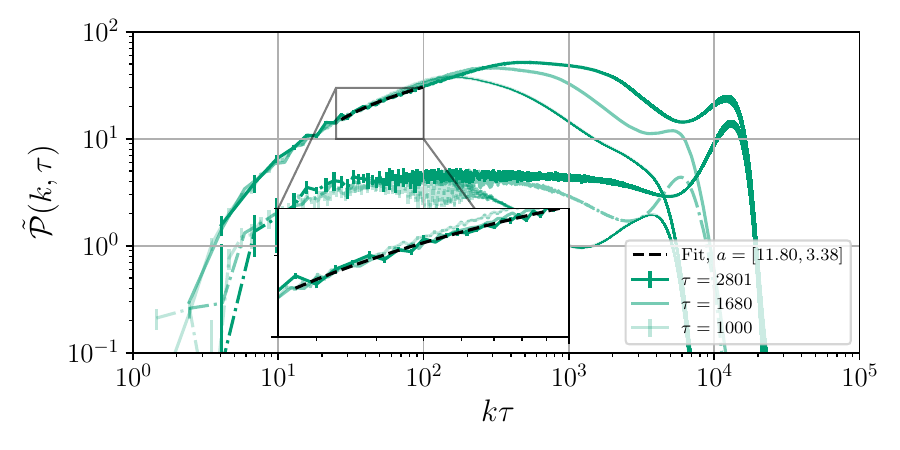}
    \includegraphics[width=0.4\textwidth]{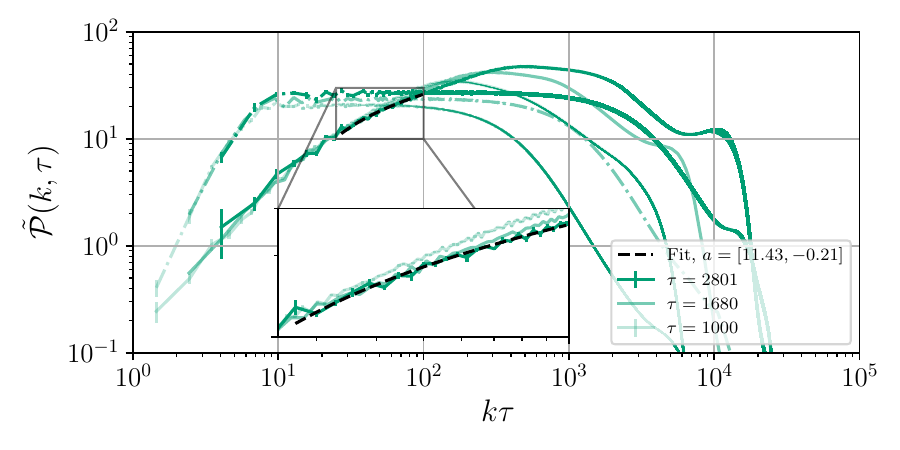}\\
    \includegraphics[width=0.4\textwidth]{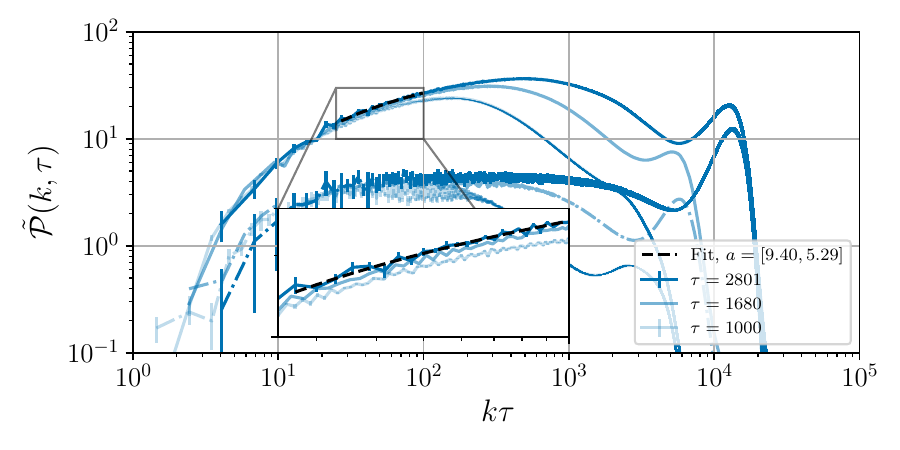}
    \includegraphics[width=0.4\textwidth]{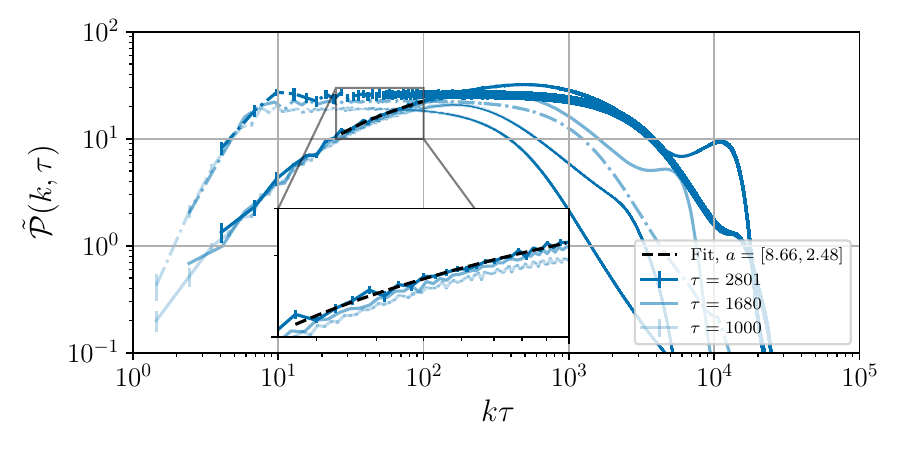}\\
    \includegraphics[width=0.4\textwidth]{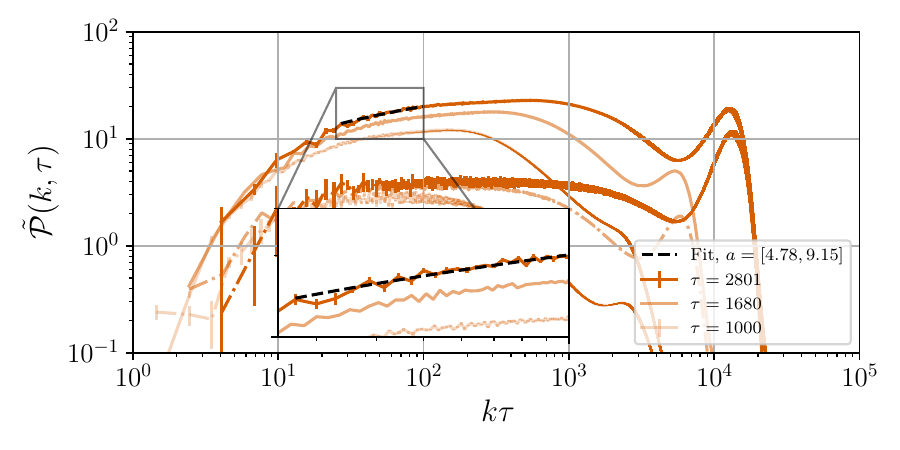}
    \includegraphics[width=0.4\textwidth]{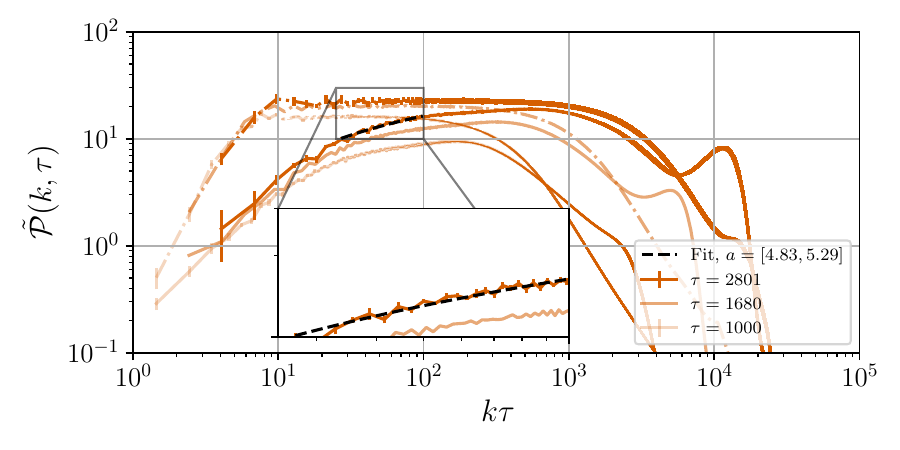}
     \caption{Left: $\ta$-scaled power spectra of $J_0$ (solid) and difference of power spectra of $J_0$ and $J_s$ (dash-dot).  Right: $\ta$-scaled power spectra of $J_s$ (solid) and sum of PS of $J_+$ and $J_-$ (dash-dot), with fit to the logarithmic function given in Eq.~\eqref{e:PowSpeFitFun} in the range $25 < k\tau < 100$. Best fit parameters are given in the legend.  Initial field correlation lengths are (top to bottom) $\l_\phi = 5, 10, 20, 40, 80$.
     \label{f:PSJ0J0_J0J0_tau}}
 \end{figure*}
\FloatBarrier

\newpage

\section{Emission spectra evolution}
\label{s:EmiSpeEvo}
\begin{figure*}[h]
    \centering
    \includegraphics[width=0.42\textwidth]{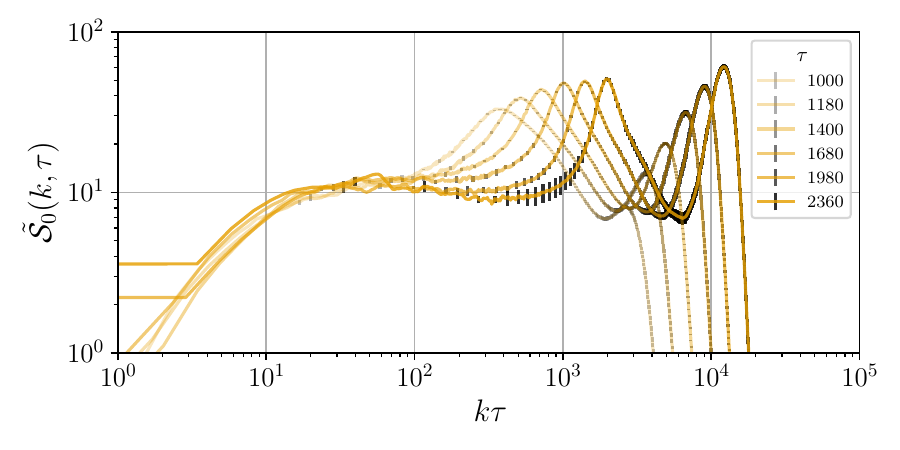}
    \includegraphics[width=0.42\textwidth]{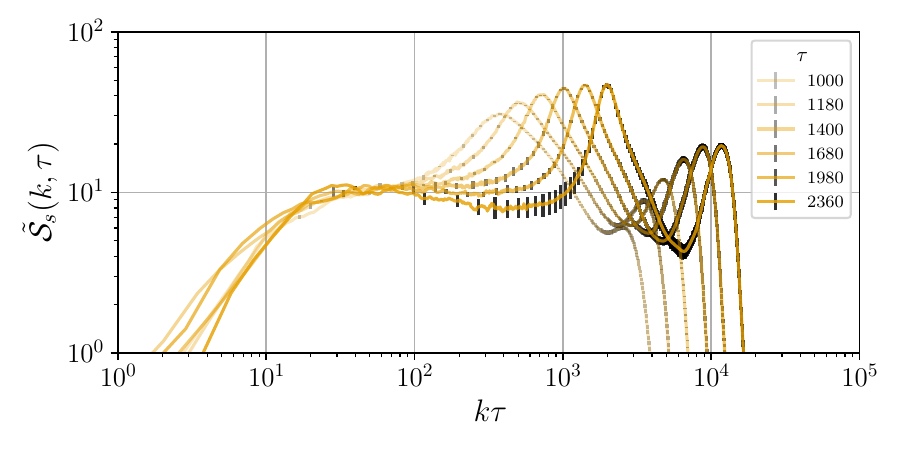}
    \includegraphics[width=0.42\textwidth]{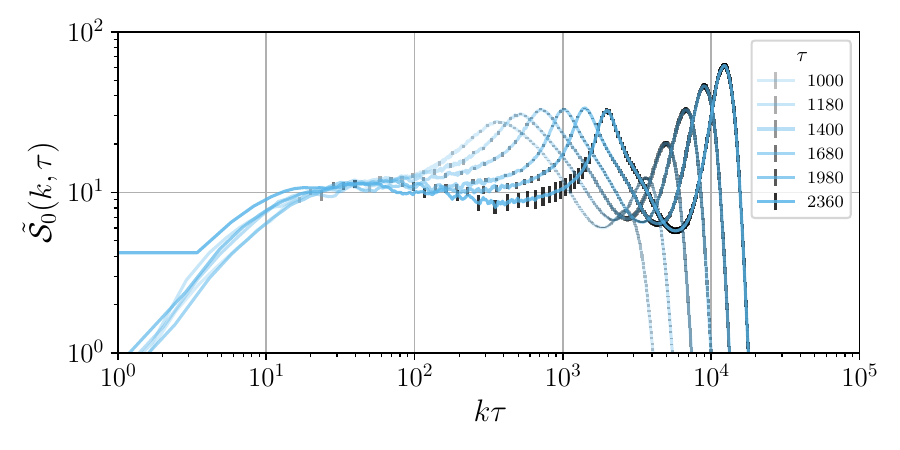}
    \includegraphics[width=0.42\textwidth]{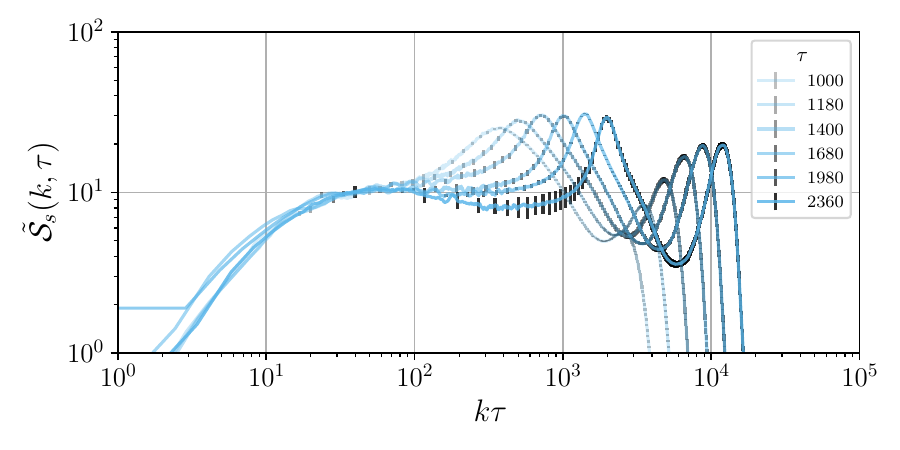}
    \includegraphics[width=0.42\textwidth]{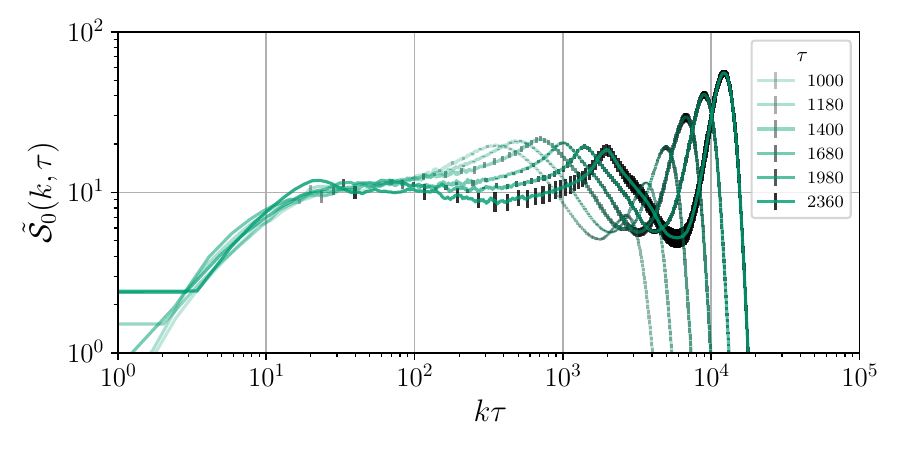}
    \includegraphics[width=0.42\textwidth]{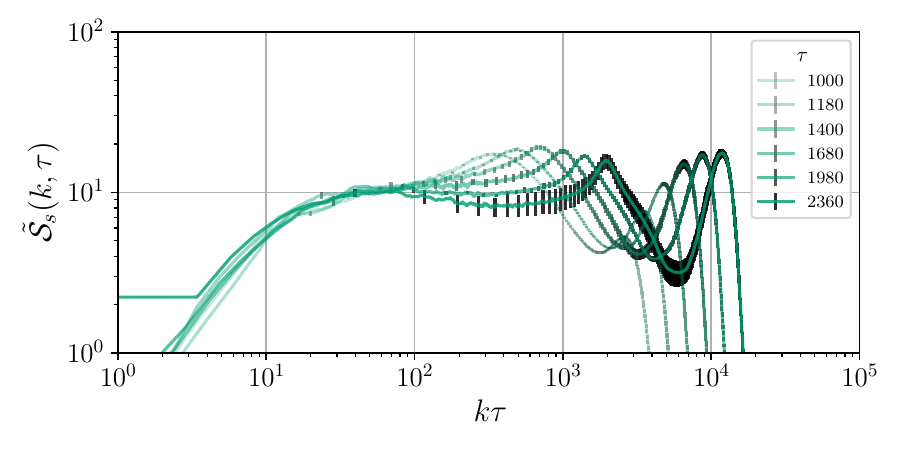}
    \includegraphics[width=0.42\textwidth]{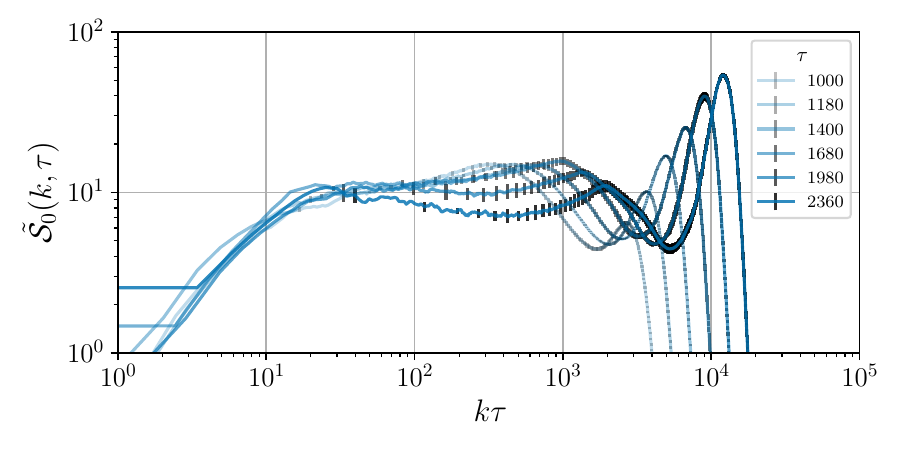}
    \includegraphics[width=0.42\textwidth]{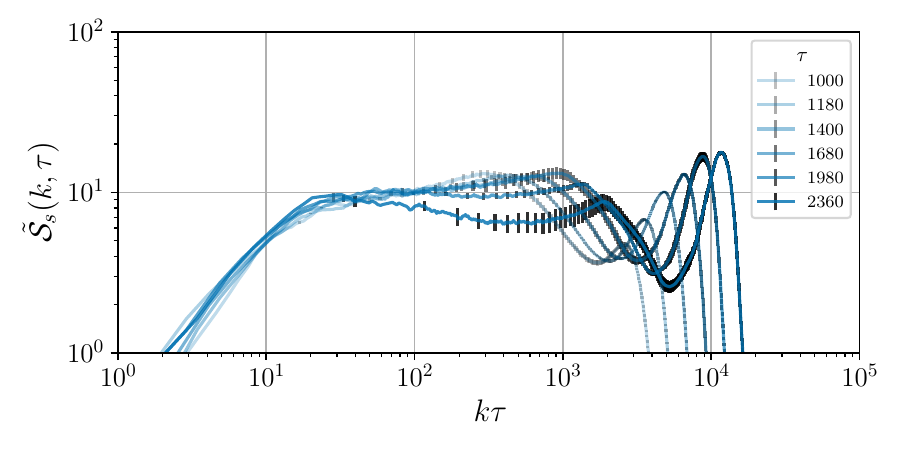}
    \includegraphics[width=0.42\textwidth]{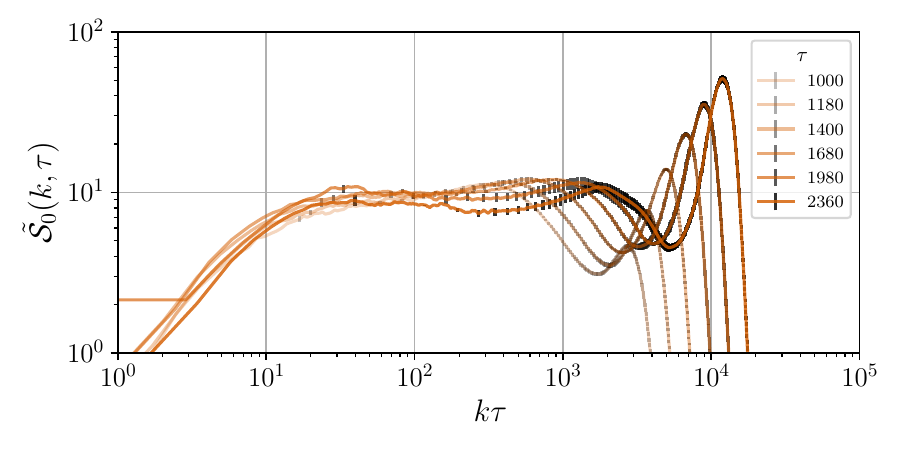}
    \includegraphics[width=0.42\textwidth]{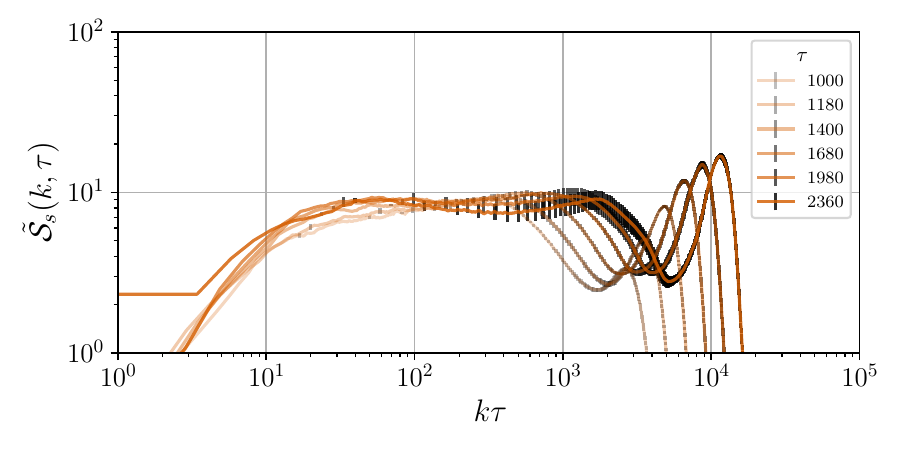}
     \caption{Scaled emission power spectra $\tilde\mcS_0$ (left) and  $\tilde\mcS_s$ (right) estimated by forward differences of scaled power spectra $\tilde\mcP_a$ between successive times shown, with a final time  $\tau = 2801$.  Initial field correlation lengths are (top to bottom) $l_\phi = 5, 10, 20, 40, 80$.
     Spectra are smoothed with a Savitsky-Golay filter, with window length 16 and polynomial order 3.
     \label{f:PSdiffJ0J0JsJs}}
 \end{figure*}
\FloatBarrier

\newpage

\end{widetext}

\newpage

\section{Oscillations of the average field as axion producers}
\label{oscillation}

\begin{figure}[!htb]
  \includegraphics[width=0.41\textwidth]{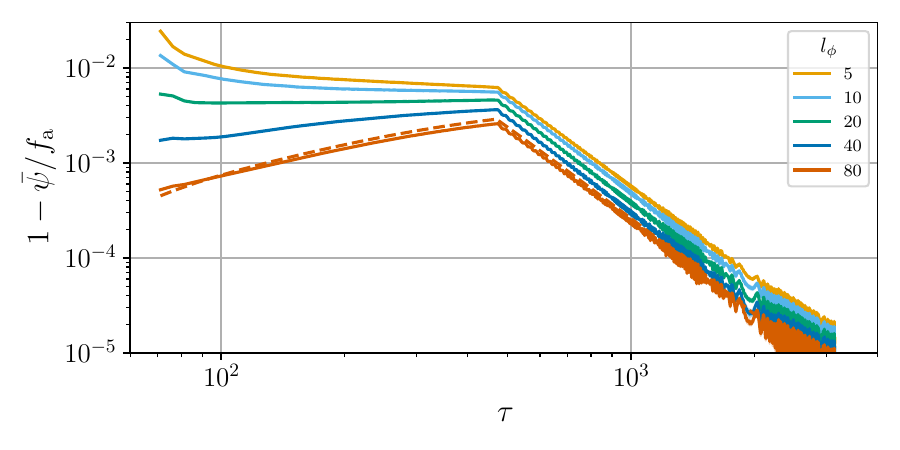}
  \includegraphics[width=0.41\textwidth]{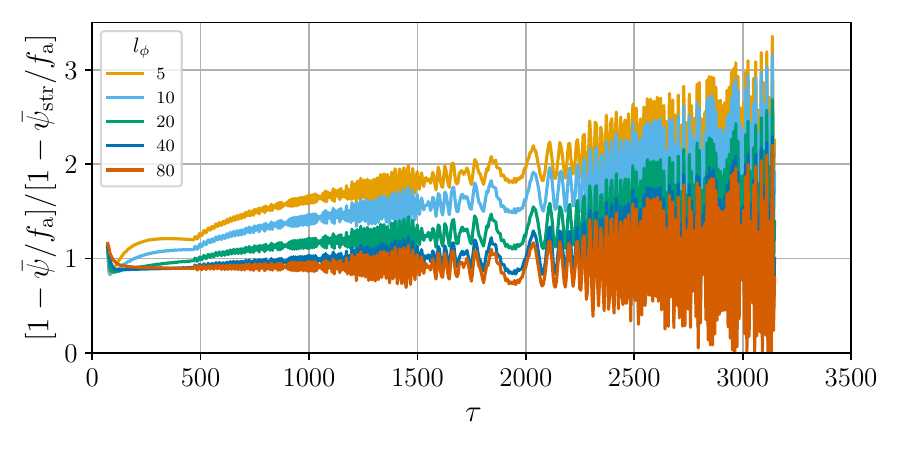}
\caption{\label{f:PsiAv} 
Top: departure of the volume-averaged scalar field $\psi$ from its equilibrium value for all initial field correlation lengths $l_\phi$, along with estimate based on Eq.~\ref{e:PsiBarStr} for $l_\phi = 80$ (dashed line). Bottom: ratio of the above to the value estimated from the string density and core field profile \eqref{e:PsiBarStr} for all $l_\phi$. 
}
\end{figure}

 In section~\ref{emissionspectra} we saw that there was a peak at about $k\ws = 0.5$ in the emission spectra $J_0$ and $J_s$ (see Fig.~\ref{f:PS_diff_all_tau_final}). One possible explanation for these are the oscillations of the average field. 
However,  they are very small with our highly cooled initial conditions. During the physical evolution the volume averaged $|\Phi|/\fa$ oscillates around a power law decay with an amplitude of order $10^{-5}$ (see Fig.~\ref{f:PsiAv}, top).

The power-law decay can be understood in terms of the decrease in the effective volume of the string core.  We approximate the saxion field at comoving distance $r$ from a straight string as 
\ben
\psi_\text{str} = \left\{
\ba{lr}
 \fa \left( 1 - {\ws^2}/{r^2} \right), \quad r > \ws, \\
0 , \quad r \le \ws,
\ea
\right.
\label{e:PsiApprox}
\een
The average saxion field produced by a network of strings with mean comoving string separation $\xiw^\text{c}$ can be estimated as the average value in a cylinder of length $\xiw^\text{c}$ and volume $(\xiw^\text{c})3$, or
\ben
1 - \bar\psi_{\text{str},0}/\fa \simeq 2 \pi \ln(\xiw^\text{c}/\ws){\ws^2}/{(\xiw^\text{c})^2},
\label{e:PsiBarStr}
\een
where we have approximated the cylinder radius as $\xiw^\text{c}$.

In Fig.~\ref{f:PsiAv} (bottom) we plot the ratio of the measured value of $1 - \bar\psi$ and $1 - \bar\psi_{\text{str},0}$, from the end of the diffusive evolution.  We see that at this time, when strings are very smooth and almost stationary, the ratio is just over 1 for all initial string densities, supporting the model behind the calculation of the average field value. Immediately afterwards, the ratio decreases, as would be expected from the Lorentz contraction in the direction of motion of the strings as they accelerate.  The ratio subsequently increases for all but the lowest initial string density.  

After conformal time $\tcg$, when the scalar potential changes to its physical form, the ratios begin to oscillate coherently, with similar amplitudes. The apparent beats are a result of aliasing due to the difference between the sample period (5.1 in conformal time units) and the oscillation period of the homogeneous mode $2\pi\ws$, which ranges from around 20 at $\tcg$ to  $\pi$ at the end of the simulation.  A period equal to the sample period is reached around conformal time $1900$, where the aliasing is most obvious. 

The fact that the relative amplitude of the coherent oscillation is approximately same for all string densities, while the string densities themselves are different by a factor of about 2, implies that the oscillation is larger for higher string densities, and therefore connected with the presence of strings, as suggested in Ref.~\cite{Saikawa:2024bta}. 
However, by delaying the onset of physical evolution, the amplitude of the oscillation in our simulations is much less than that observed Ref.~\cite{Saikawa:2024bta} (see Fig.~34). 

We interpret the increase in the departure of the field average from its expected value \eqref{e:PsiBarStr} as an increase in the effective width of the strings, which is consistent with their being in an excited state. 
The larger the width of the string, the greater the excitation in bound saxion modes with angular frequency $\ws^{-1}$, and the larger the resonant axion production. It is interesting that the initially denser string networks maintain the larger width and resonant production even at late times, when the string densities are closer to each other.  This suggests that the amplitude of the bound state oscillations is refreshed during the evolution.

\newpage

\begin{widetext}

\section{Decoherence functions 1}
\label{s:DecFun1}
\FloatBarrier
\begin{figure*}[htbp]
    \centering
    \includegraphics[width=0.41\textwidth]{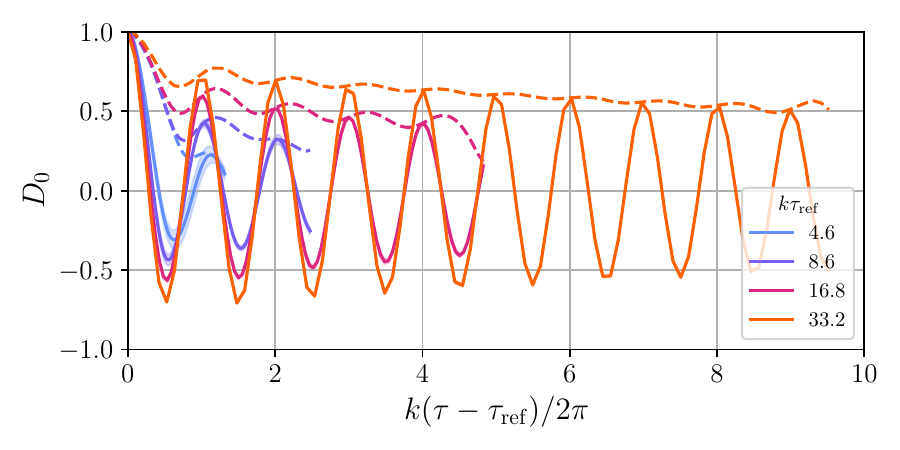}
    \includegraphics[width=0.41\textwidth]{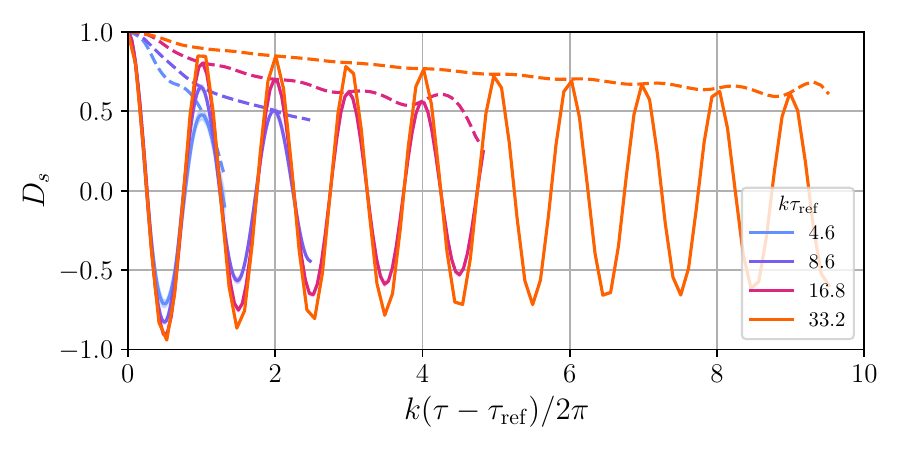}
    \includegraphics[width=0.41\textwidth]{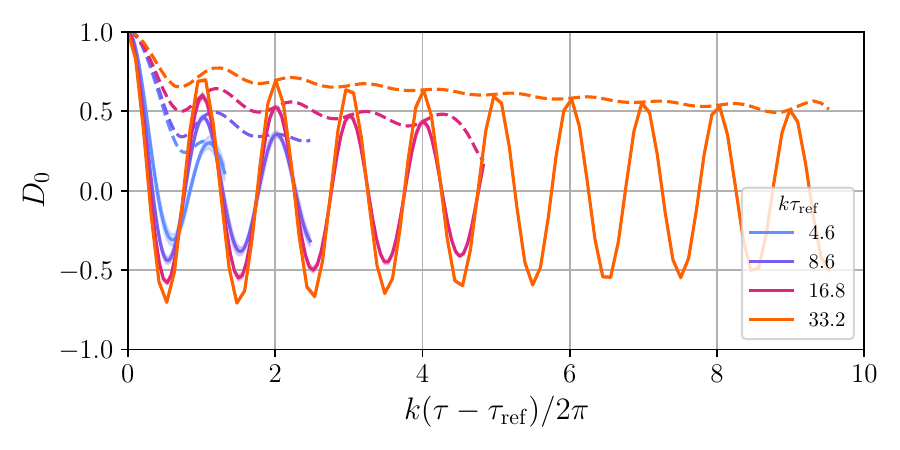}
    \includegraphics[width=0.41\textwidth]{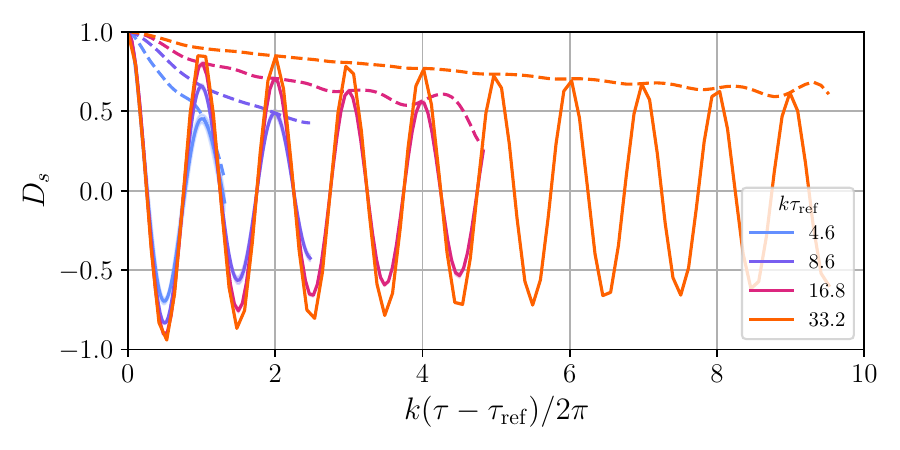}
    \includegraphics[width=0.41\textwidth]{decoh_ana_sig_J0J0_lp20.pdf}
    \includegraphics[width=0.41\textwidth]{decoh_ana_sig_JsJs_lp20.pdf}
    \includegraphics[width=0.41\textwidth]{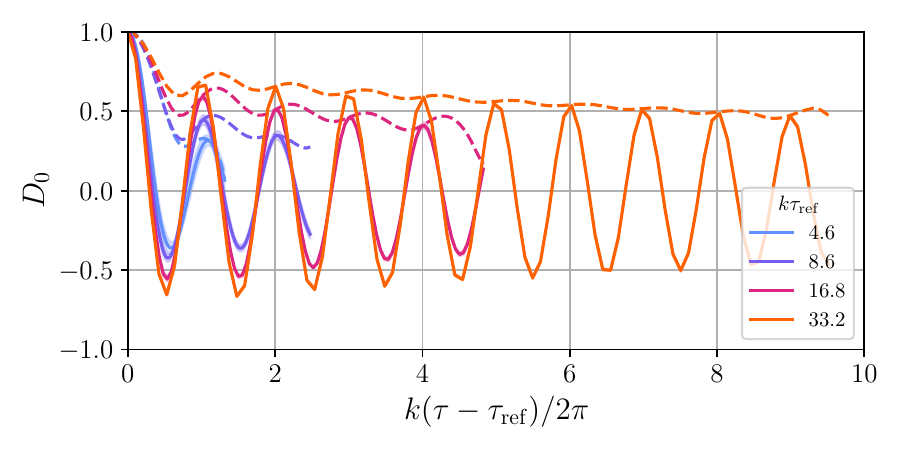}
    \includegraphics[width=0.41\textwidth]{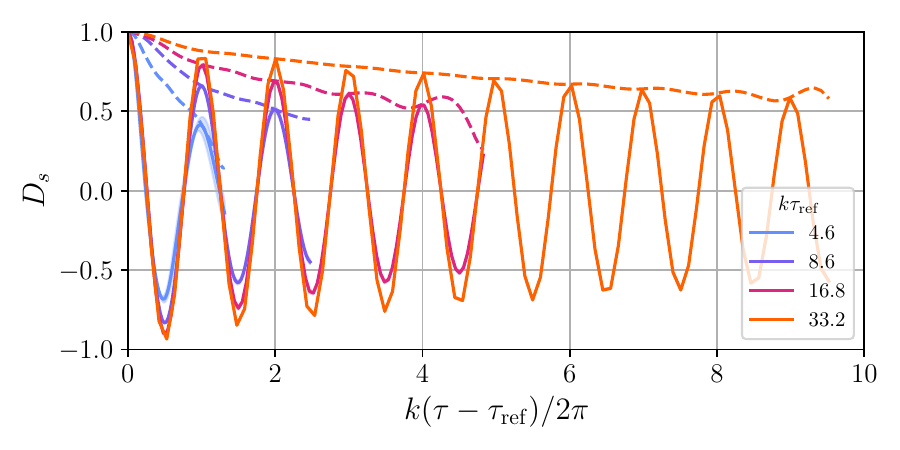}
    \includegraphics[width=0.41\textwidth]{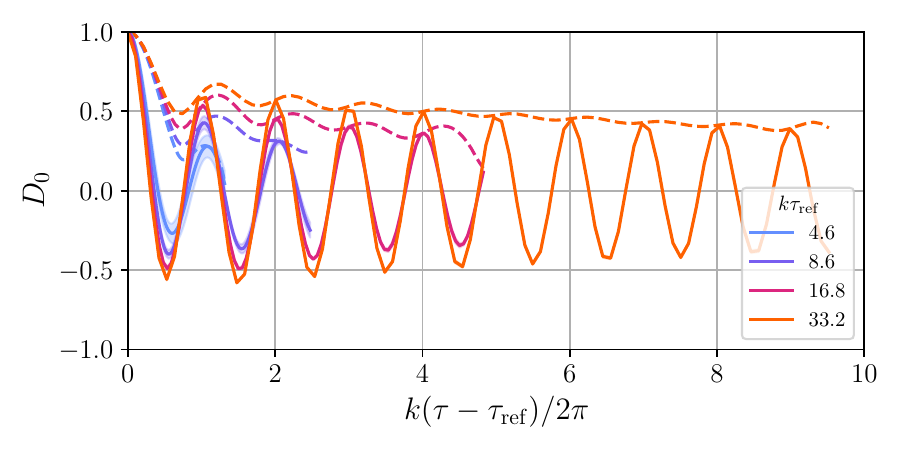}
    \includegraphics[width=0.41\textwidth]{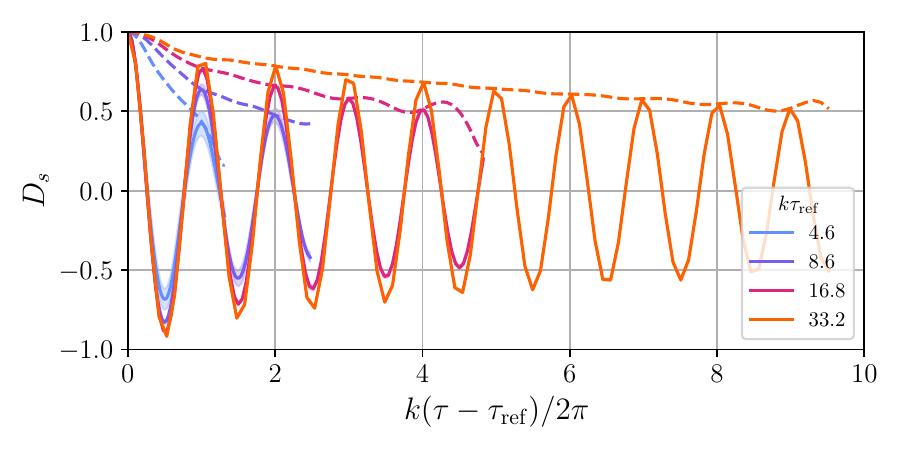}
     \caption{Decoherence functions $D_0$ (left) and $D_s$ (right) for selected $k\tRef$, as defined in Eq.~\eqref{e:DecFunDef} (solid).  Top to bottom: initial field correlation lengths $\l_\phi = 5, 10, 20, 40, 80$.
 The amplitude of the analytic signal is plotted as a dashed line. For $D_\pm$ the prediction of the USM with $\bew = 0.45$ and $\bar{v}=0.52$ is plotted.        \label{f:uetcs_axion}}
 \end{figure*}
\FloatBarrier
\newpage

\section{Decoherence functions 2}
\label{s:DecFun2}
\FloatBarrier
\begin{figure*}[htbp]
    \centering
    \includegraphics[width=0.41\textwidth]{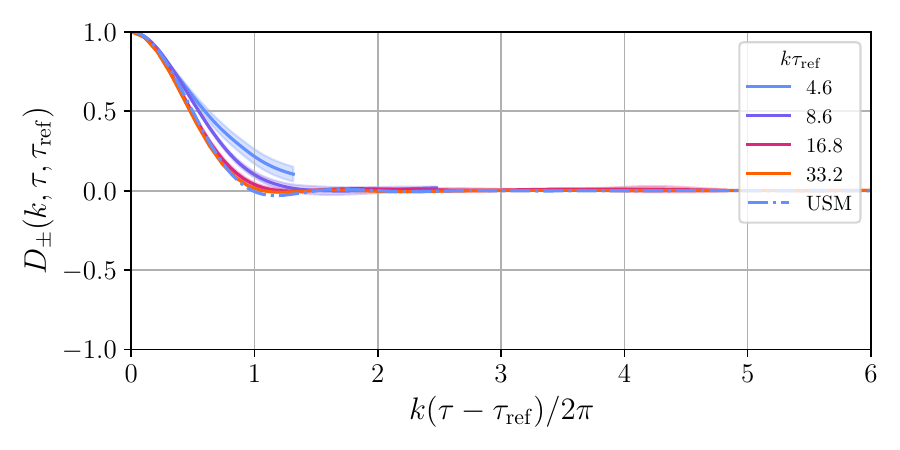}
    \includegraphics[width=0.41\textwidth]{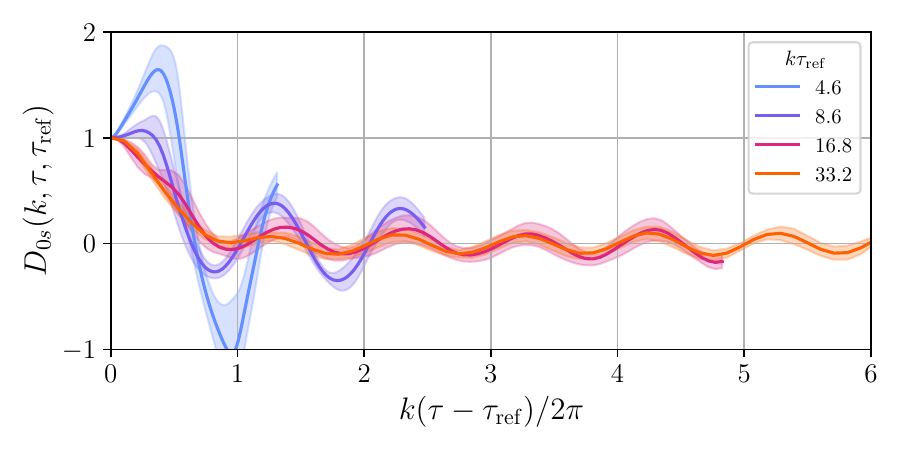}\\
    \includegraphics[width=0.41\textwidth]{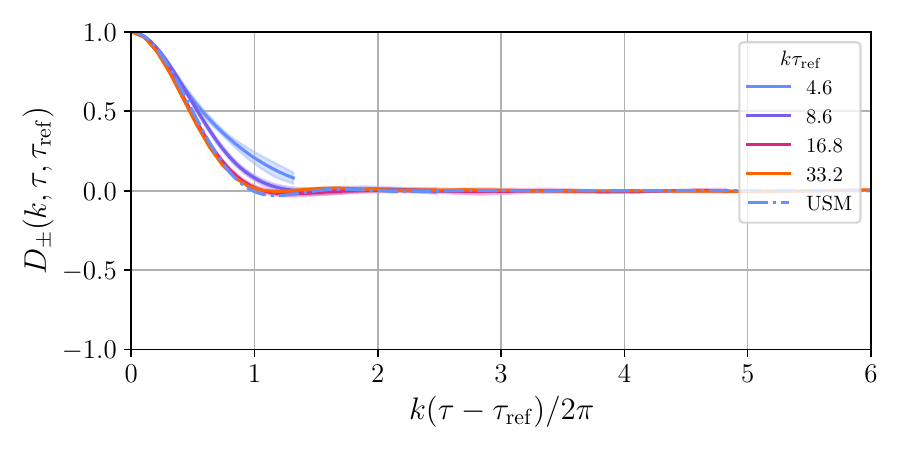}
    \includegraphics[width=0.41\textwidth]{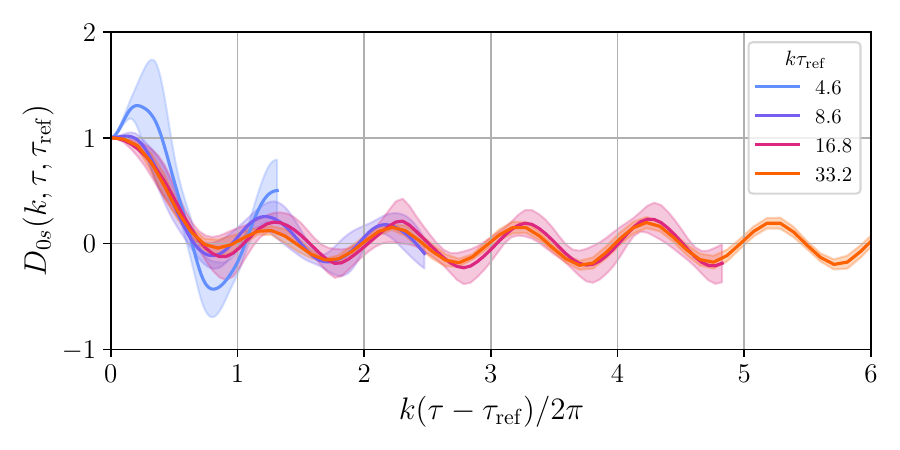}\\
    \includegraphics[width=0.41\textwidth]{decoh_Jp_Jm_sum_lp20usm.pdf}
    \includegraphics[width=0.41\textwidth]{decoh_J0_Js_diff_lp20usm.pdf}\\
    \includegraphics[width=0.41\textwidth]{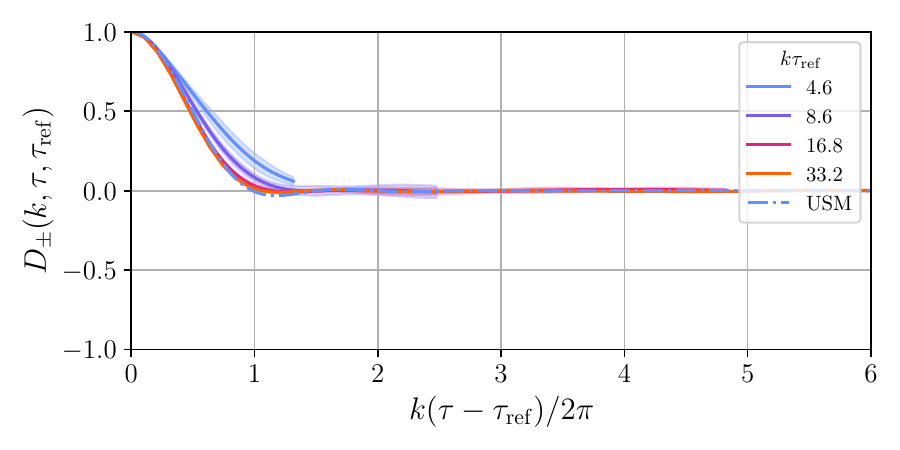}
    \includegraphics[width=0.41\textwidth]{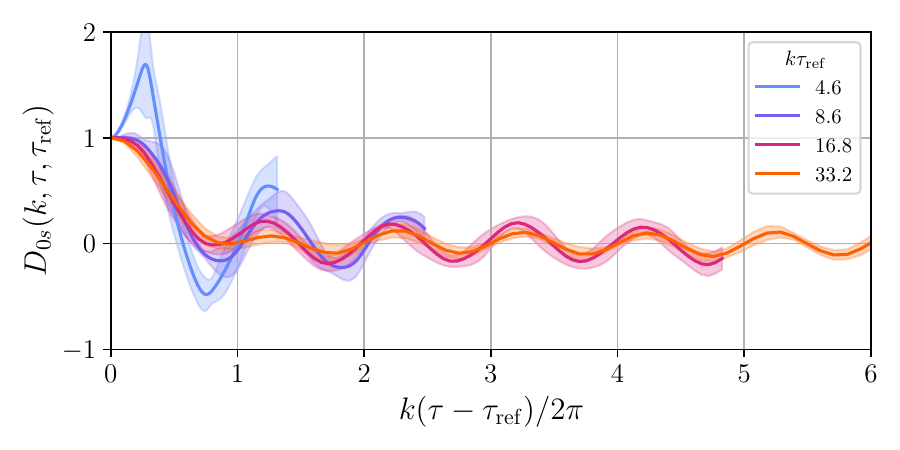}\\
    \includegraphics[width=0.41\textwidth]{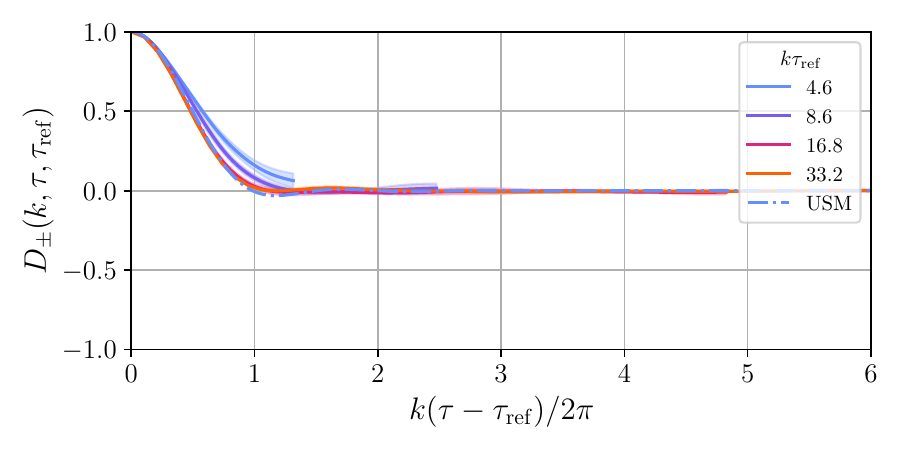}
    \includegraphics[width=0.41\textwidth]{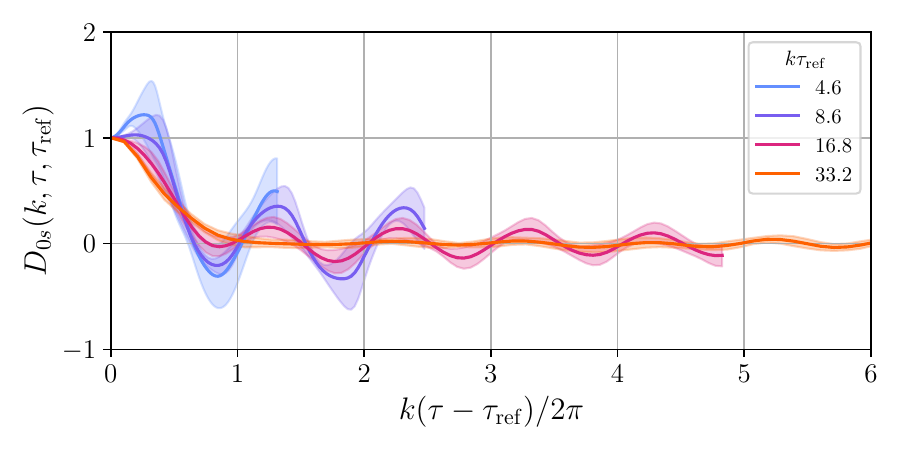}
     \caption{Decoherence functions $D_\pm$ (left) and $D_{0s}$ (right) for selected $k\tRef$, as defined in Eq.~\eqref{e:DecFunDef} (solid).  Top to bottom: initial field correlation lengths $\l_\phi = 5, 10, 20, 40, 80$.
The prediction of the USM with $\bew = 0.45$ and $\bar{v}=0.52$ is plotted as a dot-dashed line.        \label{f:uetcs_string}}
 \end{figure*}
\FloatBarrier
\newpage
\end{widetext}

\bibliography{axion} 
\end{document}